\documentclass[twocolumn,twocolappendix]{aastex63}
\usepackage{amsmath}

\newcommand\kms{km s$^{-1}$}

\newcommand\teff{$T_{eff}$}
\newcommand\logg{$\log g$}

\newcommand\msun{M$_\odot$}

\defcitealias{kounkel2019a}{Paper I}
\defcitealias{kounkel2020}{Paper II}
\usepackage{listings}

\begin{document}
\title{Untangling the Galaxy III: Photometric Search for Pre-main Sequence Stars with Deep Learning}

\author[0000-0003-2401-0097]{Aidan McBride}
\affil{Department of Physics and Astronomy, Western Washington University, 516 High St, Bellingham, WA 98225}
\author[0000-0002-9112-9314]{Ryan Lingg}
\affil{Department of Computer Science, Western Washington University, 516 High St, Bellingham, WA 98225}
\author[0000-0002-5365-1267]{Marina Kounkel}
\affil{Department of Physics and Astronomy, Western Washington University, 516 High St, Bellingham, WA 98225}
\author[0000-0001-6914-7797]{Kevin Covey}
\affil{Department of Physics and Astronomy, Western Washington University, 516 High St, Bellingham, WA 98225}
\author[0000-0002-5537-008X]{Brian Hutchinson}
\affil{Department of Computer Science, Western Washington University, 516 High St, Bellingham, WA 98225}
\affil{Computing and Analytics Division, Pacific Northwest National Laboratory, 902 Battelle Blvd, Richland, WA 99354}
\email{mcbrida5@wwu.edu, linggr@wwu.edu, \\ marina.kounkel@wwu.edu}

\begin{abstract}

A reliable census of pre-main sequence stars with known ages is critical to our understanding of early stellar evolution, but historically there has been difficulty in separating such stars from the field. We present a trained neural network model, Sagitta, that relies on \textit{Gaia} DR2 and 2MASS photometry to identify pre-main sequence stars and to derive their age estimates. Our model successfully recovers populations and stellar properties associated with known star forming regions up to five kpc. Furthermore, it allows for a detailed look at the star-forming history of the solar neighborhood, particularly at age ranges to which we were not previously sensitive. In particular, we observe several bubbles in the distribution of stars, the most notable of which is a ring of stars associated with the Local Bubble, which may have common origins with the Gould's Belt.
\end{abstract}


\section{Introduction}

Historically, the pre-main sequence (PMS) stars that have been the easiest to identify and classify are those that are the youngest and are still in possession of their natal envelopes and/or protoplanetary disks. Such sources could be identified on the basis of large infrared excess, and these dusty young stellar objects (YSOs) have been searched for using a number of all-sky surveys, such as using IRAS, 2MASS, AKARI, and WISE, \citep[e.g.,][]{prusti1992,koenig2012,toth2014,marton2016}. Furthermore, detailed infrared maps of a large number of star forming regions have been constructed with more targeted surveys, such as with Spitzer and Herschel \cite[e.g.,][]{evans2009,megeath2012,fischer2017}.

However, after a star loses its protoplanetary disk, its colors begin to resemble those of much more evolved field stars, making follow-up identification difficult. 


\subsection{\textit{Gaia} DR2 classification of YSOs}

In comparison to previously available techniques (such as spectroscopic follow up measurements, or X-ray emission), the release of \textit{Gaia} DR2 \citep{gaia-collaboration2018} has allowed a revolution in the search and characterization of young stars. Through its unprecedented precision in the measurements of parallax and proper motions, as well as its remarkable photometric quality, two new techniques became available to the community. First is the phase space clustering. Young stars form in the dynamically cold molecular clouds. These clouds commonly form anywhere from a few hundred to several thousands of stars in relatively close proximity and typically have low velocity dispersion. Thus, through searching for an overdensity in the position and velocity space it is possible to identify a young comoving group of stars. Such clustering has been employed both systematically across the entire Galactic Disk \citep[hereafter Paper I]{kounkel2019a} as well as to better constrain the membership of individual star forming regions \citep[e.g.,][]{kounkel2018a,galli2019,damiani2019}.

But, clustering requires that all of the stars in a comoving group retain the group's characteristic velocity in order to be identifiable. As these groups slowly dissolve into the Galaxy and lose coherence, an increasingly small fraction of them is recoverable. Indeed, even 1 Myr populations have some stars that have already been ejected from the clusters they inhabit \citep{mcbride2019,schoettler2020,farias2020}. Searching for such high velocity YSOs may be of interest to better characterize intracluster dynamics, but it is impossible to do through clustering. Furthermore, some young populations may be too small or too diffuse to robustly identify them with clustering at all.

The second method that \textit{Gaia} DR2 made possible is through examining the position of the stars on the HR diagram. YSOs are overluminous compared to their main sequence counterparts due to their still-inflated radii, and most are fainter and cooler than the red giants. If the distance is known accurately, it is possible to resolve the degeneracy between the nearby dwarfs and distant giants, and thus separate YSOs from more evolved stars.

Such an approach is rather simple to use when attempting to identify YSOs in a single star forming region with a known position on the sky and known distance, particularly if this region is only a few Myr old. In this case, the low mass PMS stars can be cleanly separated from the low mass main sequence counterparts, and it is possible to determine color cuts that would prevent contamination from red giant stars and or high mass main sequence stars. However, extending this to multiple populations that have different ages, distances, or extinctions is difficult. \citet{zari2018} performed a selection of YSOs in the \textit{Gaia} DR2 catalog that are consistent with being younger than the 20 Myr isochrone and that are located within 500 pc. Although the catalog effectively identifies sources throughout known nearby star forming regions, at larger distances the contamination does become significant. Thus, it is necessary to reevaluate the selection criteria if one wishes to reliably extend the catalog beyond 500 pc.

Machine learning, and, in particular, the use of neural networks, is a method that facilitates the search for complex correlations in large volumes of data. Machine learning classifiers have been used to search for young stars in a number of works, from searching for infrared excess \citep{marton2016,marton2019,chiu2020}, to using H$\alpha$ in conjunction with photometry to search for Herbig Ae/Be stars \citep{vioque2020}, to using the optical Hubble Space Telescope colors to give a probabilistic assessment of young stars in the Magellanic clouds \citep{ksoll2018}.

\subsection{Derivation of stellar ages}

Beyond classifying a star as young, extracting its properties (such as its age) can be a challenge. The way this is commonly done is through comparison of photometric colors (or age-sensitive spectroscopic features) to theoretical isochrones. While this practice has a long standing history \citep[e.g.,][]{cohen1979,greene1995,covey2010,da-rio2012}, this process is not trivial \citep[e.g.,][]{olney2020}, especially with the inconsistencies between young stars discussed earlier.

The first issue lies with the isochrones themselves. Over the years, a number of different stellar evolution models have been developed \citep[e.g.,][]{dantona1994,baraffe1998,siess2000,baraffe2015,choi2016}, and the ones that seem to have gained the most wide-spread usage in the community in the recent years are the PARSEC isorchrones \citep{marigo2017}. Due to slightly different assumptions regarding the underlying stellar physics, these models produce distinct isochrones and evolutionary tracks, and thus produce different age and mass estimates even when applied to the same stellar population \citep{hillenbrand2008}. However, no isochrones offer a perfect match to the data, especially for the low mass stars, and they may result in up to 50\% systematic deviation on the measured property, such as mass \citep{braun2021}.

For example, the M dwarfs appear to be overinflated compared to what the isochrones would suggest even in one of the best studied open clusters, the Pleiades \citep{jackson2018}. Thus, attempting to estimate their age through isochrone fitting would yield a systematically younger age than what is appropriate for the cluster. Similarly, a cluster ``birthline'' (i.e., the region of the parameter space that would correspond to a 0 Myr population) is ill-defined, such that in the young populations that are just a few Myr old, the higher mass stars appear to be systematically older than their low mass counterparts \citep{leebook,herczeg2015}. The presence of a protoplanetary disk further alters the photometry in such a way that makes it difficult to place a star onto the isochrones correctly.

The second issue lies with the physical properties of YSOs. They tend to be more complex than even the most advanced stellar evolution models can account for. Due to being mostly convective, a large fraction of their photospheres are covered with spots, resulting in a mismatch in the effective temperature \teff\ and the expected mass of the star. In addition to this, there is also the unavoidable issue of stellar multiplicity. Among the main sequence stars, the binary sequence can be clearly seen and separated from the single stars. However, in young populations with an intrinsic age spread of even a few Myr, the visual binaries may be inferred to be systematically younger than what is appropriate \citep[e.g.,][]{bouma2020}.

Third, there is an issue of the self-consistency of the fitting process. Even for the same populations of stars, using the same set of isochrones, but focusing on somewhat different features and using a different interpolation method, it is possible to produce an age estimate that is somewhat inconsistent between various works. Taking Orion as an example, particularly the region near 25 Ori cluster, while different authors were able to estimate roughly comparable ages \citep{kounkel2018a,briceno2019,zari2019}, some infer ages that are systematically older \citep{kos2019}. Such differences would only compound when comparing ages of individual stars.

To compensate for some of these issues, data driven models may perform better compared to the theoretical isochrone fitting. With distilling the previously existing estimates of ages for stars (both in the cases when the age can be assigned to all stars in a cluster on a population level, and in cases where measurements for individual stars are available), it is possible to construct a neural network that would assign ages to stars. While the predictions it would generate could only be as accurate as input data on which the network is trained on, through leveraging the ages derived by various methods, the systematic differences can be significantly reduced. This includes systematic differences between low and high mass stars, as well as differences between various stellar evolution models, resulting in a more self-consistent interpolation that is more faithful to the data.

In this work, we present a tiered deep learning model, that we refer to as Sagitta. This model identifies the PMS stars using \textit{Gaia} DR2 and 2 Micron All-Sky Survey (2MASS) photometry and astrometry and estimates the ages of these stars. In Section \ref{sec:data}, we describe the data that were used to train Sagitta, as well as the data on which we perform the evaluation. In Section \ref{sec:model}, we detail the process of constructing and training the model. In Section \ref{sec:validation}, we test the results benchmarked against other catalogs of PMS stars, as well as known star forming regions. In Section \ref{sec:discussion}, we discuss the features in the data, such as their implications on the star forming history of the solar neighborhood and the origin of the Gould's belt. Finally, we conclude in Section \ref{sec:concl}.

\section{Data}\label{sec:data}
Neural networks are reliant on labelled training data in order to generate a model able to generalize; that is, make accurate predictions for new sources. The bulk of the training data was obtained from \citet[hereafter Paper II]{kounkel2020}. This catalog contains almost 1 million stars that have been clustered together into more than 8000 different moving groups, extending up to parallax limit of $\pi>0.2$ mas. Average ages (ranging from $<1$ Myr to $1$ Gyr) has been inferred through the photometry of each moving group's members using the Auriga neural network, which is described in full in \citetalias{kounkel2020}.

To train its sister network, Sagitta, presented in this paper, we ingest 8 parameters for each star in the sample. Six of them are the photometries in different bandpasses, namely $G$, $G_{BP}$, $G_{RP}$, $J$, $H$, and $K_s$, with the 2MASS passbands queried using the precomputed \textit{Gaia} DR2 crossmatch table. We also rely on parallax, as well as an approximation of extinction $A_V$. The latter is not necessarily the intrinsic extinction the star may have (e.g., it does not account for the presence of a protoplanetary disk), but rather, it is an estimate along the line of sight to the star based on its galactic positions $l$, $b$, and $\pi$ (Section \ref{sec:av}).
 
The tasks of both classification (i.e., blind selection of PMS stars) and regression (i.e., interpolation of the ages for the identified young stars) are reliant on the different features of this catalog; furthermore, they require different augmentation procedures to improve the homogeneity of coverage. 

\subsection{Classification training sample}\label{sec:classsample}

\begin{figure} 
\epsscale{1.2}
\plotone{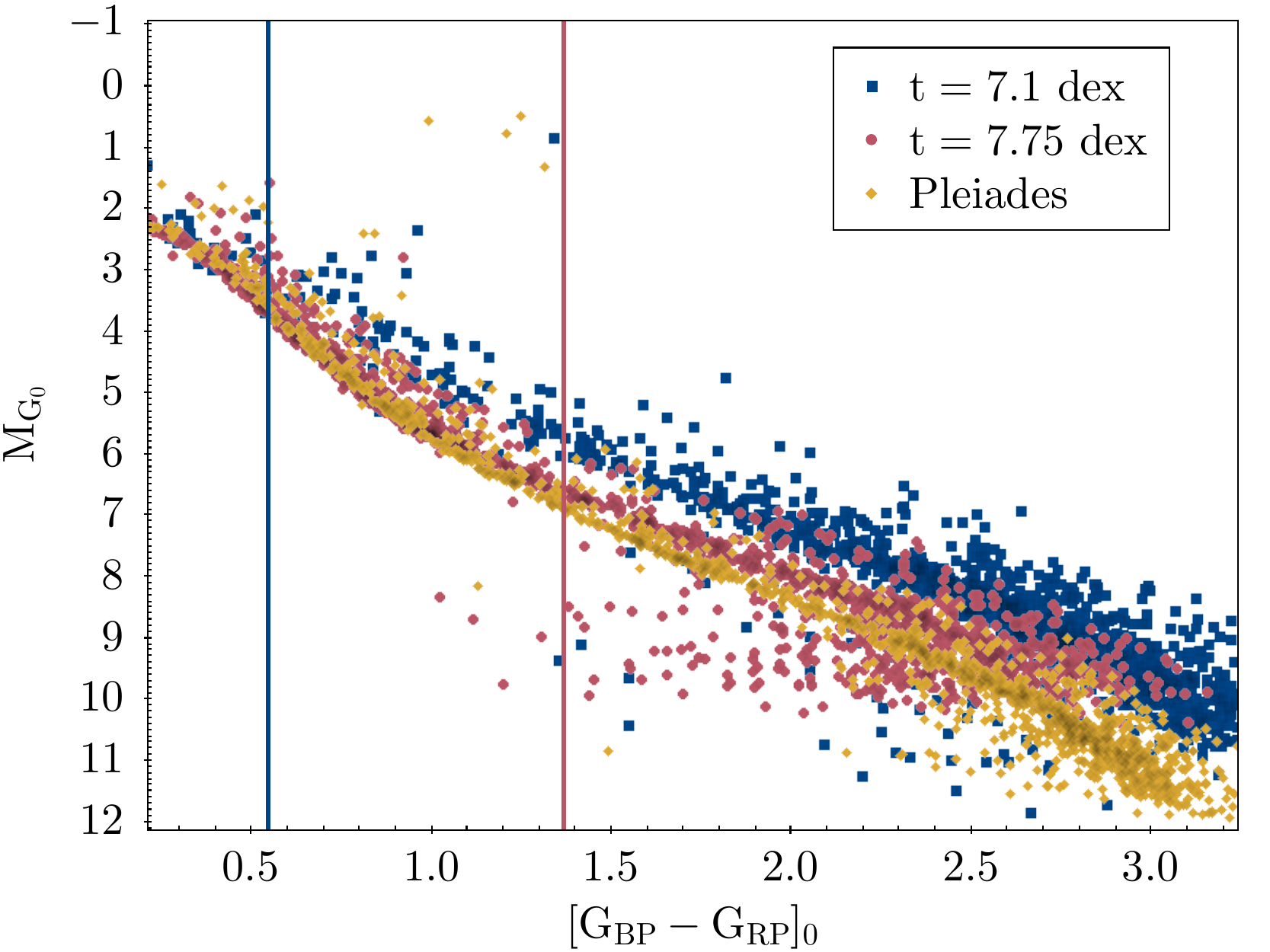}
\caption{A demonstration of a color-dependent cut-off for pre-main sequence stars as a function of age in comparison to the Pleiades. Note the presence of the binary sequence in each of the populations. The sources below the Pleiades tend to have poor photometry at these bands, and may be recovered in other bands. The fraction of such sources is relatively low, and it has no significant impact on the subsequent analysis. \label{fig:ageinterp}}
\end{figure}

The first necessary step is to select the PMS stars in the catalog from \citetalias{kounkel2020}. Massive YSOs will reach the main sequence sooner than the low mass ones: e.g., OB stars will be born directly on the main sequence, whereas M dwarfs may take as long as 100 Myr to reach it. As main sequence stars of similar mass are difficult to distinguish from one another, regardless of their age, a simple cut in age is not sufficient to reliably separate PMS stars. Rather, we compared the HR diagrams of populations in each 0.1 dex age bin to that of the Pleiades to determine the most massive/bluest star that can still be considered pre-main sequence. We then assign the extinction-corrected $[G_{BP}-G_{RP}]_0$ color corresponding to such star as the location of the turn-off for that age bin.  The extinction was estimated on a population level in \citetalias{kounkel2020}. We then interpolated across all the age bins to obtain the cut of
\[[G_{BP}-G_{RP}]_0>49.3686-14.3347\times t+1.05042\times t^2\]
\noindent where $t$ is the age of the population in dex. This relation is valid for $t<7.85$ dex (Figure \ref{fig:ageinterp}). Using this age-dependent relation to identify the color at which the MS-PMS transition should occurs at a given age, we assign a preliminary PMS designation to all low-mass stars redward of the critical color in each population, according to its age.

Furthermore, we imposed a cut of
\[M_{G_0}>2.8\times[G_{BP}-G_{RP}]_0-2\]
\noindent to separate possible contamination from red giants.

A total of 62,484 sources (about 6\%) of the catalog in \citetalias{kounkel2020} meet these criteria, and are identified as the training sample of likely PMS stars for Sagitta. The remaining sources are considered to be evolved, i.e., main sequence or post-main sequence sources. All sources receive a binary numerical flag to indicate their evolutionary state, with 1 assigned to all sources satisfying the YSO criteria and 0 assigned to the remaining evolved sources.

While the catalog from \citetalias{kounkel2020} offers a comprehensive coverage of PMS stars, it is incomplete at the oldest age bins for stars that are more evolved. All moving groups eventually dissolve into the Galaxy, and after this happens, they are no longer recoverable through clustering. Therefore, the distribution and properties of the red giants that are found in the field are not represented by the stars in the \citetalias{kounkel2020} catalog. This introduces a bias that may result in reddened red giants that scatter into PMS parameter space (or even located well above it) to erroneously be classified as PMS, as the classifier would not know enough about red giants to reject them. To correct for this bias we took 3 million randomly selected stars from \textit{Gaia} DR2 (using the \texttt{random\_index}), with the same quality cuts as the ones specified in Section \ref{sec:testsample} to better train the network on the colors of older field stars to be able to discriminate against them. Any stars that happened to also be included in the clustered catalog were excluded from this random selection. We assume that the remaining stars in this random catalog can all be considered to be evolved. If any true PMS stars remained in this random sample, their fraction, as well as the overall number is expected to be so small as not to make a substantial difference for the classifier, as PMS are comparatively rare relative to the older stellar classes. 

The spatial position of the stars (i.e., $l$ and $b$) was not used directly in the training sample to avoid introducing spatial bias. However, some positions can be inferred through the combination of $\pi$ and $A_V$. To prevent spatial bias, we used augmentation, i.e., the process of making the training catalog larger through artificially modifying it to reduce various biases. Augmentation is also beneficial to improve performance beyond 1 kpc, where the training sample is highly incomplete, as few PMS stars at those distances are found in the \citetalias{kounkel2020} catalog due to the sensitivity limits. To augment the sample, we first modified the photometry of the real stars (both the clustered and random catalogs) to simulate the effects of them being found at different distances (up to 5 kpc) and with different $A_V$ (up to 10 mag), both of which were randomly generated. Both of the effects of distance and extinction combined often exceeded the typical magnitude at which a source could be reliably detected with Gaia, especially for the low mass stars. Then to complete the process, random errors in flux were drawn from a normal distribution and were added to all passbands. The extinction coefficients for different passbands were taken from the web interface of the PARSEC isochrones \citep{marigo2017} as they cover a wide range of stellar masses and ages, with various passbands applied to the photometry. Each real star was drawn multiple times, with this multiplier treated as a hyperparameter in the model for each subset (Section \ref{sec:classifier}). These `synthetic' stars were then passed on to the classifier alongside the photometry of the real stars to improve generalization.

\subsection{Regression training sample}\label{sec:regressionSample}

The training sample for the regression to estimate ages was limited only to the stars identified as PMS stars in the catalog from \citetalias{kounkel2020} from above, excluding the sources considered to be evolved. This was performed because retaining them in the sample could introduce considerable biases to the PMS stars since the main sequence stars of similar mass share the same parameter space in fluxes.

Most of the PMS stars in \citetalias{kounkel2019a} and \citetalias{kounkel2020} are found in stellar strings (i.e., extended populations spanning several tens or even hundreds of pc). Each individual region can sustain star formation for up to $\sim$10 Myr. Such a duration is hardly noticeable within the errors of the ages assigned to populations older than 100 Myr (e.g, the colors and fluxes of a 90 Myr and 100 Myr stars are not going to be very different). However, the differences in fluxes between the youngest and the oldest generation of stars in the same region are more pronounced in regions that are still forming. Therefore, assigning a single age to all the stars in a single string, if there is any sort of underlying age gradient (as is the case in Orion or Sco Cen, for example), can introduce some biases. In estimating ages for the moving groups in \citetalias{kounkel2020}, Auriga preferentially considers the oldest stars in a region, overestimating the ages of the younger stars. Without correcting for this effect in the training sample, Sagitta cannot accurately estimate ages of stars younger than a few Myr.

Thus, to compensate, several steps were taken. First, many of the strings can be subdivided into populations more homogeneous in age. In \citetalias{kounkel2019a}, some of the strings had to be manually assembled from smaller subgroups based on the coherence in phase space. Furthermore, the HR diagrams of the strings were visually examined in trying to identify populations of different age sequences in close proximity (e.g., as is the case with $\rho$ Oph and Upper Sco), and some attempt was made to separate them into subgroups. We used Auriga on these subgroups to generate a somewhat more granular distribution of ages in the training sample.

\begin{deluxetable}{ccc}
\tablecaption{Ages different from \citetalias{kounkel2020} assigned to young populations in the training set. \label{tab:ageresets}}
\tabletypesize{\scriptsize}
\tablewidth{\linewidth}
\tablehead{
\colhead{Region} & \colhead{Source} &  \colhead{Age (Myr)}}
\startdata
Ara OB1a & \cite{Wolk2008} & 3 \\
Carina: Tr 16 & \cite{Smith2008} & 3\\
Cep OB2a & \cite{kun2008b} & 7 \\
Cep OB2b & \cite{kun2008b}  & 3.7 \\
Cep OB3b & \cite{kun2008b} & 4 \\ 
Cep OB6 & \cite{kun2008b}  & 38 \\
Chamaeleon & \cite{luhman2008cha} & 2 \\  
CrA & \cite{Neuhauser2008} & 6 \\
Cyg OB1 & \cite{Reipurth2008cygnus} & 7.5 \\
Cyg OB2 & \cite{Reipurth2008cygnus} & 5 \\
Cyg OB3 & \cite{Reipurth2008cygnus} & 8.3 \\
IC 348 & \cite{bally2008a} & 2\\
IC 1396 & \cite{Walawender2008} & 1 \\
IC 5146 \& W4 & \cite{herbig2008} & 1 \\
LK H$\alpha$ 101 & \cite{Andrews2008} & 0.5 \\
Lagoon Nebula & \cite{tothill2008} & 1 \\
Lower Cen/Crux & \cite{preibisch2008} & 16\\
Lupus & \cite{Comeron2008} & 3.2 \\
Monoceros & \cite{Carpenter2008} & 6 \\
NGC 1333 & \cite{Walawender2008} & 1 \\
NGC 2264 & \cite{dahm2008} &  3 \\
NGC 6383 & \cite{Rauw2008} & 2 \\
NGC 6604 & \cite{reipurth2008serob2} & 4.5 \\
NGC 6823 & \cite{prato2008} & 5 \\
Per OB2 & \cite{bally2008a} &  6 \\
Rosette Nebula & \cite{Roman-Zuniga2008} & 3 \\
Serpens & \cite{herczeg2019} & 3 \\
Sh 2-234-Stock 8 & \cite{Reipurth2008anticenter} & 2 \\
Taurus/Auriga & \cite{kenyon2008} & 1 \\
Upper Cen/Lup & \cite{preibisch2008} & 17 \\
Upper Sco & \cite{preibisch2008} & 5 \\
$\rho$ Oph & \cite{Wilking2008} & 0.3 \\
\enddata
\end{deluxetable}

Second, to achieve a better consistency with the ages of well-studied star forming regions, we identified the moving groups from \citetalias{kounkel2020} that correspond to the populations listed in the Handbook of Star Forming Regions \citep{reipurth2008a,reipurth2008b}, and assigned them the more appropriate ages that are reported in the Handbook (see Table
\ref{tab:ageresets} for group identifiers and corresponding ages).

Third, individual ages of some young stars are available in the literature. Namely, we've included the sources in the catalogs from \citet{palla2002,kun2009,delgado2011,lopez-marti2013,fang2013,kumar2014,herczeg2014,getman2014a,erickson2015,azimlu2015,fang2017,suarez2017,prisinzano2018,panwar2018} that have reliable parallaxes and that meet the age-dependent criteria for a source to be identified as PMS from Section \ref{sec:classsample}. This added 6,248 stars. 

Finally, the ages have been reevaluated in the Orion Complex. As this region singularly contains the largest number of stars out of any other population in the catalog, and it contains stars that span in age from $<$1 to 12 Myr, it is of particular importance for the training sample from Orion to achieve good accuracy across different age bins. The region does have a rather complicated morphology that could not be broken into subgroups in \citetalias{kounkel2019a}. However, a different, more granular analysis was performed by \citet{kounkel2018a}, using hierarchical clustering of the 6-dimensional phase space to segment the Complex into 190 different groups, and an average age was estimated for each group. The sample in that work, however, is limited to the stars almost a magnitude brighter than the sample in \citetalias{kounkel2020}. Excluding the fainter stars does introduce a bias in the training process. Thus, to shuffle these low mass stars into the most appropriate group, we created a simple fully connected neural network that has one layer with 300 neurons, taking in $\alpha$, $\delta$, $\mu_\alpha$, $\mu_\delta$, and $\pi$, and outputting a probability of belonging to each one of the 190 groups. The members of the Orion Complex from \citetalias{kounkel2020} that were not included in the catalog from  \citet{kounkel2018a} were then assigned to the group with the highest probability. Then, each star was given a label of the average age of the group it was in.

In a similar fashion to the classification training sample, the age regression training sample was augmented to help negate some of the potential spatial biases. The synthetic samples included were generated from the real data by simulating the effects of changing the distance and extinction.

\subsection{Evaluation sample}\label{sec:testsample}

To evaluate the performance of the classification and regression models, we downloaded the \textit{Gaia} DR2 data that satisfied the following quality criteria:
\[\texttt{phot\_bp\_rp\_excess\_factor}>1+0.015\times\texttt{bp\_rp}^2\]
\[\texttt{phot\_bp\_rp\_excess\_factor}<1.3+0.06\times\texttt{bp\_rp}^2\]
\[\texttt{ruwe}<1.4\]
\[\texttt{phot\_g\_mean\_flux\_over\_error} > 10\]
\[\texttt{phot\_bp\_mean\_flux\_over\_error} > 10\]
\[\texttt{phot\_rp\_mean\_flux\_over\_error} > 10\]
\[\texttt{parallax}>0.2\]
\[\texttt{parallax}/\texttt{parallax\_error}>10~\textrm{or}~ \texttt{parallax\_error}<0.1\]

These criteria are adapted from the quality cuts by \citet{lindegren2018}. They also ensure that the sources are nearby enough to have a complete coverage of the volume of space over which low mass PMS stars are detectable. All three fluxes from \textit{Gaia} DR2 are required to be detected with high signal to noise. 2MASS fluxes can be undetected, in which case they are set to a constant outside of their maximum range (See Section \ref{sec:limits}).

The resulting sample consists of $\sim$139.3 million stars. We note that the models presented in this paper can be expected to work even if some of the selection restrictions are relaxed, although it is not done in this paper for the purposes of the computational expediency.

\begin{figure*}
    \gridline{\fig{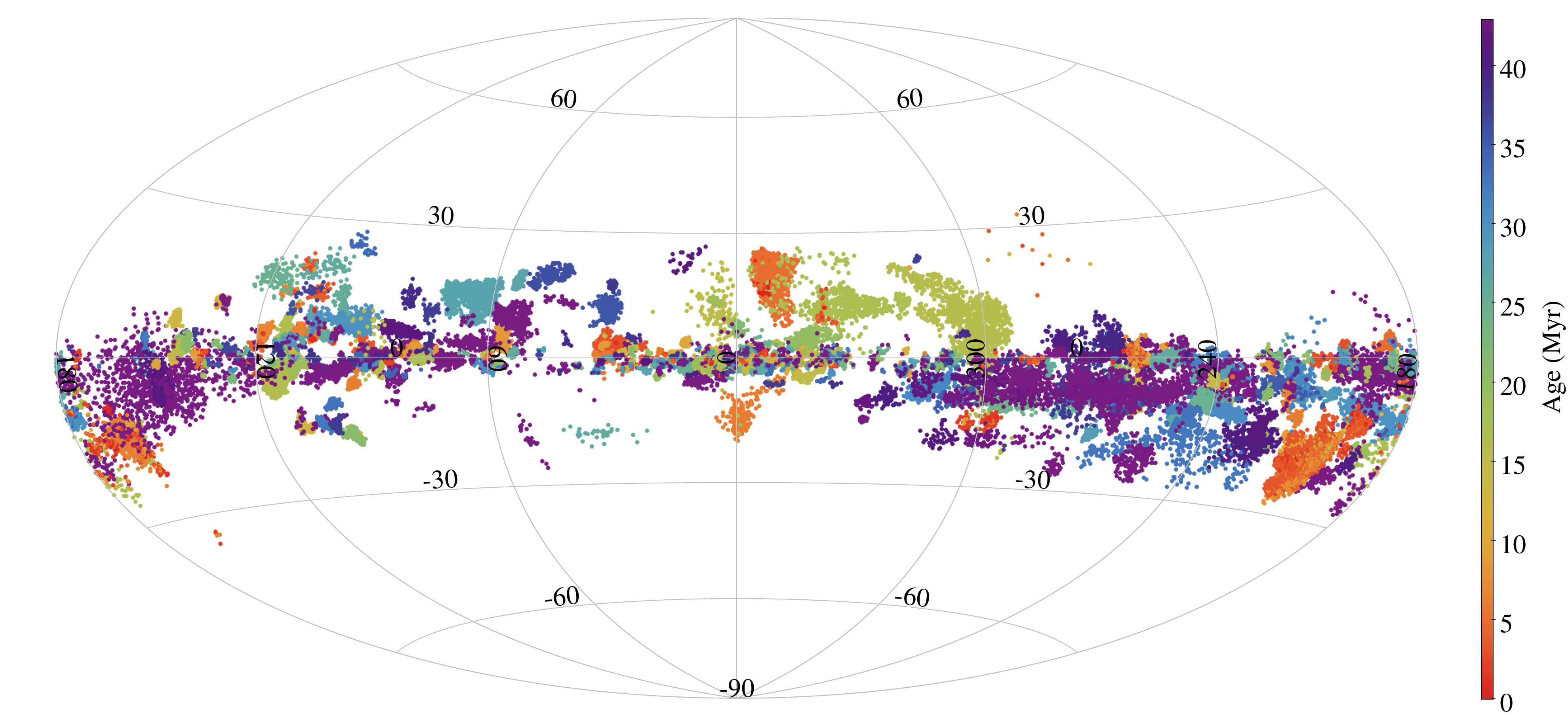}{\textwidth}{}}
    \gridline{\fig{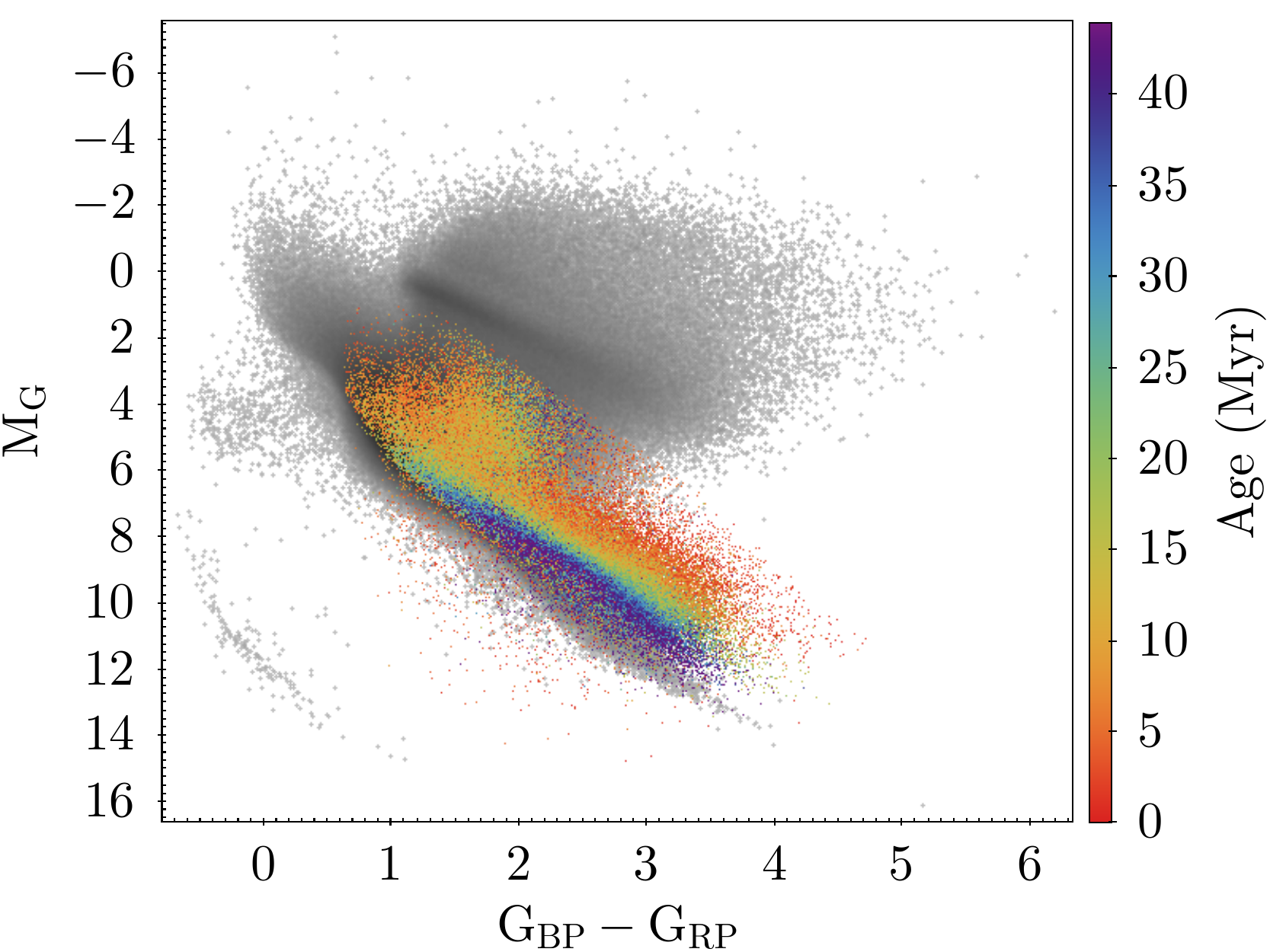}{.5\textwidth}{}
    \fig{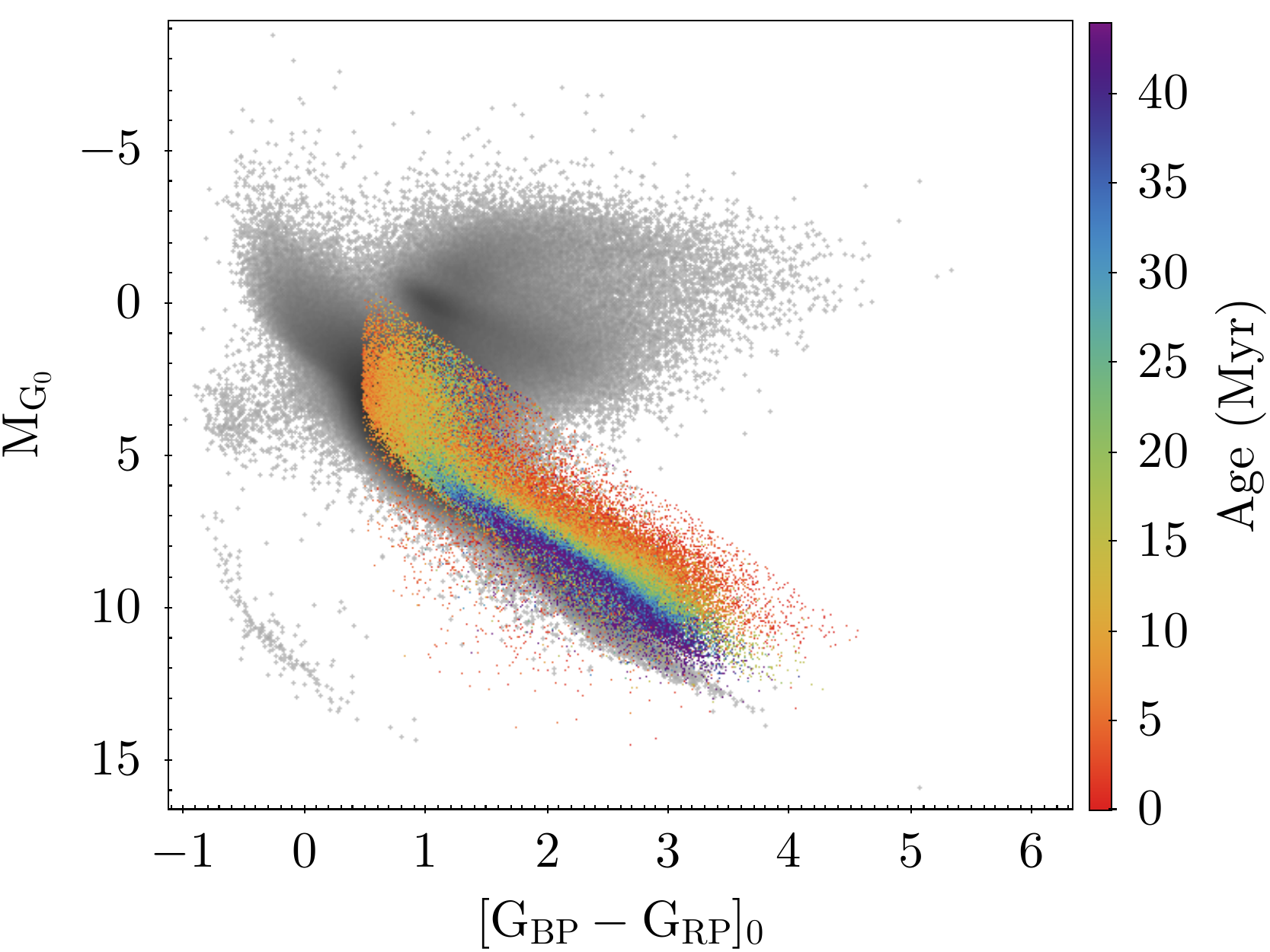}{.5\textwidth}{}}
    \caption{Top: The distribution of PMS stars on the sky that were used as part of the training process. The sources are color coded by the age assigned to them. Bottom: HR diagram of stars in the input dataset, uncorrected (left) and corrected (right) for extinction. Grey points represent evolved stars used in the classifier training, while color coded points represent pre-main sequence stars weighted by isochrone age, used to train the age regressor model. \label{fig:dataset}}
\end{figure*}

\section{Sagitta neural network}\label{sec:model}

\subsection{Network Architecture}

\begin{figure} 
\epsscale{0.625}
\plotone{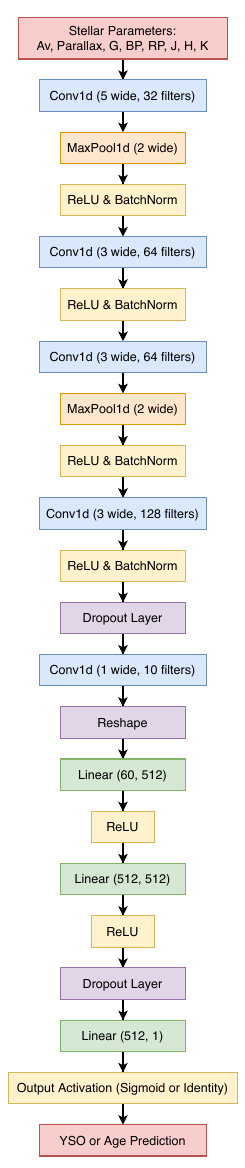}
\caption{Architecture of the model used by Sagitta. This architecture is used by each of three networks in this work, independently of each other. \label{fig:model}}
\end{figure}

For flexibility and effectiveness in model structure, we implemented a set of three convolutional neural networks (CNNs) where each network serves a distinct purpose. The first network generates the extinction map based on galactic position. The second network assigns each star a probability of being pre-main sequence based on its photometry (using the training sample described in Section \ref{sec:classsample}). And the third network approximates the ages of PMS stars, typically ranging from $<1$ Myr to $>40$ Myr (using the training sample described in Section \ref{sec:regressionSample}).

All three networks within Sagitta share a common architecture adapted from the MNIST configuration of an online repository of models\footnote{\url{https://github.com/eladhoffer/convNet.pytorch/blob/master/models/mnist.py}}, which was further adapted in \citetalias{kounkel2020} for Auriga. However, there are important  differences between these models. The aforementioned examples use 2-dimensional arrays as inputs (e.g., for Auriga the inputs are the the fluxes across various bands for each star in a cluster). In the case of Sagitta though, the input is only a 1-dimensional array representing parameters of a single star, so the operational dimensions (such as convolution and pooling) in the network have been reduced.

The networks were implemented using PyTorch \citep{pytorch}. The model's architecture broadly consists of three main segments (Figure \ref{fig:model}). The first segment is made up of of four 1-dimensional convolutional layers that gradually increase the number of channels to a maximum size of 128. In this segment, only the first and third convolutional layers are followed by max-pooling layers, but each of the convolutional layers or convolutional and max-pooling layer pairs is followed by ReLU and Batch-Normalization operations. The second segment consists of a convolution layer that reduces the number of channels from the output of the first segment followed by a data reshaping operation that transforms the convolution layer's output into a 1-dimensional list (width $\times$  channels $\rightarrow$ width only). Once reshaped, the data are then fed into the third segment which consists of three fully connected linear layers where each layer is followed by a ReLU operation. These final layers expand and then contract the processed data down to a single scalar output value. For the $A_V$ estimation and age estimation models the output value is kept as is (i.e., linear output activation), however in the case of the YSO classifier model this final value is then fed into a logistical sigmoid function to bound the output probabilities between 0 and 1.

\subsection{Data Handling}\label{sec:limits}

In the process of passing the data through the model, it is beneficial to first normalize all of the input parameters to a similar range with mean close to zero. This creates a more comparable dispersion of input parameter magnitudes and mitigates potential issues with numerical stability or inherently biasing any input because of its original scaling. Thus, all the parameters for both classification and regression were linearly scaled to the range of $[-1,1]$ based on the lower and upper bounds specified in Table \ref{tab:normvals}. Although the overall distribution of either parallaxes or fluxes is not Gaussian, thus resulting in a skewed distribution that is dominated by values towards one of the ends of the normalization, the overall bounds were considered to be sufficiently effective.

\begin{deluxetable}{ccc}
\tablecaption{Normalization constants used in training. \label{tab:normvals}}
\tabletypesize{\scriptsize}
\tablewidth{\linewidth}
\tablehead{
\colhead{Parameter} & \colhead{Lower} &  \colhead{Upper}}
\startdata
$\pi$ (mas) & 0 & 29 \\
$A_V$ (mag) & 0 & 5 \\
$G$ (mag) & 4 & 20 \\
$G_{BP}$ (mag) & 0 & 21 \\
$G_{RP}$ (mag) & 0 & 19 \\
$J$ (mag) & 0 & 18 \\
$H$ (mag) & 0 & 18 \\
$K$ (mag) & 0 & 18 \\
Age (dex) & 6 & 8 \\
\enddata
\end{deluxetable}

A number of sources in the catalog may have one or more fluxes missing (most commonly in the 2MASS bands). The non-detection usually carries meaningful information, as it shows that the source is fainter than the detection limit, and limiting the catalog to only the sources for which complete data across the different bands are available would not be optimal. Nonetheless, neural networks are unable to handle null values as inputs. 

Thus, to allow the inclusion of sources with incomplete data, we set the missing values of both the training and evaluation set to the upper limit specified in Table \ref{tab:normvals}. These upper limits are typically somewhat fainter than the detection limit in each band, allowing the network to learn to give these fluxes an appropriate weight.

Another aspect that was considered when structuring the data was the exact ordering of the stellar input parameters given to the YSO Classifier and Age Regressor networks. Convolutional layers in a CNN operate with the use of a sliding filter component, wherein only data spatially close enough together to fit inside the filter can be used to directly detect patterns in that layer. The input features were ordered as $A_V$, $\pi$, $G$, $G_{BP}$, $G_{RP}$, $J$, $H$, and $K$ to preserve the rough ordering of the bandpasses with the wavelength, and to keep $\pi$ and $G$ adjacent as they are measured in the same dataset, and attaching it at the end would have resulted with $\pi$ being associated with occasionally incomplete 2MASS data. Similarly, $A_V$ has the strongest effect in the optical portion of the spectrum. However, as the initial width of the sliding filter is five elements being convolved together, the order should not have a very strong effect on the final results.
Given that the features are not spatial in the traditional sense, we also tried using a fully connected deep neural network instead of the CNN, but found it to perform somewhat worse.

\subsection{$A_V$ estimate}\label{sec:av}

In order to help with differentiating the PMS stars from reddened massive main sequence stars and red giants that scatter into the parameter space PMS stars inhabit, one parameter that can help is an estimate of $A_V$ along the line of sight.

This estimate was obtained from the neural network used to generate the completeness map in \citetalias{kounkel2020}. The model was not modified in any way; rather, it is ported into Sagitta directly as the first step.

The $A_V$ estimator uses the same architecture as the classifier and the age regressor. It was trained on 3 million randomly chosen stars from \textit{Gaia} DR2 that had the same quality constraints as the ones imposed in \ref{sec:testsample} and with measured $A_G$ reported in the catalog \citep{andrae2018}.

The network used $l$, $b$, and $\pi$ to predict $A_V$ \citep[scaled from $A_G$ by a factor of 0.859,][]{marigo2017} corresponding to the particular 3d spatial position. Although transformation extinction from one bandpass to another can be a complex process, the linear transformation was done for the sake of nomenclature. As all of the parameters are also normalized through linear scaling, the net result is comparable to training on $A_G$ directly.

In training, positions were normalized from 0 to 1 for $l$ from 0 to 360$^\circ$, $b$ from -90 to 90$^\circ$, $\pi$ from 0 to 5 mas, and $A_V$ from 0 to 5 mag, and the individual measurements of $\pi$ or $A_V$ were allowed to exceed the maximum to be $>1$ after the normalization. The training was done using the Adam optimizer, mean square error loss, and a learning rate of $10^{-3}$.

\begin{figure} 
\epsscale{1.2}
\plotone{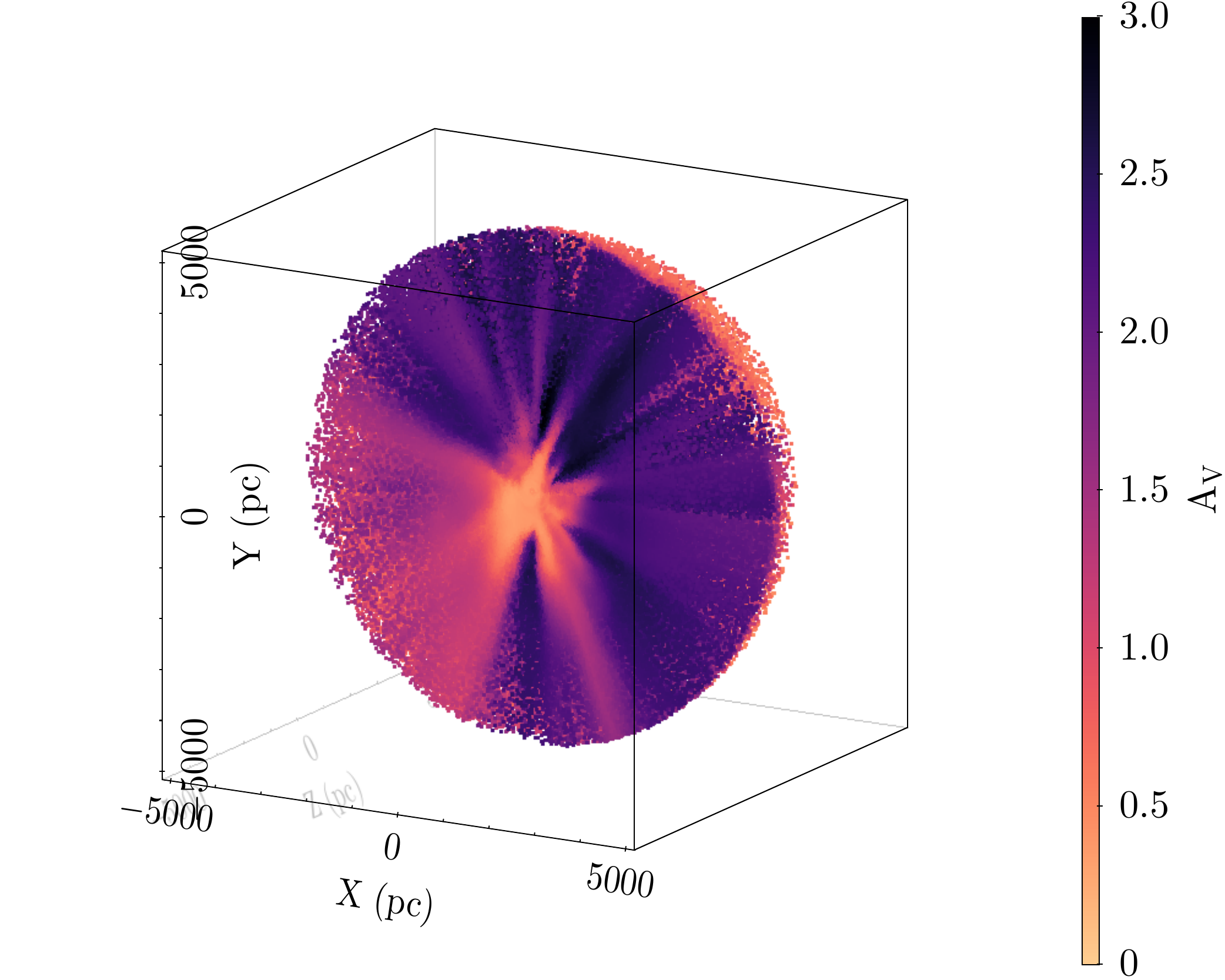}
\caption{A slice of sphere across the galactic plane, showing the 3-dimensional distribution of the sources in the evaluation sample, color-coded by the predicted $A_V$ values. An interactive 2d figure traced by random distribution of points along the sky is made available at \url{http://mkounkel.com/mw3d/avmap.html}; the circle in the interactive plot has a radius of 40$^\circ$, correspondng to the 30 Myr Bubble (Section \ref{sec:30MyrBubble}). \label{fig:av}}
\end{figure}

The resulting estimate of $A_V$ is consistent to within 0.3 mag with the extinction map from \citet{green2019} over the applicable volume, as well as to the cluster $A_V$s estimated on the population level in through pseudo-isochrone fitting with Auriga \citepalias{kounkel2020}. The resulting 3-dimensional extinction map is shown in the Figure \ref{fig:av}.

This is sufficiently precise for the purpose of this paper, as both the classifier and the age regressor do not depend on the absolute magnitude of $A_V$ (or $A_G$) directly. Rather, they rely on the non-linear correlations that are present in the data that can be inferred with a help of this parameter, and they can learn to compensate for color-dependent systematic differences that may be present.

The advantage of the resulting spacial extinction map is that it is available across the entire sky; however it is less robust than the more detailed maps derived through the use of multi-color optical-NIR photometry, such as the map from \citet{green2019}. Different extinction maps are not interchangeable within Sagitta - as $A_V$ is one of the input variables for both classifier and the regressor, supplying it with an unfamiliar map would skew the weights. However, it is possible to train a different model using the same architecture, using the extinction map from \citet{green2019} instead. This limits the spatial coverage (as their map is incomplete in the Southern hemisphere), but, qualitatively, this does not create a significant difference in the sources that are being selected by the classifier, or in the features that are discussed in Section \ref{sec:validation}.

\subsection{YSO Classifier}\label{sec:classifier}
\subsubsection{Training}
The classification network was trained to perform a binary classification task, with classes 0 (not PMS) and 1 (PMS). It maps each stellar parameters $A_V$, $\pi$, $G$, $G_{BP}$, $G_{RP}$, $J$, $H$, and $K$ to a single scalar output in $(0,1)$ representing the star's probability being PMS, using a logistic sigmoid output activation. The model was trained to maximize the log-likelihood of the training data (equivalent to cross-entropy loss).

The data from \citetalias{kounkel2020} was partitioned into three disjoint sets where each served a distinct purpose in classifier development. The training set, containing the first $80\%$ of the sources, was comprised of the stars for the model to train off of. The development set, containing the next $10\%$ of the sources, was used during training to evaluate model performance but was never shown to the model as examples. The testing set, containing the last $10\%$ of sources, was only used after tuning to confirm generalization.

With the training set in place, we augmented the catalog in order to increase performance. The subsets that were sampled for augmentation included PMS stars from \citetalias{kounkel2020}, the non-PMS stars from \citetalias{kounkel2020}, and the randomly selected 3 million star sample from Gaia DR2 (see Section \ref{sec:classsample}. Due to the difference in size of these subsets, all the stars in each of these subsets were sampled certain amount of times, changing $\pi$ and $A_V$ to affect the flux, to produce the augmented sample. The number of times each star was sampled was treated as a hyperparameter in the training process with the tested sample rates listed in Table \ref{tab:classparams}. Multiple models were trained on all permutations of the augmentation ratios, however adjusting the ratios did not appear to have a strong impact on the performance of the model.

Finally, the evaluation sample described in section \ref{sec:testsample} (that consisted mostly of sources for which we did not have explicit a-priori labels) was used for final testing and performance comparison of various models through examining known star forming regions, regions of high extinction, and other features of the solar neighborhood (Section \ref{sec:classifiervalidation}).

To improve the classifier's performance in distinguishing PMS stars, during the data augmentation phase we oversampled PMS stars (i.e., used stratified sampling), yielding 15\% PMS stars during training (compared to $1.6\%$ in the training set). By improving class balance, the model had to focus more on correctly identifying and reducing contamination in the PMS star class. However, while the initial augmentation has improved sensitivity to more distant PMS stars compared to the unaugmented sample, continuing to grow their number in training through augmentation would not necessarily result in a better classifier, as it provides diminishing returns. Once the training had completed, the output detection threshold used for extracting the PMS stars was then selected based on methods described in section \ref{sec:classifiervalidation}. Fine tuning of the ratio for PMS to non-PMS stars was found to not have a significant impact on the performance of the model.

\begin{deluxetable}{cc}
\tablecaption{Classifier Hyperparameter Tuning Values
\label{tab:classparams}}
\tabletypesize{\scriptsize}
\tablewidth{\linewidth}
\tablehead{
\colhead{Hyperparameter} & \colhead{Values}}
\startdata
$Optimizer$ & Adadelta, Adagrad, Adam, RMSProp, SGD \\
$Learning$ $Rate$ & 0.001, 0.01, 0.1 \\
$Dropout$ &  0\%, 10\%, 30\%, 50\%, 70\% \\
$Minibatch$ $Size$ & 5000, 10000, 25000, 50000\\
$Weight$ $Decay$ & 0, 0.00001, 0.0001, 0.001 \\
\hline\hline
Subset & Star Sample Rate \\
\hline\hline
PMS stars from \citetalias{kounkel2020} & 20, 25, 50\\
Non-PMS stars from \citetalias{kounkel2020} & 3, 5, 10\\
Random stars from Gaia DR2 & 1, 2\\
\enddata
\end{deluxetable}

Hyperparameter tuning and early stopping were also employed to help improve classifier predictions. The list of possible settings for each of the hyperparameters tuned are listed in Table \ref{tab:classparams}. Each hyperparameter configuration instance in the hyperparameter sweep was trained until the development set loss failed to beat its best loss for 20 successive epochs, at which point that instance's training stopped. During each instance's training, only the snapshot with the best development set performance was saved. Once the sweep finished, each instance's predicted outputs on the development set were used to visually confirm that the model was predicting desired values. The final model used in the pipeline was chosen based off its low development set loss and qualitatively good predictions.

Through the hyperparameter sweep, it was found that using very little or no weight decay consistently provided models with the best development set performance. All of the other hyperparameters tuned on seemed to not produce any significant improvement in model performance one way or another. The configuration of hyperparameters that produced the best model was comprised of Adagrad for the optimizer, $0\%$ dropout, a batch size of $5000$, a learning rate of $0.01$, and a weight decay of $10^{-5}$.

\subsubsection{Classifier validation}\label{sec:classifiervalidation}

The Upper Sco region, which contains the $\rho$ Oph dark cloud,
is particularly useful in evaluation of how reliably the classifier can discriminate between bona fide YSOs and the contamination from evolved stars that have been reddened to the PMS parameter space. It is located nearby, with $\pi\sim7$ mas. No other star forming regions are known to be located behind it, nor are there likely to be any distant undiscovered populations behind it, given its high elevation above the Galactic Plane, at $b\sim20^\circ$. Therefore, if any sources identified as PMS are located far beyond, e.g., 200 pc, they are most likely to be false positives. The precision in distance that can be inferred from the parallax decreases the further away a star is, while the number of field stars in each parallax bin increases. Therefore, the large parallax of Upper Sco makes it particularly easy to separate the stars associated with this region from the false positives, to the degree that even other star forming regions along the Gould's Belt do not. 

\begin{figure}
    \fig{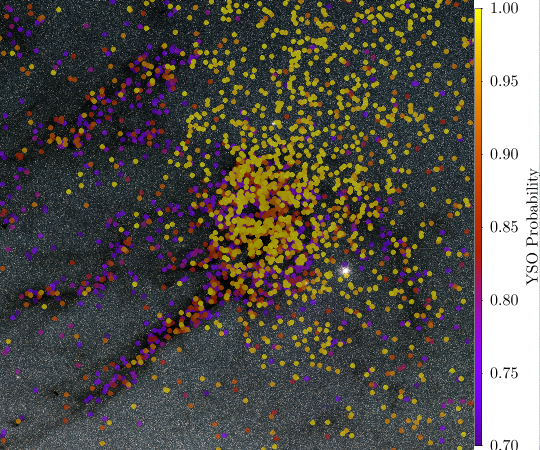}{\linewidth}{}
    \caption{Identified PMS sources $>70$\% probability towards $\rho$ Oph and Upper Sco, plotted over the \textit{Gaia} DR2 map of the sky \citep{gaia-collaboration2018}. As the $\rho$ Oph dark cloud has high extinction, it is clearly visible in this map. Note the highest confidence PMS sources are tracing the known regions of star formation. On the other hand, sources with lower probability tend to be co-located along the pertruding filaments that are not actively forming stars. \label{fig:oph_dust}}
\end{figure}

\begin{figure*} 
\plottwo{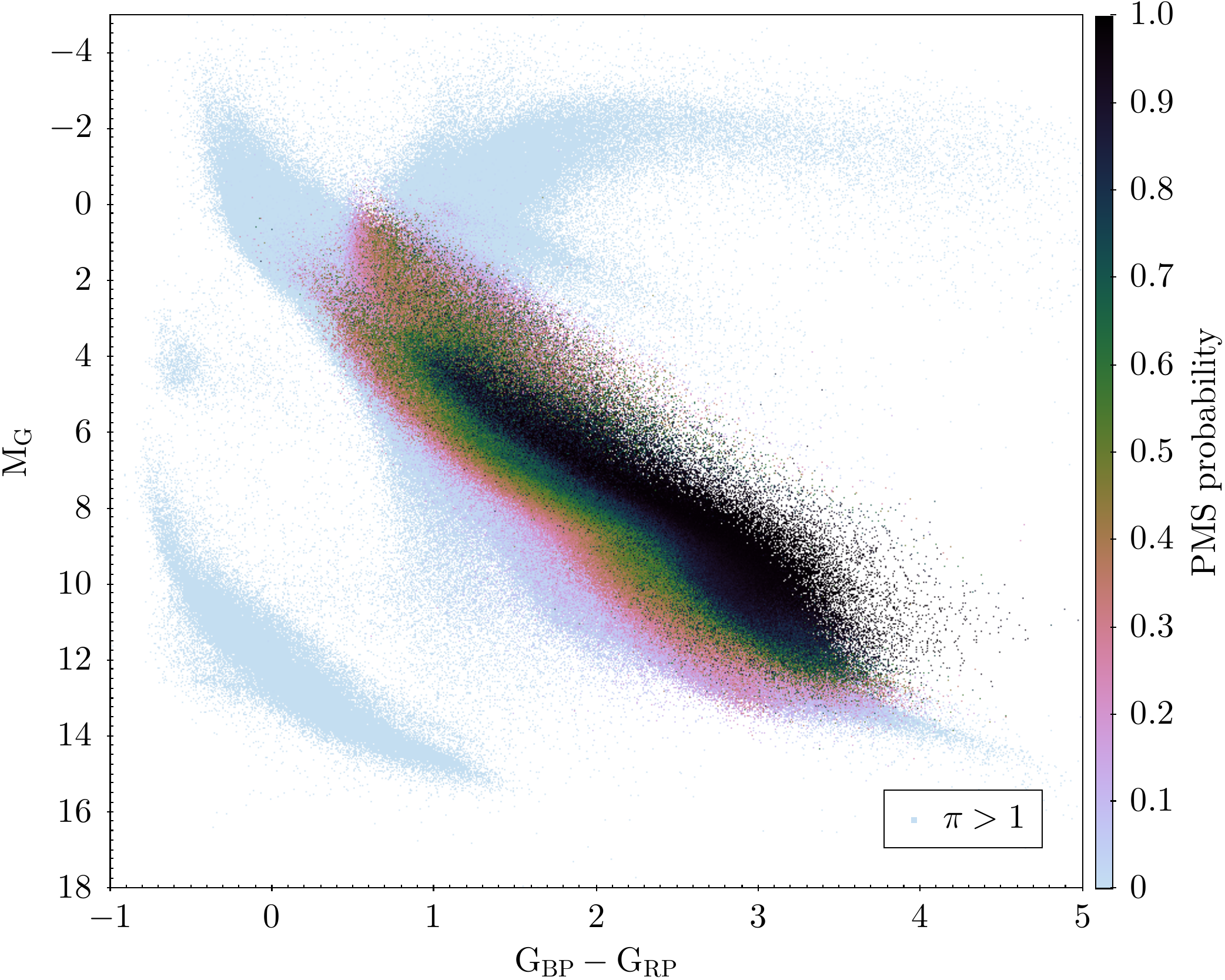}{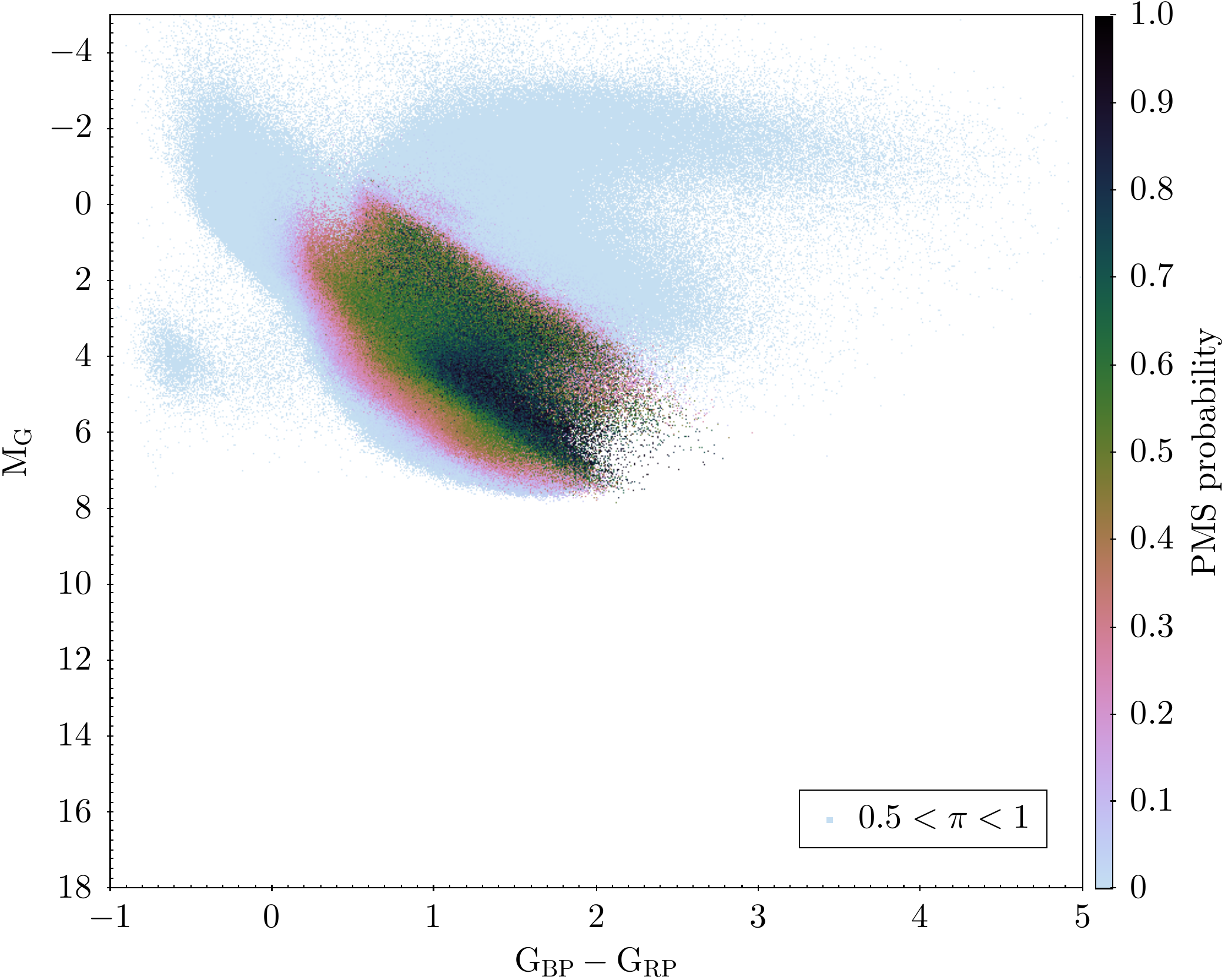}
\plottwo{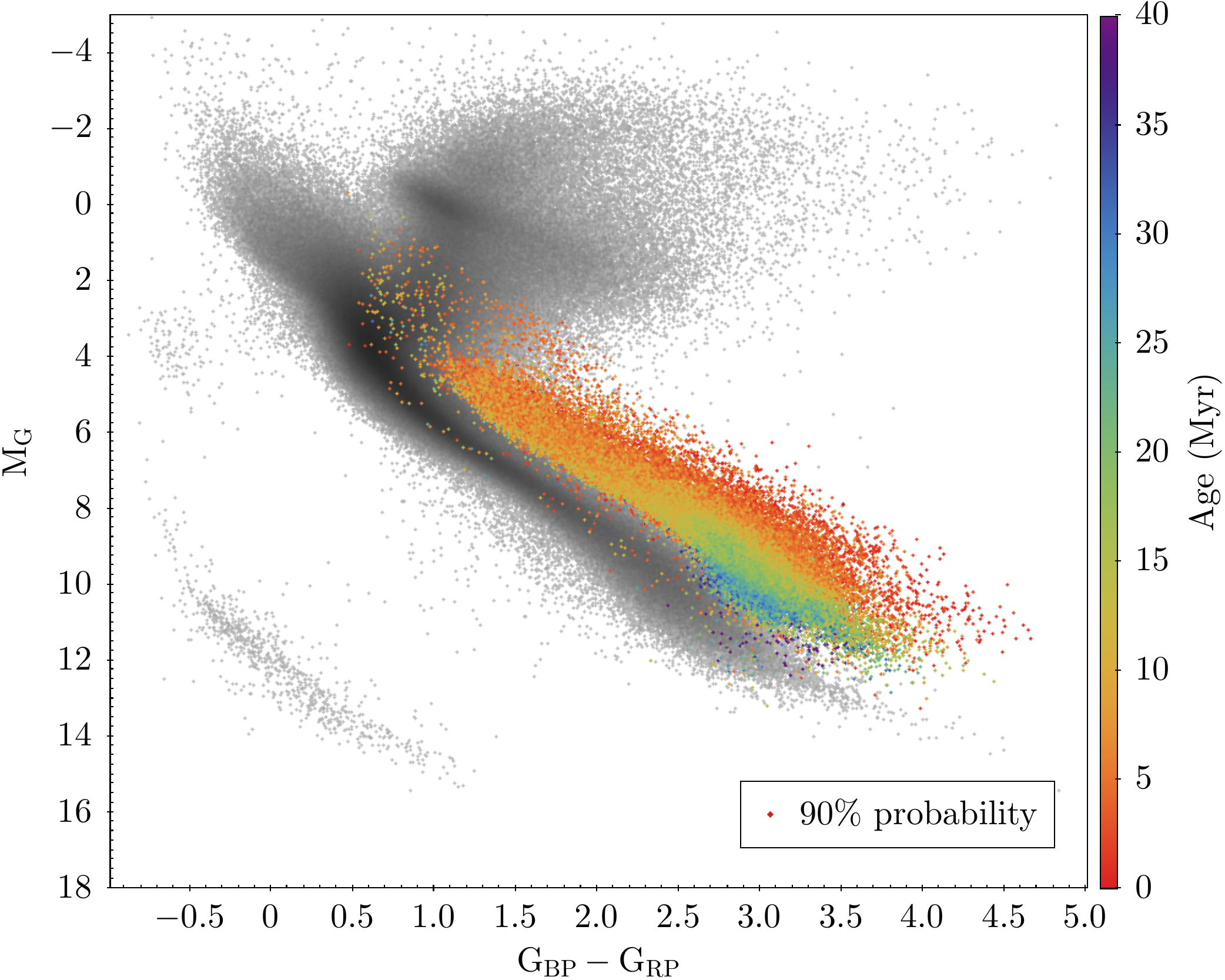}{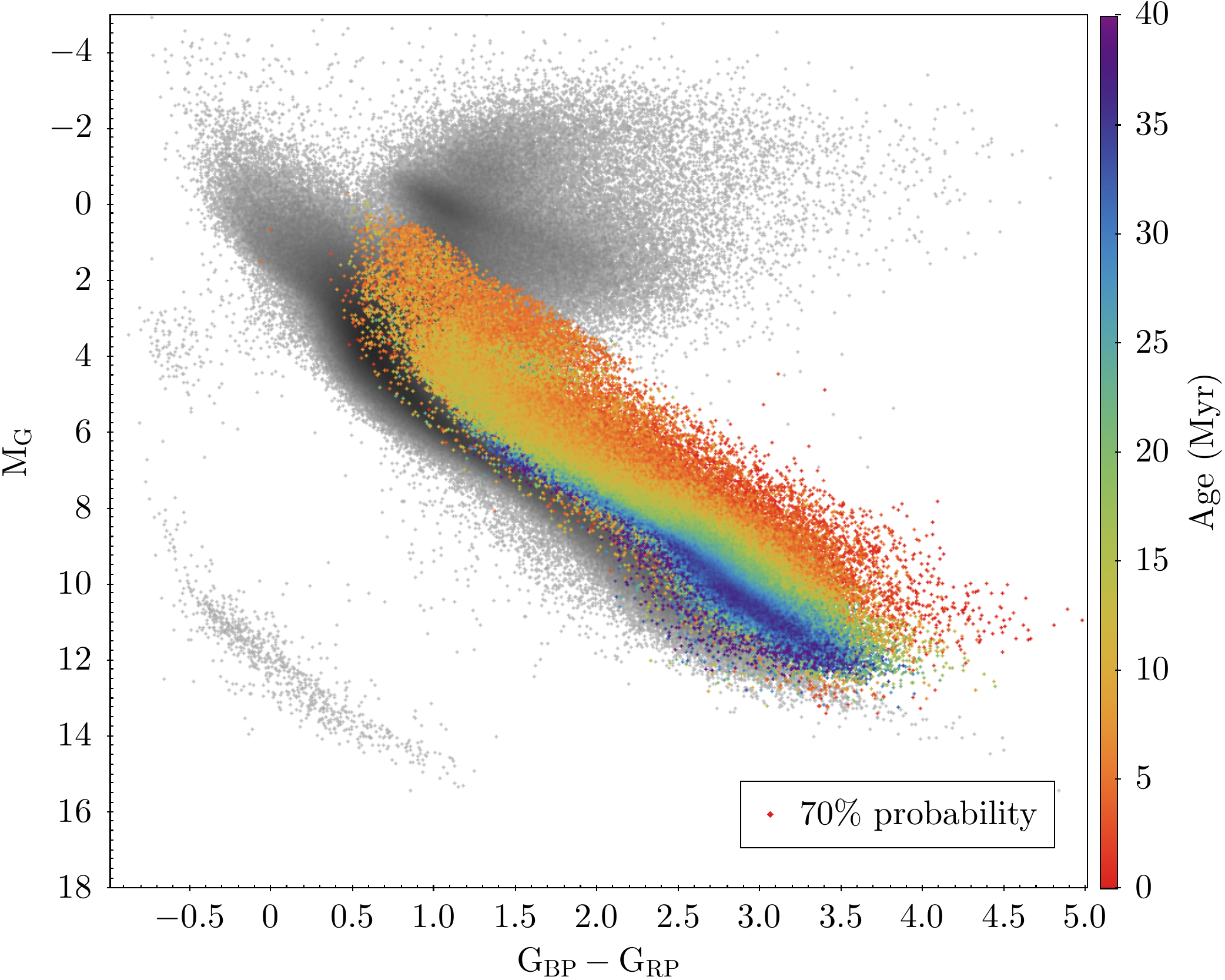}
\caption{HR diagram of the evaluation sample. Top: Color coded by the maximum probability of identifying a star as PMS in each 2-dimensional bin, in two distance slices. Bottom: Color coded by the mean age in each 2-dimensional bin. The bottom left panel shows the sample with PMS probability $>$90\%, the bottom right panel shows the sample wihth $>70$\% probability. They are plotted over the greyscale HR diagram of a random subset of the full evaluation sample. \label{fig:hrprob}}
\end{figure*}

As $\rho$ Oph is very young and deeply embedded in a dusty cloud, the line of sight extinction rises significantly behind the cloud, offering an excellent test of the sensitivity of contamination to reddening (Figure \ref{fig:oph_dust}). Similarly, the cloud has a very particular shape, with several long filaments protruding away from the center that are not actively involved in star formation. Even without considering their distance, contamination due to extinction can be apparent if the identified PMS candidates follow the outline of the cloud too well.

With this in mind, we imposed several criteria to evaluate the different trained classifier models with different hyperparameters. Ordering the model predictions from the highest PMS probability to the lowest, for sources within the box of $345<l<360^\circ$ and $15<b<25^\circ$, we identified the typical probability threshold for each model where the number of sources with $\pi<5$ begins to match the number of sources with $\pi>5$ in a given probability bin (i.e., the point where the rate contamination/false positives is comparable to the rate of adding bona fide PMS stars/true positives). The best model needs to:

\begin{itemize}
  \item Maximize the overall number of $\pi>5$ sources that the model identifies up to that point
  \item Minimize the ratio [number of sources with $\pi<5$]/[number of sources with $\pi>5$], i.e., minimize the overall contamination fraction up to that point.
\end{itemize}

Although these criteria have been optimized for the selection of the Upper Sco sources, they generally yield a good selection of PMS candidates in other nearby star forming regions as well. 

Selection of regions beyond 1 kpc presents a bigger challenge, as their lower mass members tend to be too faint to be within the sensitivity limits. Thus, members of various star forming regions located at those distances tend to have overall lower probability than their more nearby counterparts. Because of this, it is difficult to estimate the contamination among them.

However, almost all of the identified YSOs beyond 1 kpc should be located close to the Galactic plane. While this is generally the case, due to the scanning law of \textit{Gaia}, there is a slight excess of (most likely false positive) PMS candidates that are found towards the ecliptic poles in almost all models. Such false positives are usually fainter stars that do not have 2MASS photometry. Depending on the exact limiting threshold, this usually amounts to a few hundred stars. Therefore, in evaluation of the best model, we also consider maximizing the number of sources with $\pi<1$ and $|b|<15$ and minimizing the number of sources with $\pi<1$ and $|b|>20$.

With all of these considerations, we tested more than $100$ different models that were trained using different hyper-parameters and with slightly different architectures. Most models had comparable performance in the evaluation sample, although some had a greater difficulty in separating false positives from true positives.

Of all of them, however, one model had almost an order of magnitude better performance than the rest in the combined evaluation metric, although it is unclear if it was due to the most optimal tuning of the hyperparameters, or luck in the process of the stochastic gradient descent. Regardless, this classifier model was chosen to be implemented into Sagitta. The HR diagram showing the outputs of this model is shown in Figure \ref{fig:hrprob}.

We note that despite the chosen model being more optimal for purpose of identifying young stars across the entire sky in comparison to a numerous other experiments, it is not devoid of contamination, particularly at large distances and in the background of dusty clouds. Most of these contaminants tend to have at lower probabilities but a small fraction of false positives can be found even at relatively high probability thresholds. Across the entire sky, the reported probabilities are quasi-Gaussian, as such, even 90\% or 95\% thresholds are susceptible to some fraction of false positives. The situation can be somewhat more extreme in individual regions. For example, Pipe Nebula has distance and extinction comparable to what is found in $\rho$ Oph, but it has few true members. As such, most of the sources in the catalog observed towards this region tend to be more distant and are contaminants. Thus, in evaluating membership of each individual region it is important to consider the known priors, such as age, distance, and foreground opacity.

We note that while some data driven approaches (e.g., decision trees) may be reduced to a human-readable set of conditions by which classification takes place, this is not the case with deep learning. As such, while the model can differentiate PMS and non-PMS stars based on their fluxes (most likely noting in some fashion that PMS stars tend to be redder and/or over-luminous than the main sequence stars, but not in the parameter space inhabited by red giants), it is difficult to express precisely how these fluxes are utilized by the model. Although beyond the scope of the current study, one could employ model interpretability methods (e.g. \cite{zeiler2014}) in an attempt to gain some insight into model behavior.

\begin{figure} 
\plotone{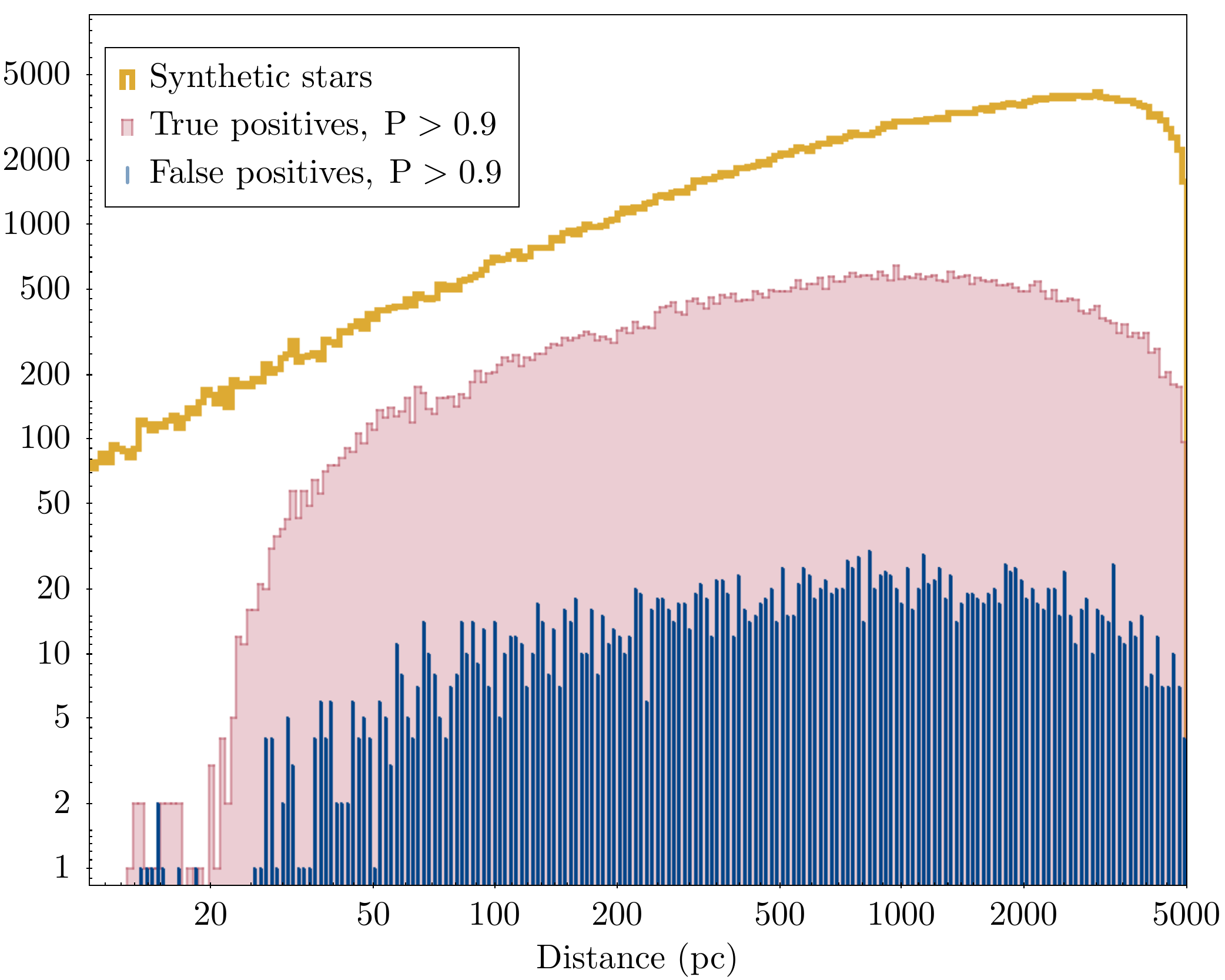}
\caption{Recovery of PMS stars in a synthetically generated sample as a function of distance. A similarly smooth distribution (with a higher recovery fraction at a cost of more false positives) can be seen at lower thresholds as well. \label{fig:dist}}
\end{figure}

To ensure that no residual biases in distance from massive populations propagate to the model, we examine a synthetically generated sample of stars based on the sample from \citetalias{kounkel2020}, similar to the one described in Section \ref{sec:classsample}. All of the synthetic stars, including both the stars labeled as PMS as well as those that were more evolved, have randomly drawn distances and extinctions. There are some difference in the fraction of sources recovered at the distances between 20 to 5000 pc relative to the input sample (e.g., smaller fraction of more distant sources is recovered at the same probability threshold, in part due to a smaller fraction of low mass stars). However, in a uniform sample the model does not systematically favor a specific set of distances corresponding to, e.g., the distance of the Orion Complex or Sco Cen OB2, either in a form of better recovery of true positives than for stars at other distances, or in form of contamination from evolved stars (Figure \ref{fig:dist}).

To further test the reliability of Sagitta's PMS probabilities and assigned ages, we performed a series of checks using empirical catalogs of candidate Upper Sco members \citep{luhman2020}, older (30-300 Myr) open clusters \citep{meingast2020}, and field stars \citep[low membership probability sources in the DaNCE catalog of IC 4665][]{miret-roig2018}. Sagitta was used to assign PMS probabilities and ages to the sources in each catalog, both at their true distances, and after artificially adjusting their apparent distances by adjusting their parallaxes and applying appropriate distance moduli to their magnitudes. 

\begin{figure} 
\plotone{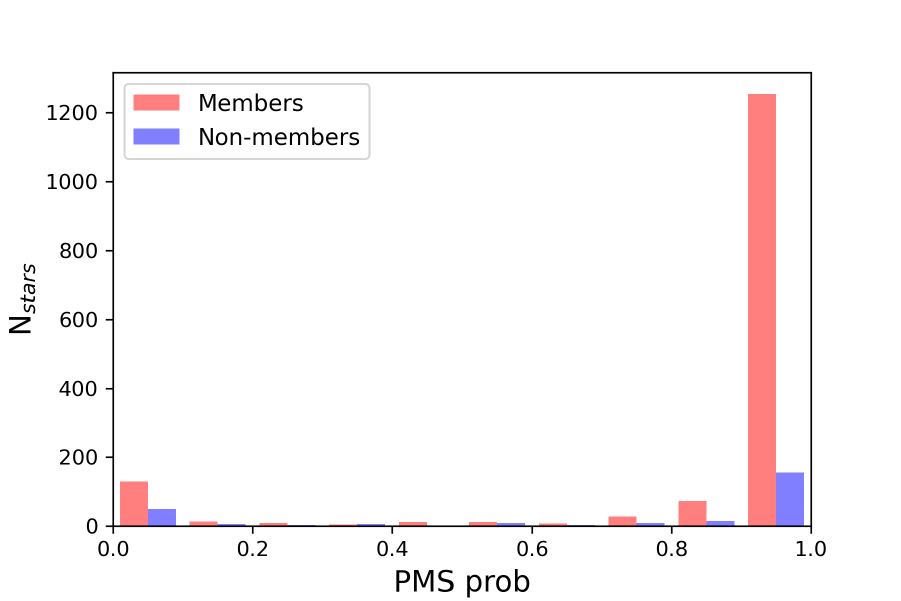}
\plotone{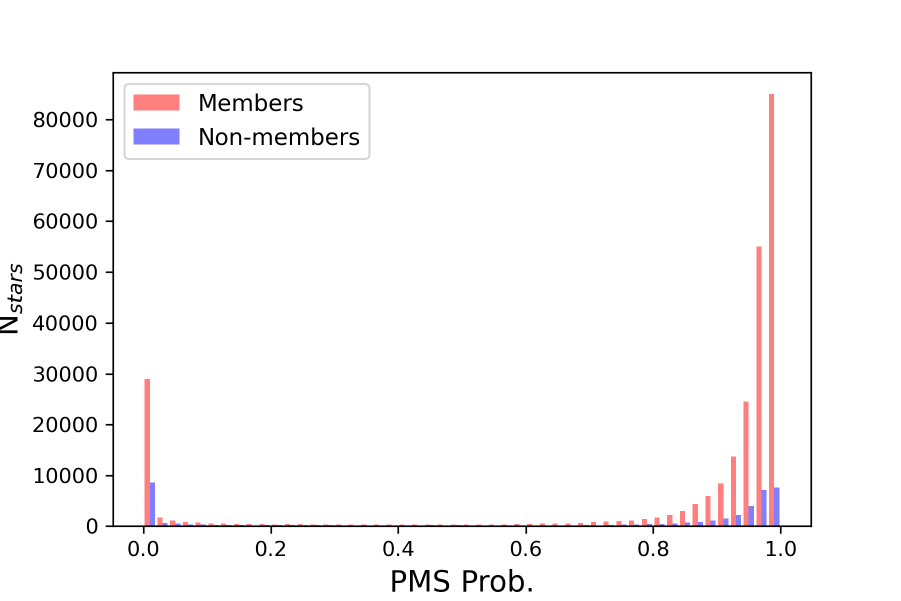}
\caption{Histograms of Sagitta's PMS probabilities for Upper Sco members and non-members, as determined by \citet{luhman2020}, at Upper Sco's true distance (top panel), and as artificially shifted to a range of distances from 30-500pc (bottom panel). While non-members exhibit some potential for pre-main sequence status to be selected for analysis by \citet{luhman2020}, Sagitta nonetheless assigns high PMS probabilities to a significantly larger fraction of the bona fide members, both when stars are considered at their true distances, and also when synthetically shifted to distances between 30-500pc.  \label{fig:upScoProb}}
\end{figure}

\begin{figure} 
\plotone{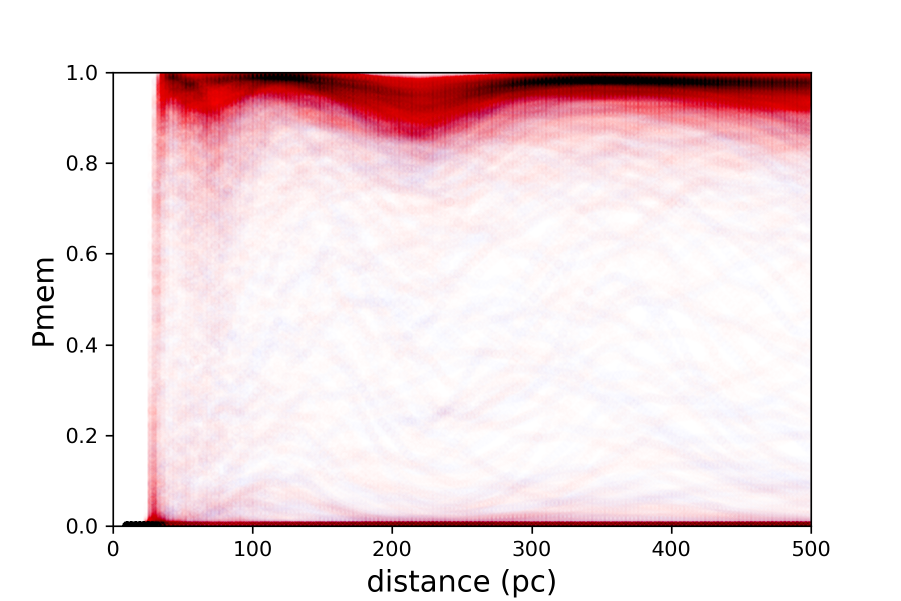}
\caption{Pre-main sequence probabilities as a function of distance, as assigned by Sagitta to synthetically shifted candidate Upper Sco members \citep{luhman2020}. Beyond 30pc, Sagitta consistently assigns high ($> 90\%$) or low ($\sim$0\%) pre-main sequence probabilities to each source at all distances; sources do not switch from high to low probabilities as a function of distance. \label{fig:upScoDistProb}}
\end{figure}

As a first check, we examined the pre-main sequence probabilities Sagitta determines for candidate Upper Sco members. We find that Sagitta returns higher probabilities for sources assessed to be true members by \citet{luhman2020}, with only modest differences as a function of (synthetically shifted) distance. At Upper Sco's true distance, 85\% and 63\% of the \citet{luhman2020} determined members and non-members, respectively, are assigned PMS probabilities $>$85\% (see Fig. \ref{fig:upScoProb}, top panel); both fractions drop by 12\%, to 73\% and 51\% respectively, when the PMS probability threshold is raised to $>$ 95\%. Once the distances to these YSO candidates are synthetically shifted (see Fig. \ref{fig:upScoProb}, bottom panel), slightly lower fractions of each population pass each PMS probability threshold: 79\% and 59\% of members and non-members meet the 85\% threshold for distances between 30-500pc, while 61\% and 40\% meet the 95\% threshold. As seen in Figure \ref{fig:upScoDistProb} the pre-main sequence probabilities show only modest ($\pm$5\%) changes with distance: sources are consistently assigned a high ($>$90\%) or low ($\sim$0\%) pre-main sequence probability at all distances. These tests indicate that Sagitta successfully recovers a larger fraction of bona fide YSOs than non-members, over a wide range of distances, even within a sample of sources explicitly selected on the basis of CMD positions indicative of youth. We note that the selection of members in \citep{luhman2020} is conservative and may exclude some YSOs in cases of e.g., onset of Li I depletion. As such, a number of sources identified as non-members may still be bona fide YSOs, which inflates their fraction of ``false postitives'' at given thresholds.

\begin{figure} 
\plotone{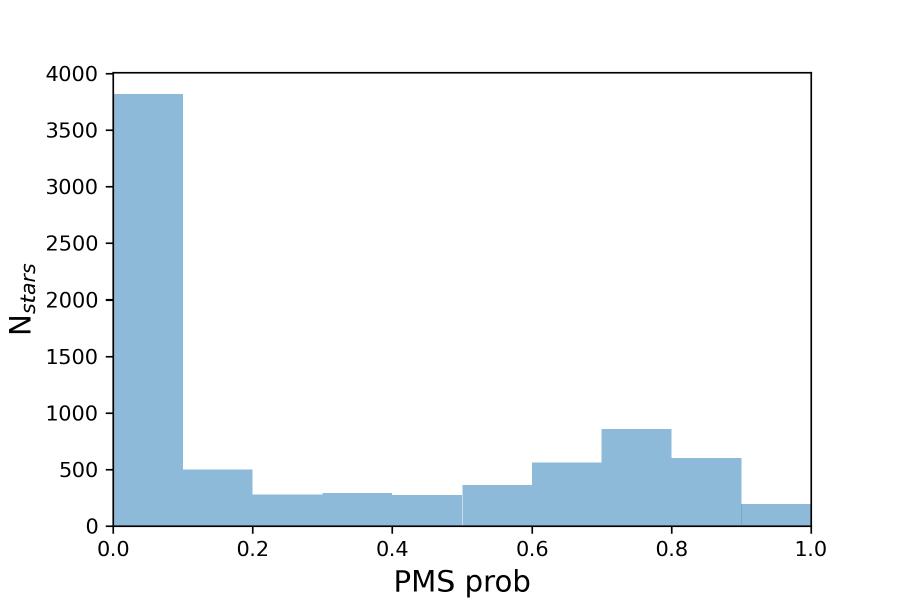}
\plotone{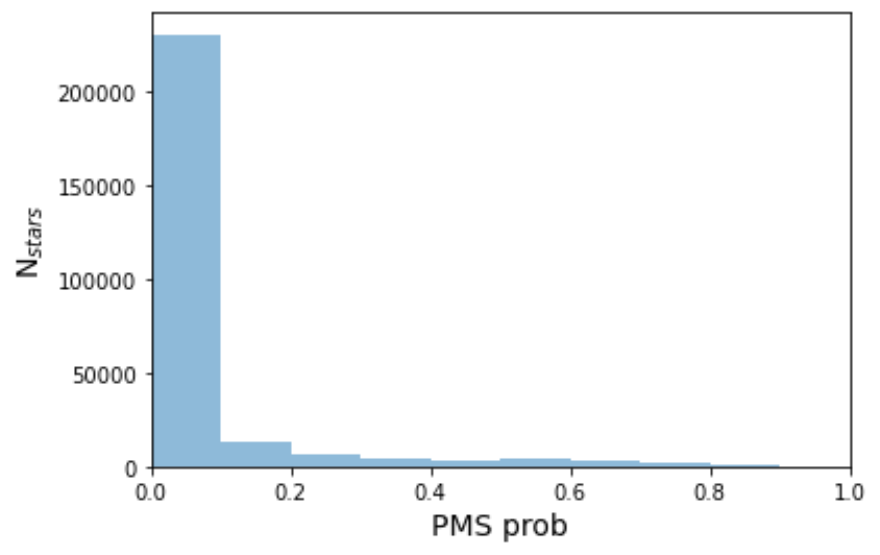}
\caption{Histograms of Sagitta's PMS probabilities for candidate open cluster members, as determined by \citet{meingast2020} (top) and likely field stars in the region around IC 4665 from \citet{miret-roig2018} (bottom). 5\% and 1\% of open cluster members, preferentially those from the youngest open clusters in the catalog, pass PMS probability thresholds of 85\% and 95\%, respectively.  In the field star sample, the fraction of sources that pass those thresholds drop further, to 0.1\% and 0.001\% respectively. \label{fig:openfield}}
\end{figure}

Similarly, Sagitta assigns much lower PMS probabilities to candidate members of older open clusters and background field stars. Analyzing the candidate open cluster members catalogued by \citet{meingast2020} at their true distances, Sagitta only identifies 5\% as having PMS probabilities greater than 85\% (see Fig. \ref{fig:openfield}, top panel). Increasing the PMS probability threshold to 95\% trims the vast majority of these marginal candidates: only 0.5\% of the open cluster members are assigned PMS probabilities $>$95\%. In both cases, the sources most likely to exceed the threshold are also preferentially members of the youngest clusters in the sample, indicating that Sagitta is correctly identifying the sources with the highest elevations above the main sequence.  Applying distance shifts to the cluster members produces similar fractions of sources above each PMS threshold: 6\% and 1\% of the distance shifted samples meet the 85\% and 95\% thresholds, respectively. 

As the bottom panel of Fig. \ref{fig:openfield} indicates, field stars show even lower PMS probabilities, at their true distances and when shifted in distance. Only 0.1\% and 0.001\% of the catalog satisfies the 85\% and 95\% PMS thresholds, respectively.

\subsection{Age Regressor}\label{sec:regressor}
Similarly to the classifier, the regression network to predict ages was trained on the six photometric bands, $\pi$, and $A_V$. In addition to the \textit{Gaia} and 2MASS bands, we originally considered including the photometry from the AllWISE catalog as well, but it was determined to be too noisy.

In constructing the catalog, we used 30\% real data, and 70\% augmented data scattered across different distances and extinctions, for a total of a total of $\sim 187,000$ sources. 80\% of this catalog was used as a training set, 10\% was used as the development set, and 10\% was withheld as a test set. The training was done using the Adam optimizer, mean square error loss, a learning rate of $10^{-3}$, and a batch size of 20,000 sources. Every ten epochs we evaluated the performance on the development set to ensure that the network is learning patterns that generalize to previously unseen data, rather than overfitting to the training set (e.g., `memorizing' it).

The training continued for $\sim$10,000 epochs. Afterwards, we continued to train the model for $\sim$2,000 epochs on real data only, to minimize potential artefacts that may be present in the augmented sample. However, we note that in evaluating the ages on the test set, there were no significant systematic differences between the models with and without the additional 2,000 epochs on real data. Similarly, in various experiments, few combinations of hyperparameters were tested, but they tended to have comparable outputs with few obvious differences in performance (in contrast to the experiments with hyperparameters in the classifier). Instead, for the age regression, the biggest gains in performance were a result of careful vetting of the labels in the training sample.

In general, the trained model is able to qualitatively reproduce the average ages of the populations in which stars are found (Section \ref{sec:sfrs}). It does improve on the ability to infer the star forming histories of different regions compared to the training sample, where usually only a single age per population was available.

\begin{figure} 
\plotone{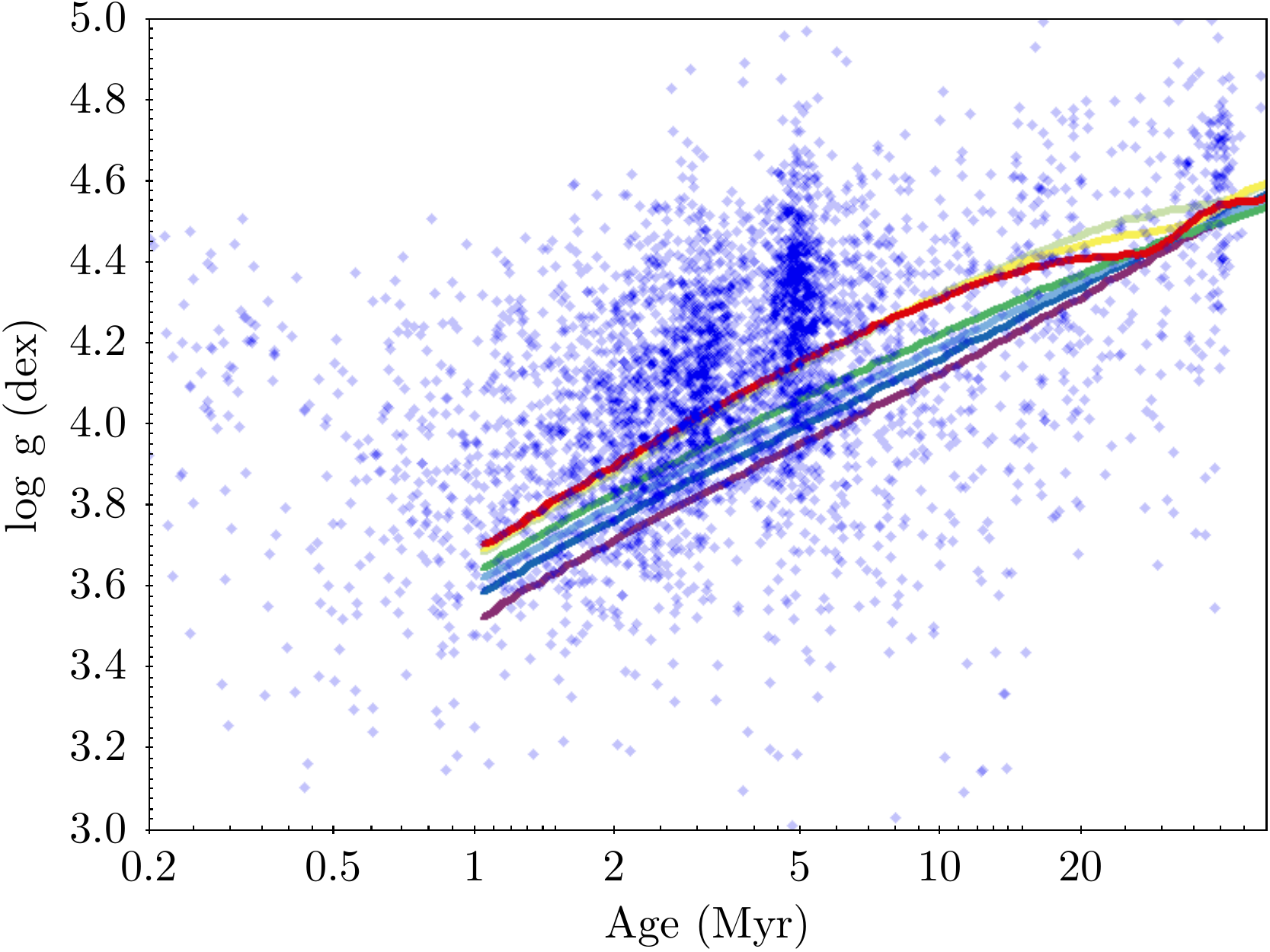}
\caption{Comparison of the estimated ages from Sagitta vs \logg\ inferred from the APOGEE spectra from \citet{olney2020}. The overdensities correspond to specific discrete clusters targeted by APOGEE. The lines show the theoretical PARSEC isochrones \citep{marigo2017} for stars with mass from 0.4 (purple) to 1 \msun (red).} \label{fig:logg}
\end{figure}

We are able to benchmark the estimates of ages for some of the stars that have been previously observed by APOGEE spectrograph. A recent study by \citet{olney2020} has been able to extract calibrated \logg\ estimates for the pre-main sequence stars, which can be used as a proxy of age. The overall trend in Figure \ref{fig:logg} does show that, as expected, \logg\ is increasing as stars evolve and approach the main sequence.

\subsection{Uncertainties}


\begin{figure*}
		\gridline{
            \fig{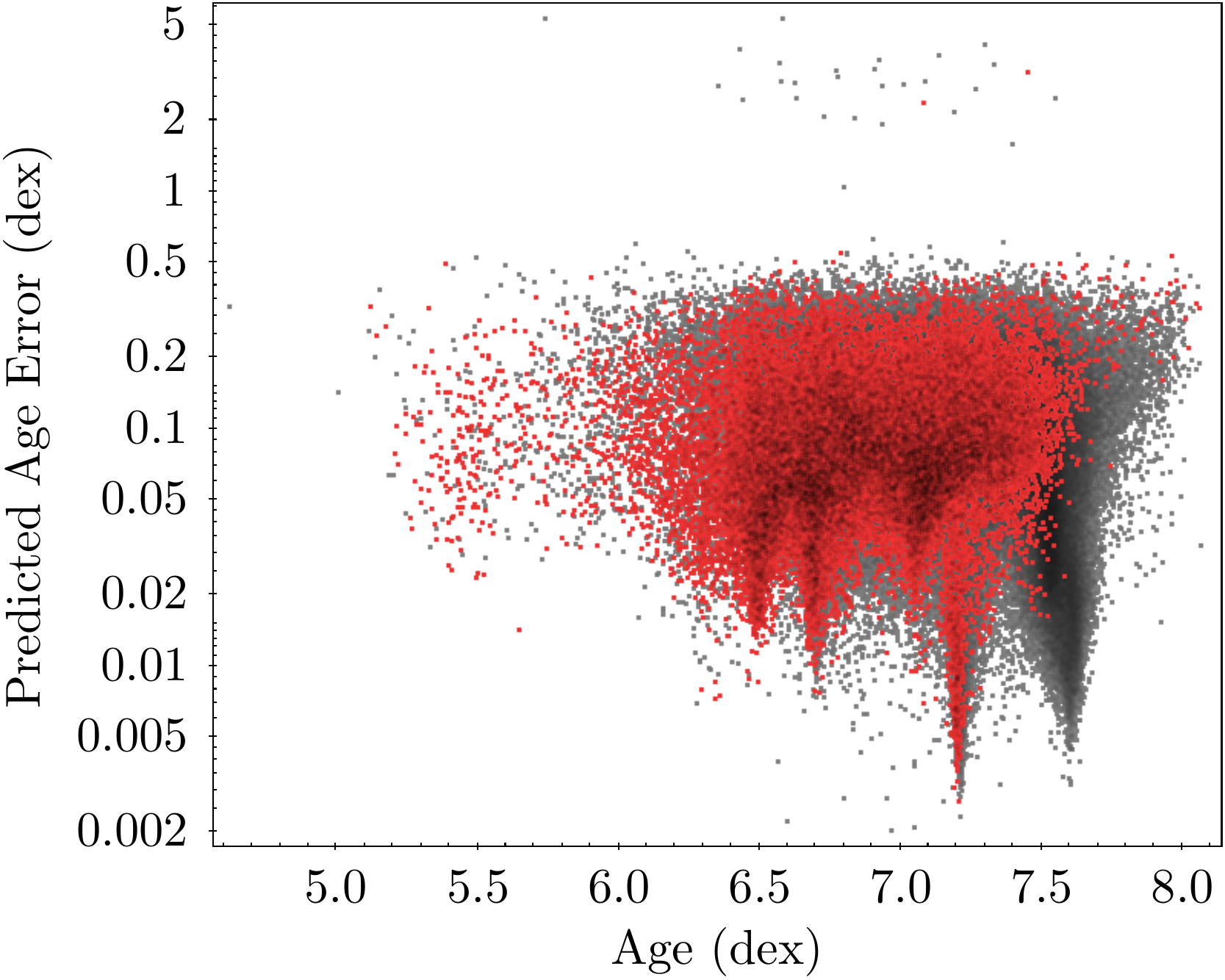}{0.33\textwidth}{}
            \fig{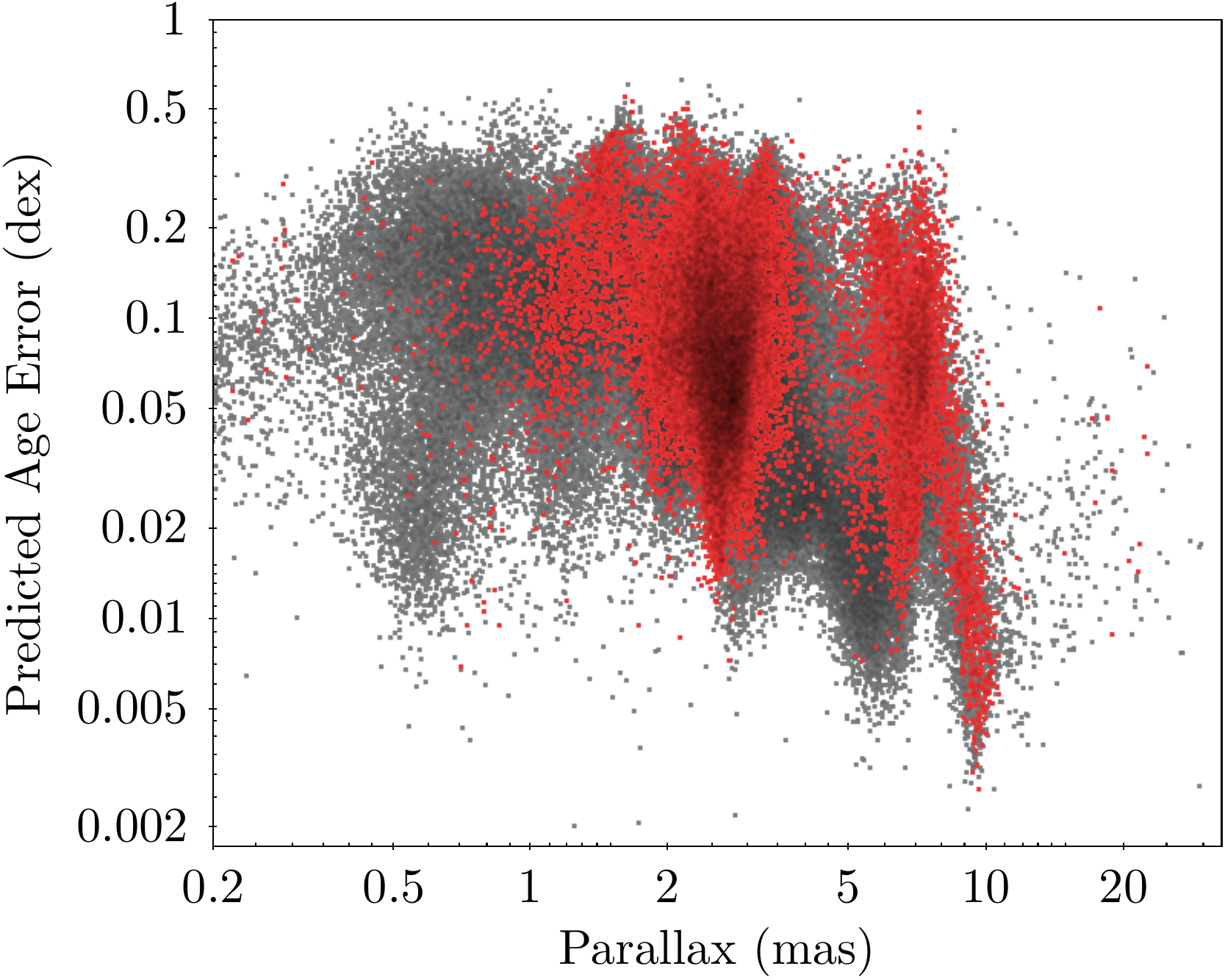}{0.33\textwidth}{}
            \fig{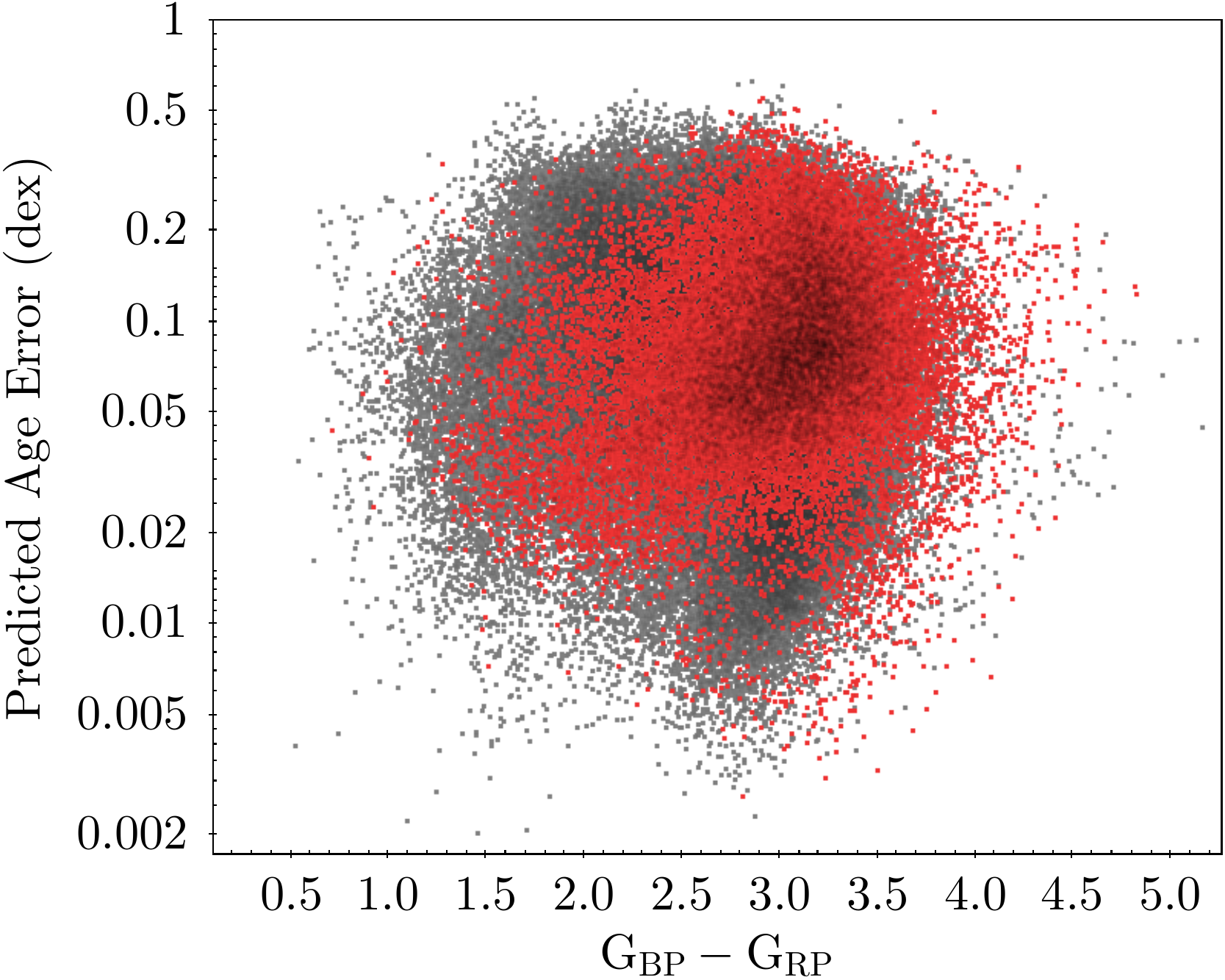}{0.33\textwidth}{}
            }
        \gridline{
            \fig{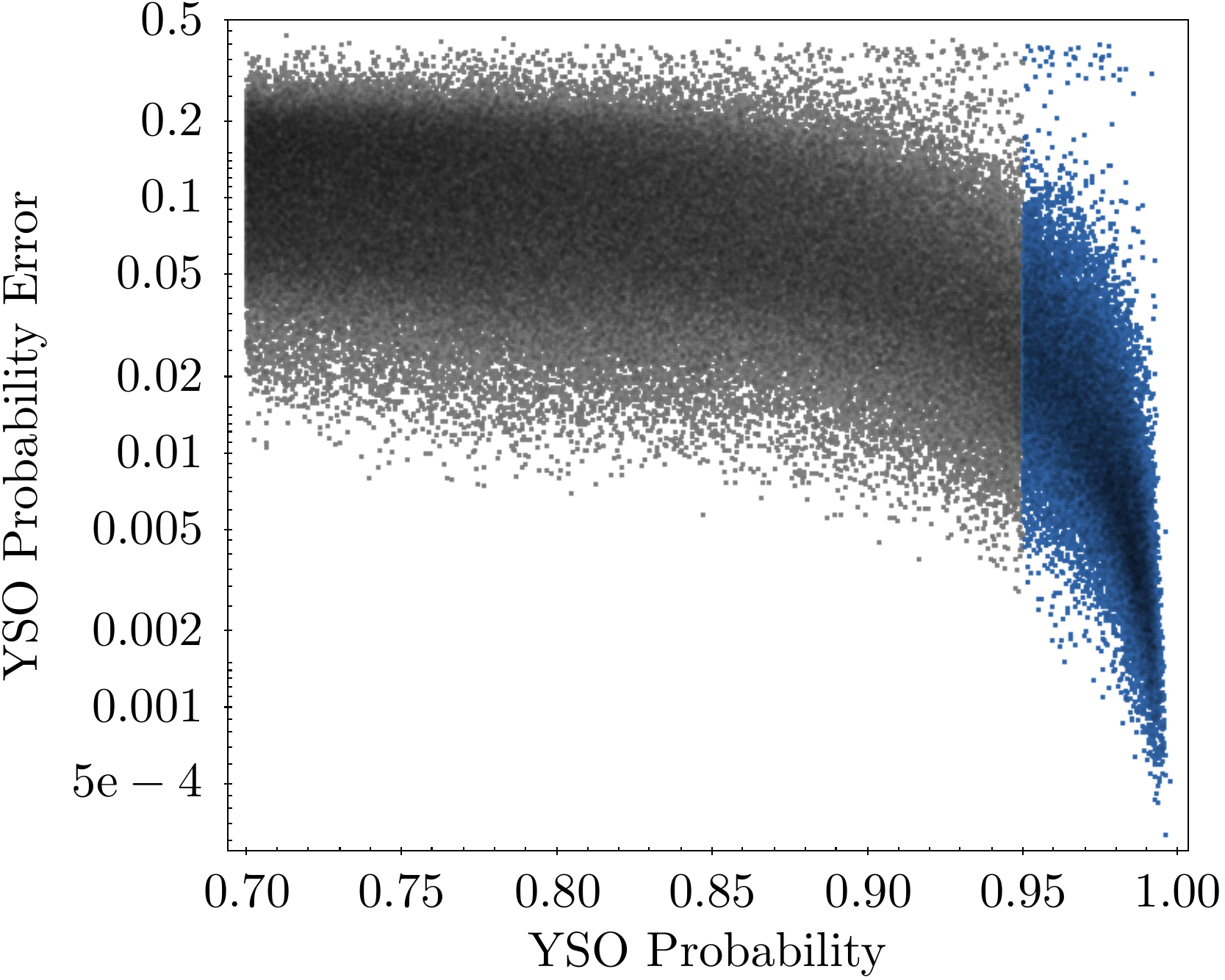}{0.33\textwidth}{}
            \fig{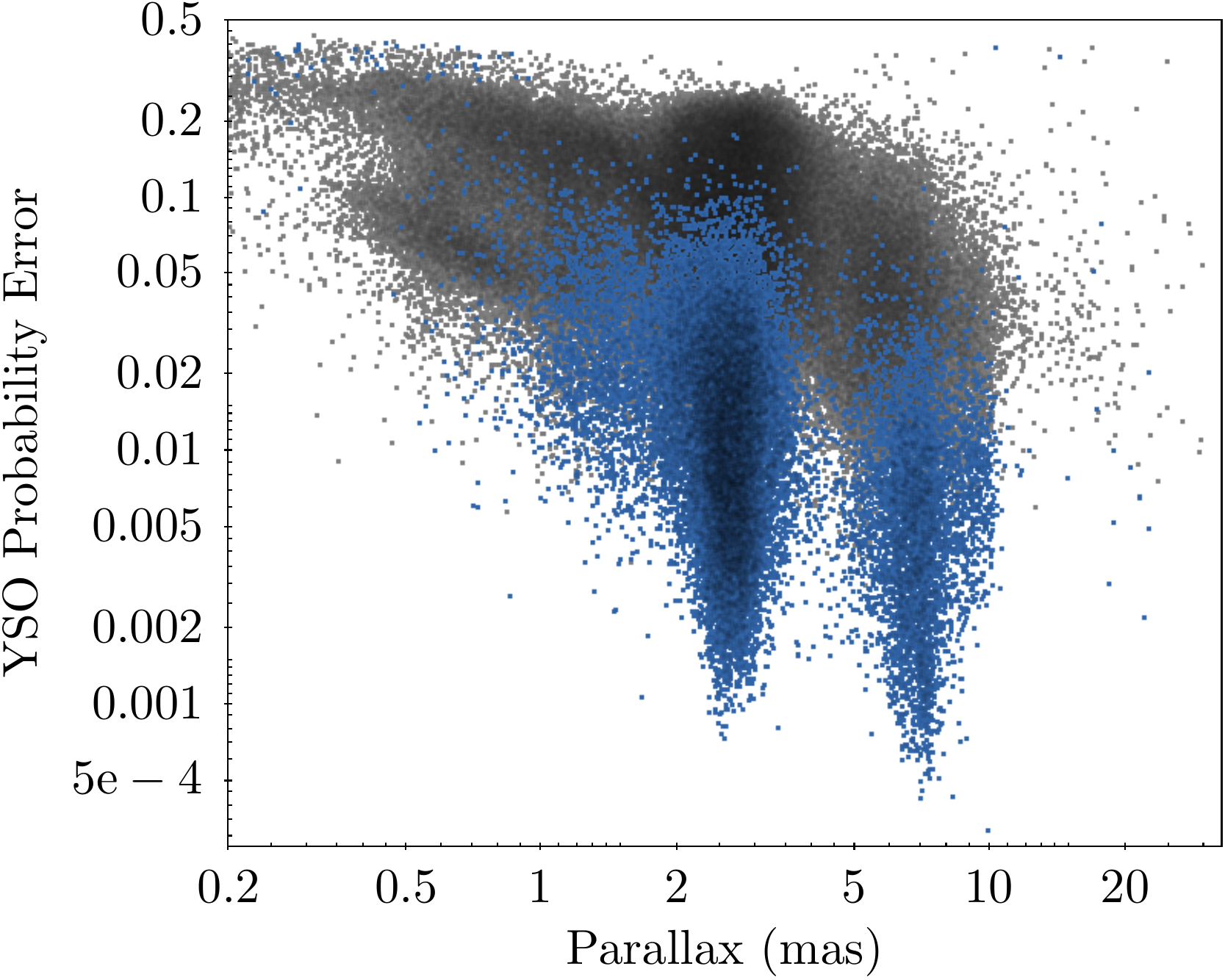}{0.33\textwidth}{}
            \fig{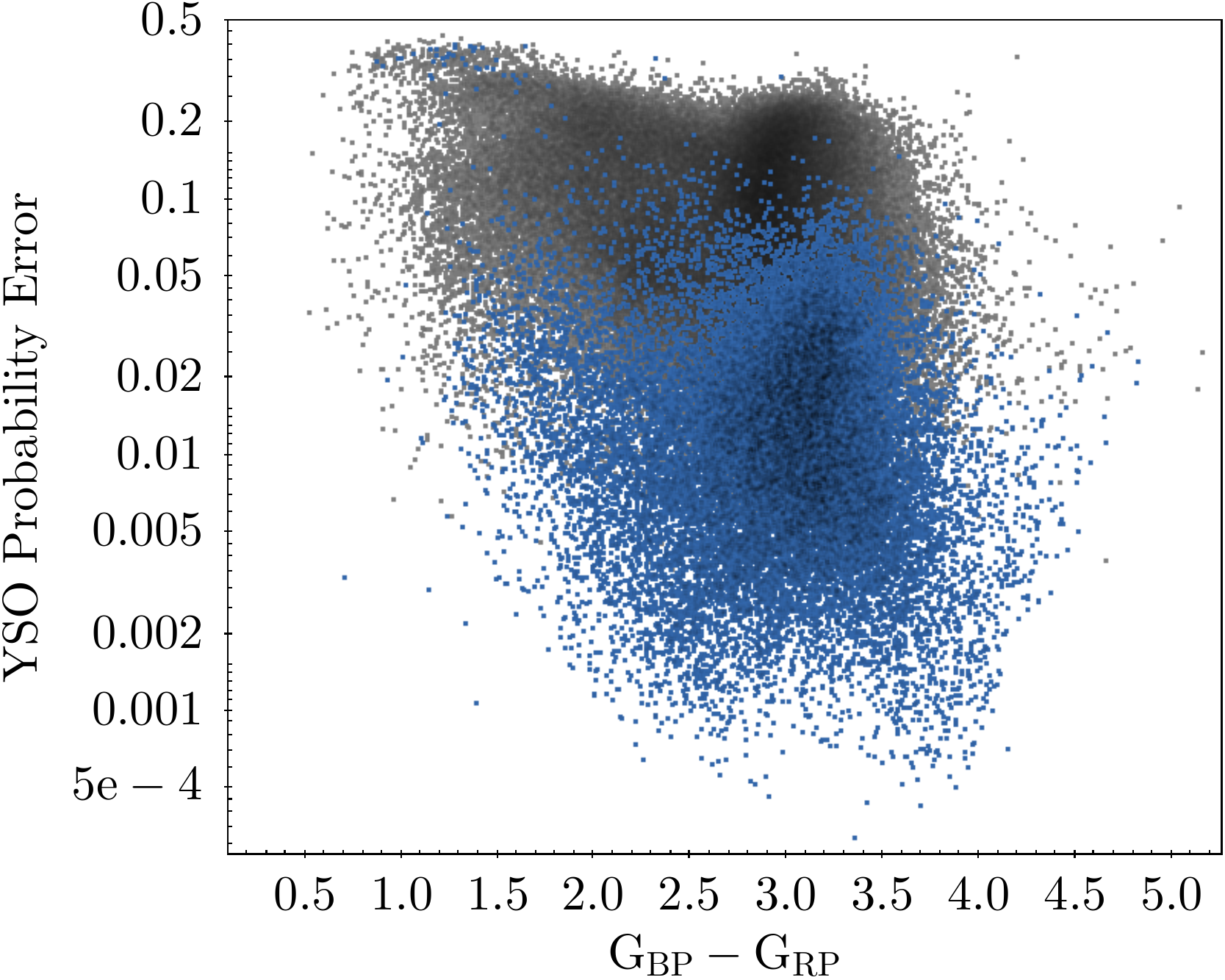}{0.33\textwidth}{}
            }
        \gridline{
            \fig{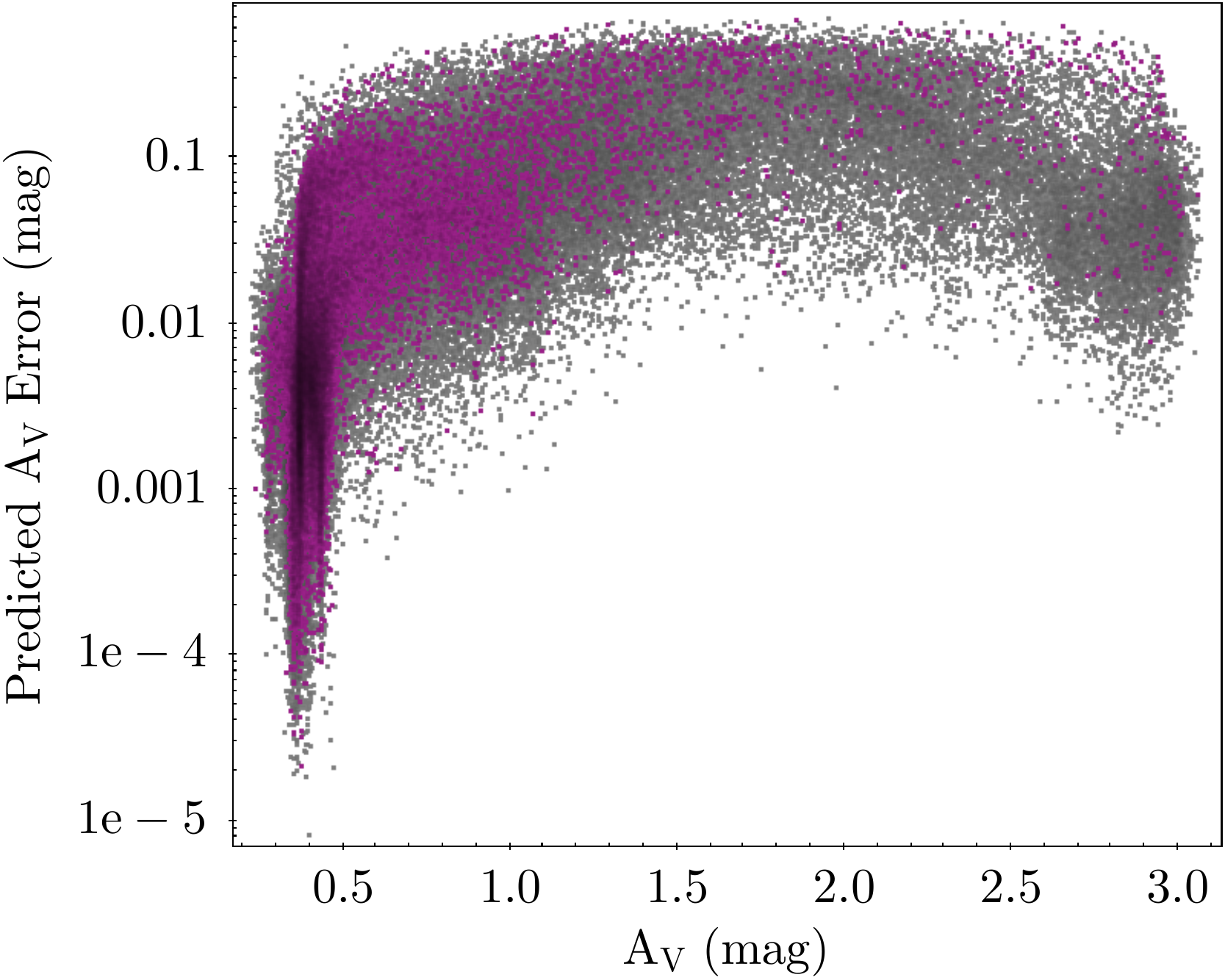}{0.33\textwidth}{}
            \fig{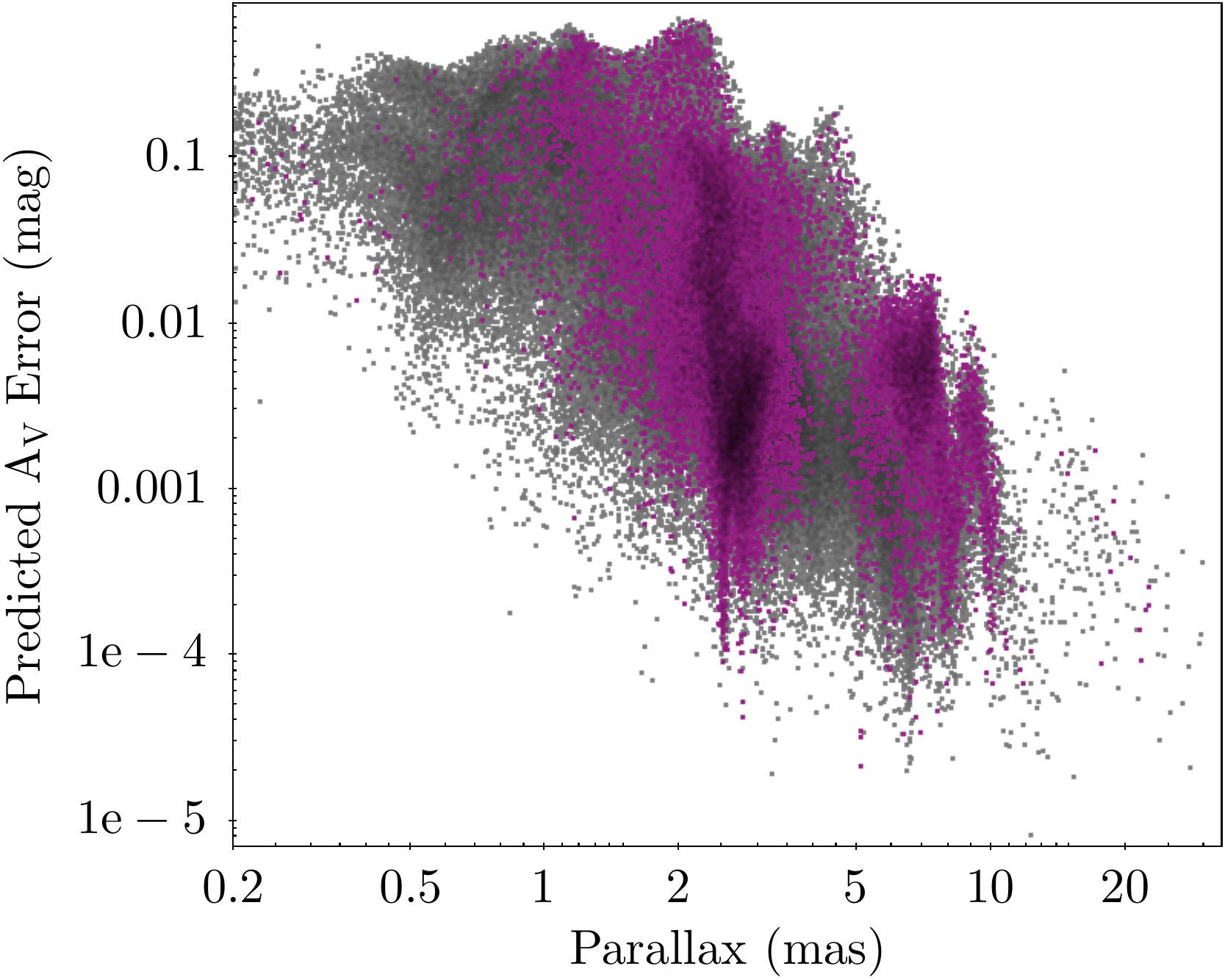}{0.33\textwidth}{}
            \fig{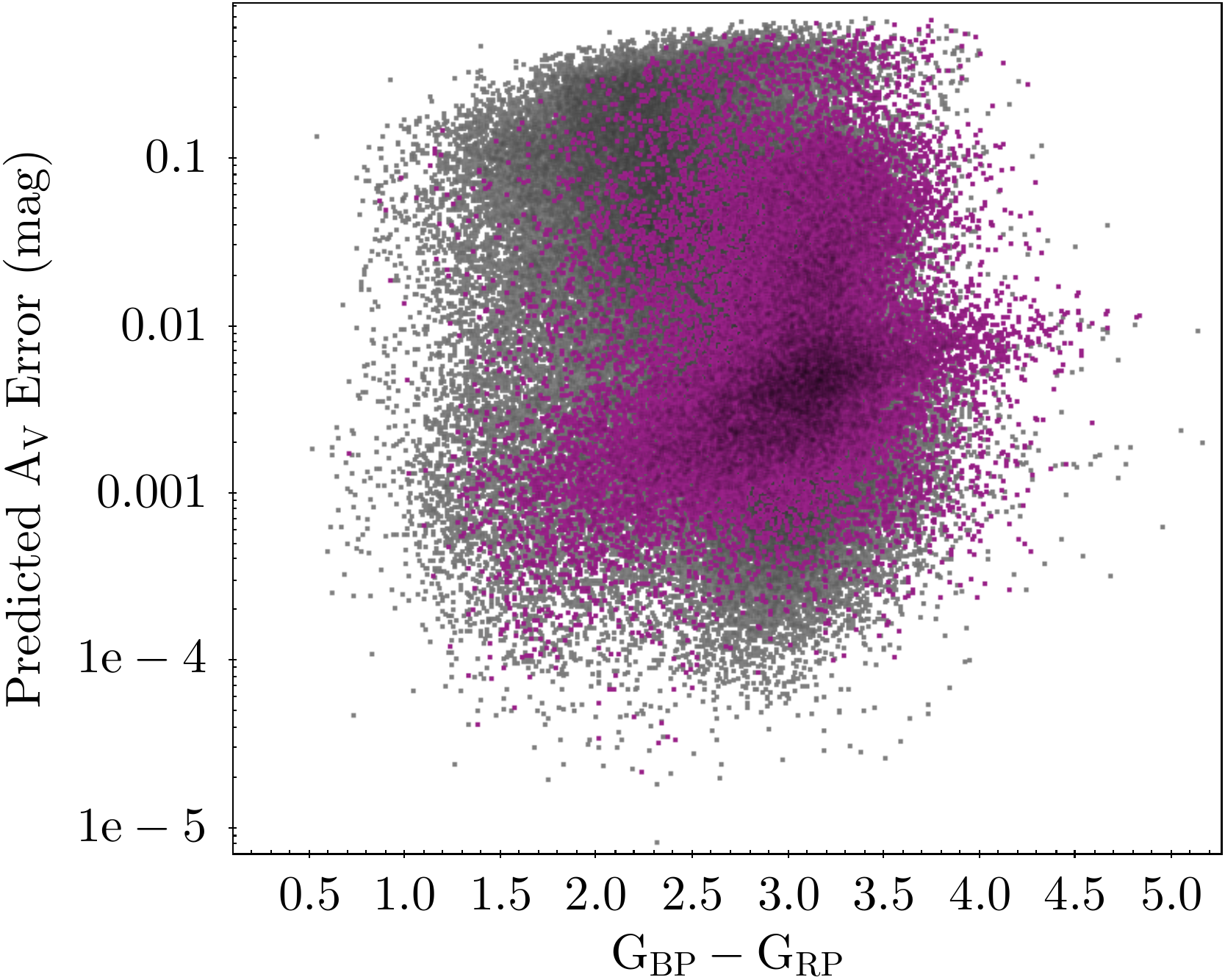}{0.33\textwidth}{}
            }
\caption{Errors produced for sources with YSO probability $>$70\% (in grey) and $>95$\% (in color) by scattering model inputs by their reported uncertainties.  \label{fig:errors}}
\end{figure*}

Convolution neural networks by default do not account for the uncertainties in the data, nor do they output the corresponding uncertainties in the predictions. Although some other machine learning architectures may be able to better positioned to learn the average uncertainty in the data and provide a resulting Bayesian posterior distribution, even they struggle to accurately parse individual source-by-source input-by-input uncertainty, such that can be present in e.g., photometry or in parallaxes.

As an alternative, we utilized the method used by \citet{olney2020} and that was used in \citetalias{kounkel2020}, by generating 1000 samples per each star for the sources in Table \ref{tab:sources}, where all the inputs are scattered by adding errors to them drawn from a normal distribution with the width corresponding to the reported uncertainty. Each one of these realizations of the same star was passed through the network. Uncertainties were estimated by calculating the standard deviation of the outputs.

By using this method, our uncertainties are indicative of both the model's stability at any given photometric regime, and the underlying photometric errors present in the input data. If photometric errors were not available in any bandpass (this is occasionally common for 2MASS data, even if fluxes themselves are available), they were assigned an uncertainty of 0.1 mag.

Despite the efficiency of neural networks, it is still time consuming to process the entire \textit{Gaia} catalog even once, let alone several times, particularly on the machines without GPU acceleration. Thus the statistics were generated only on the subset of the evaluation sample that has been classified with unaltered inputs with probability of being PMS $>70$\%.

When comparing the classification outputs for the unaltered sample to the mean classification from 1000 altered samples considering the uncertainties, the mean classifications do appear to be somewhat more accurate and are better able to filter out the suspicious sources (such as those described in Section \ref{sec:classifiervalidation} as likely false positives). The mean classification probability is also typically lower, and thus we are not likely to miss a significant number of sources for which the mean classification returns a higher confidence.

As expected, the scatter/errors increase for sources with fainter and more uncertain fluxes, as well as for those that are more distant and have more uncertain parallaxes. The scatter in the classification outputs also increases for sources with lower certainties (as a consequence, those that are older). 

The computed uncertainties in age are typically on the order of 0.1 dex (Figure \ref{fig:errors}). They do not strongly depend on age or color of a star, although there is a slight dependence on distance, with the nearby populations having somewhat lower uncertainties. 

\begin{figure} 
\plotone{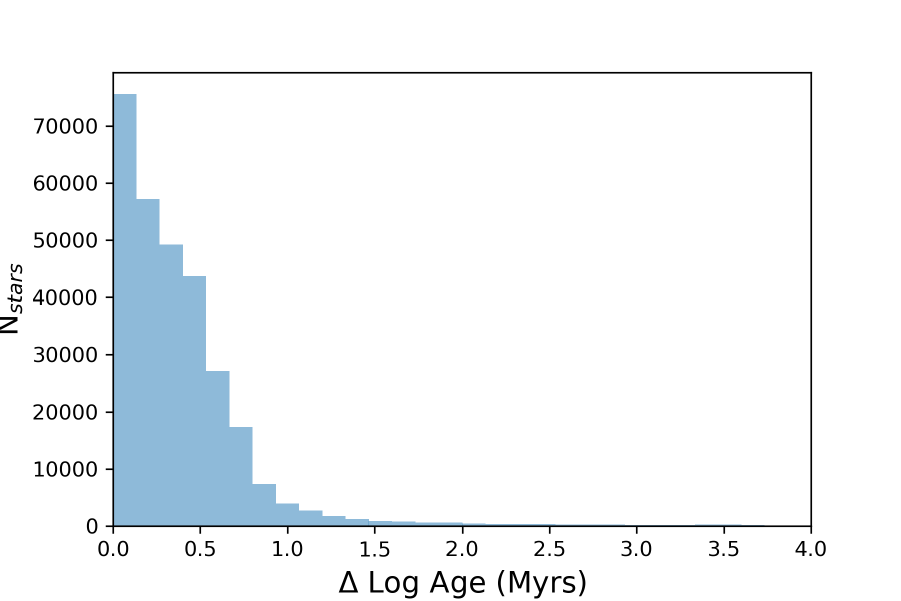}
\caption{ \label{fig:UpperScoAgeDiffs} Histogram of differences in Sagitta's age inferences for candidate Upper Sco members, at their true distance and as synthetically shifted to distances between 30 and 500pc. These log age differences appear gaussian in distribution, with a characteristic width of $\sigma = 0.3$ dex.}
\end{figure}

The ages Sagitta assigns to the synthetically shifted Upper Sco candidates provide an additional check on the uncertainties associated with these estimates. Sagitta infers a mean Log age of 6.8 dex to the candidates, as averaged over all distances; this compares to the nominal age of 10 Myrs as adopted by \citet{luhman2020}.  Some distance dependence is present in the age assignment, particularly at distances less than 150 pc; calculating the dispersion in the ages assigned to each source over all distances, and then examining the median of those dispersions, provides an empirical estimate of the stability of Sagitta's age assignments. The distribution of the dispersions in age assignments are shown in Fig. \ref{fig:UpperScoAgeDiffs}, and suggest that Sagitta's age assignments have a characteristic uncertanity of 0.3 dex.

\section{Evaluation and Validation}\label{sec:validation}
\subsection{Overall performance}

\begin{figure*} 
\gridline{\fig{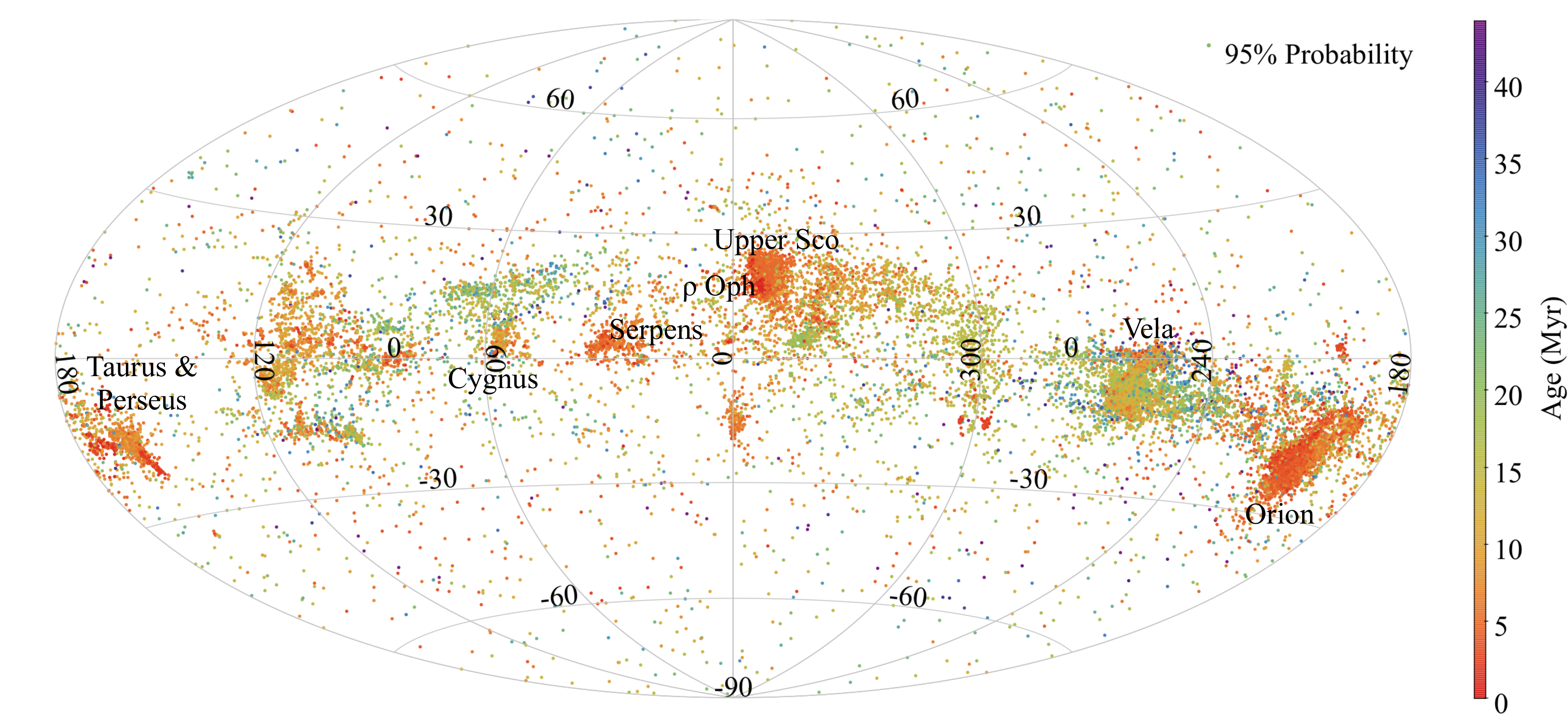}{.75\textwidth}{}}
\gridline{\fig{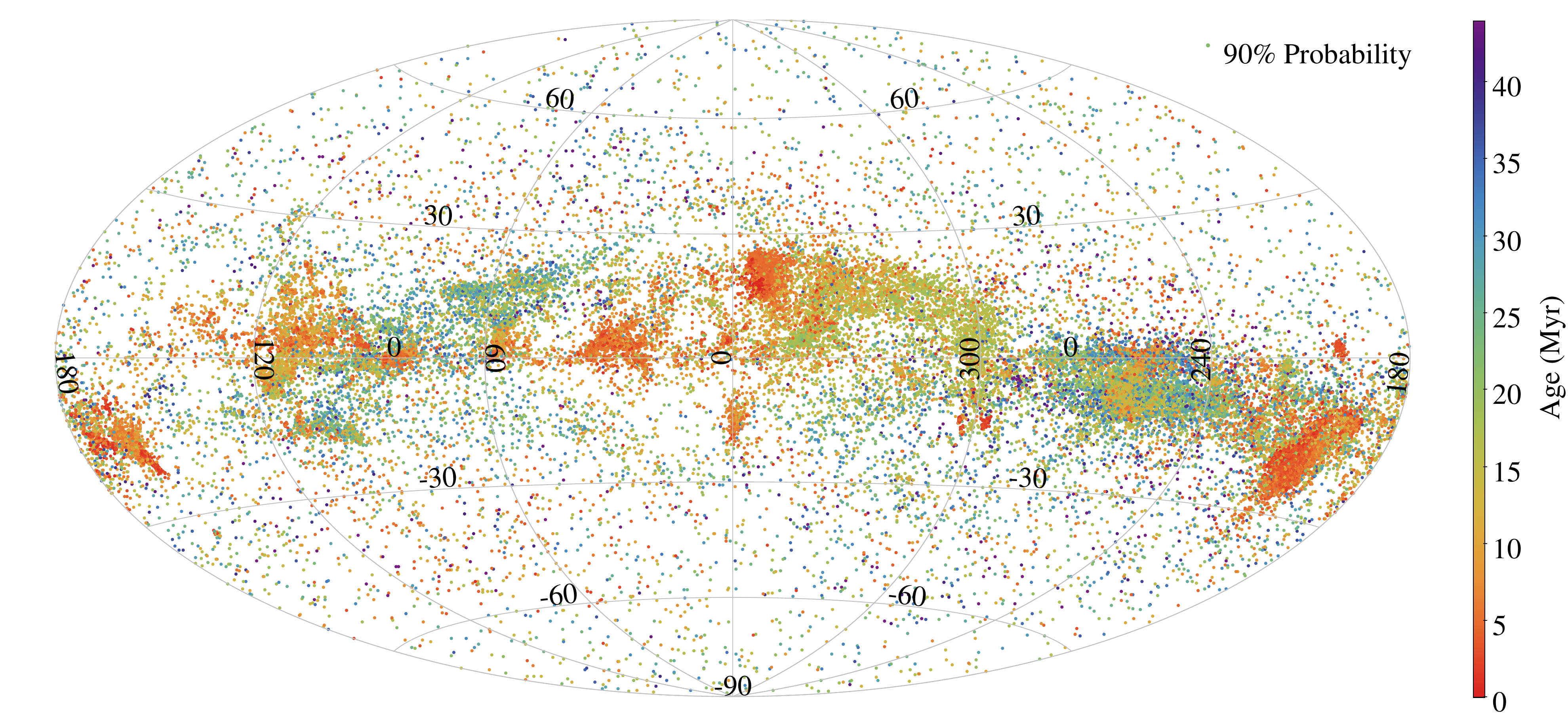}{.75\textwidth}{}}
\gridline{\fig{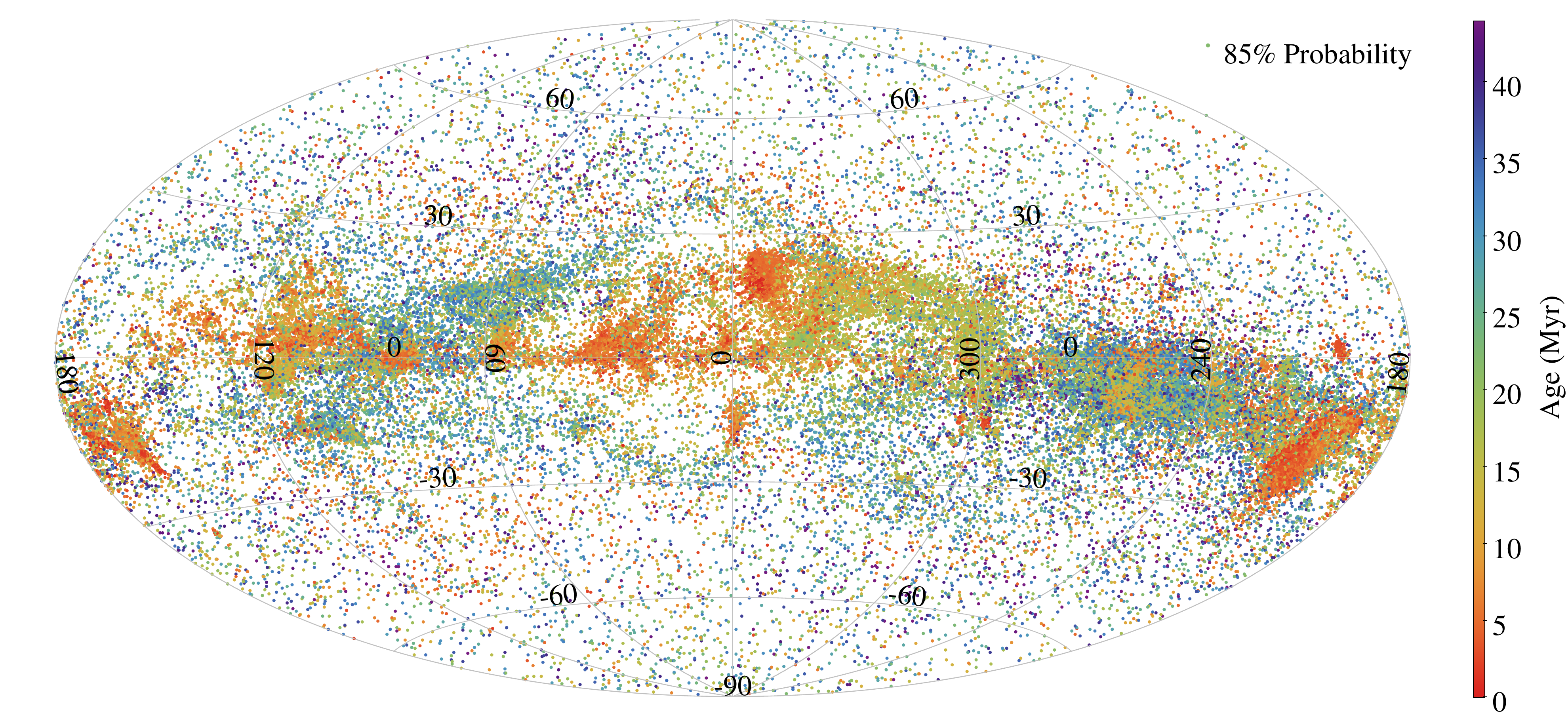}{.75\textwidth}{}}
\caption{The distribution stellar positions of the evaluation sample up to PMS probabilities of 95\%, 90\%, and 85\%. The plots are in the Galactic coordinates and are color coded by the predicted age, and the first panel has been annotated to indicate notable star forming regions. Note that older sources are more apparent at lower certainty thresholds, as they are located closer to the main sequence. \label{fig:outputs}}
\end{figure*}

\begin{figure*}
\gridline{\fig{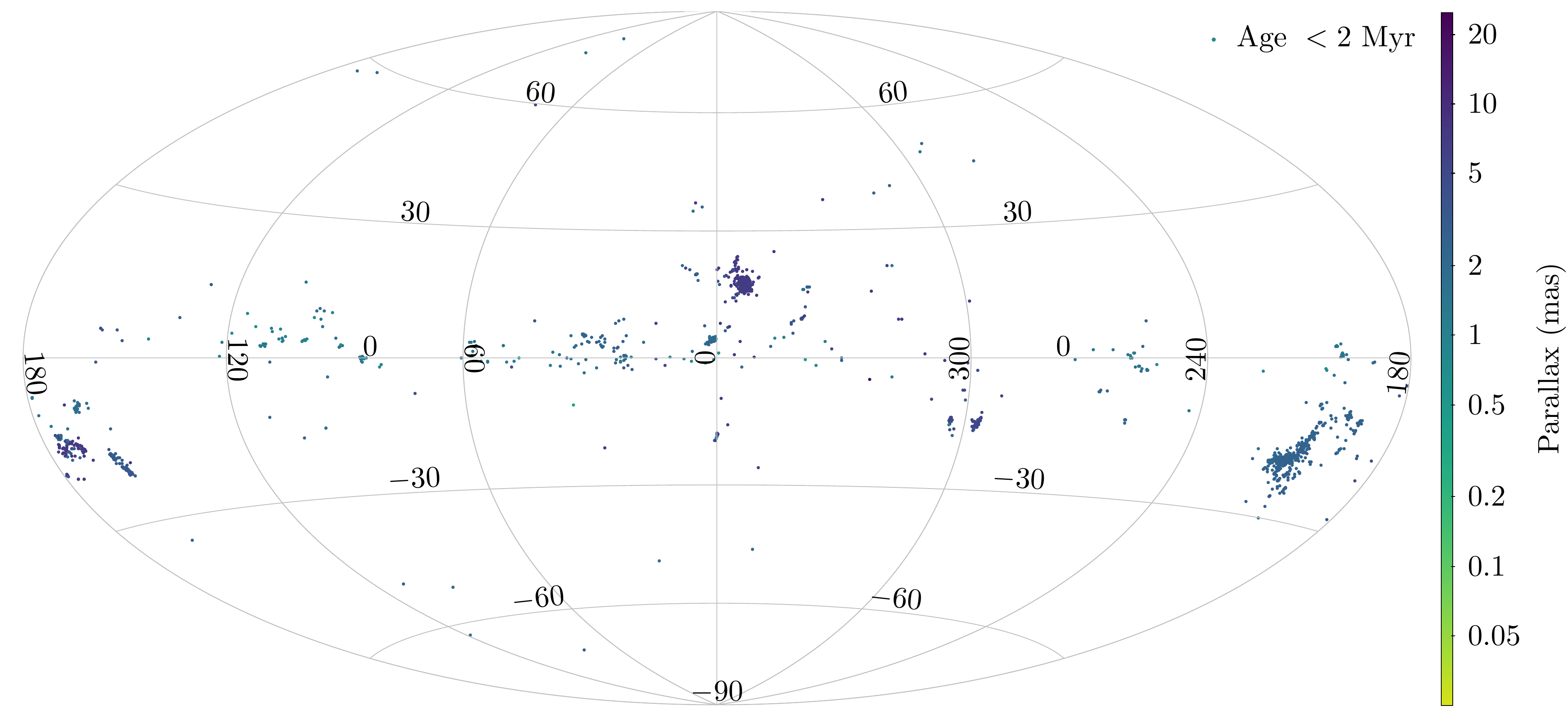}{.5\linewidth}{}
\fig{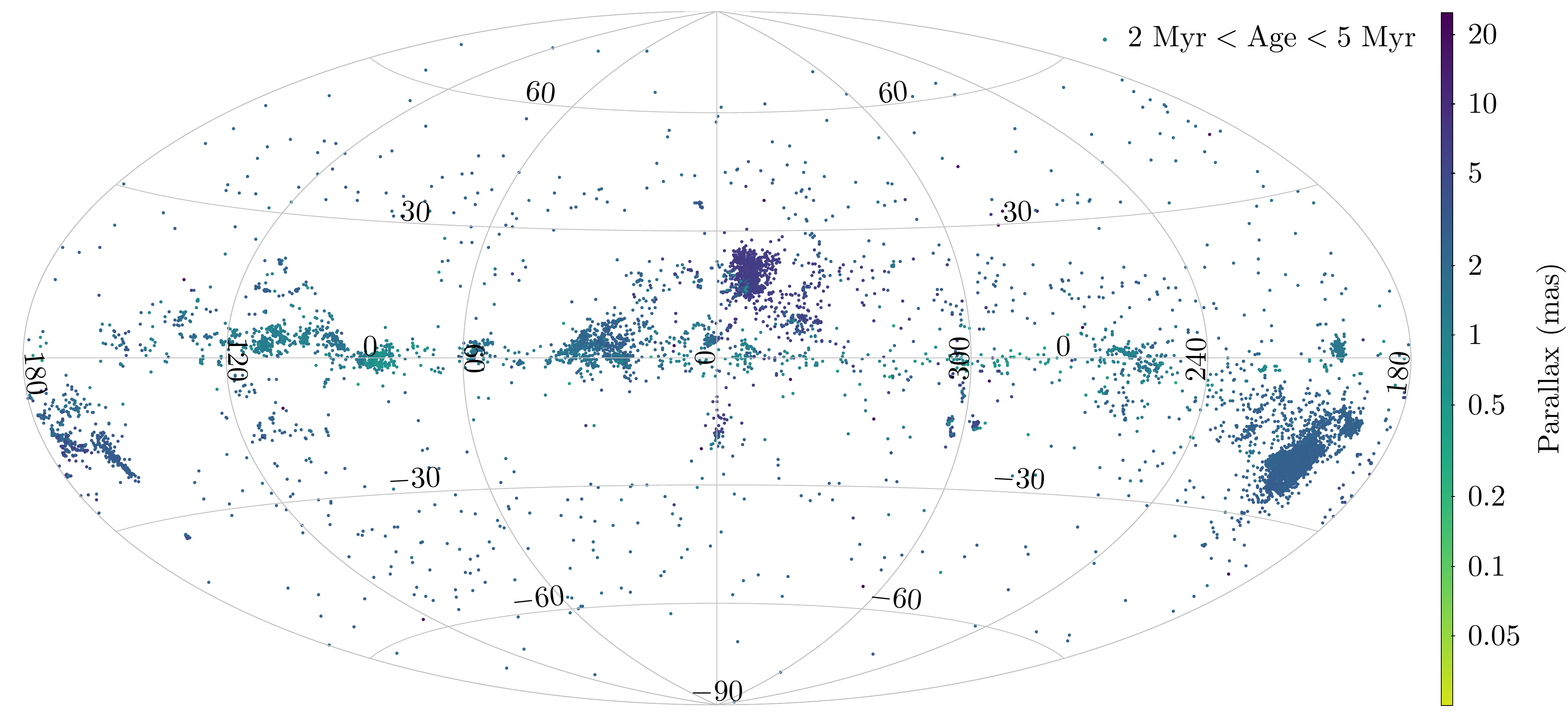}{.5\linewidth}{}}
\gridline{\fig{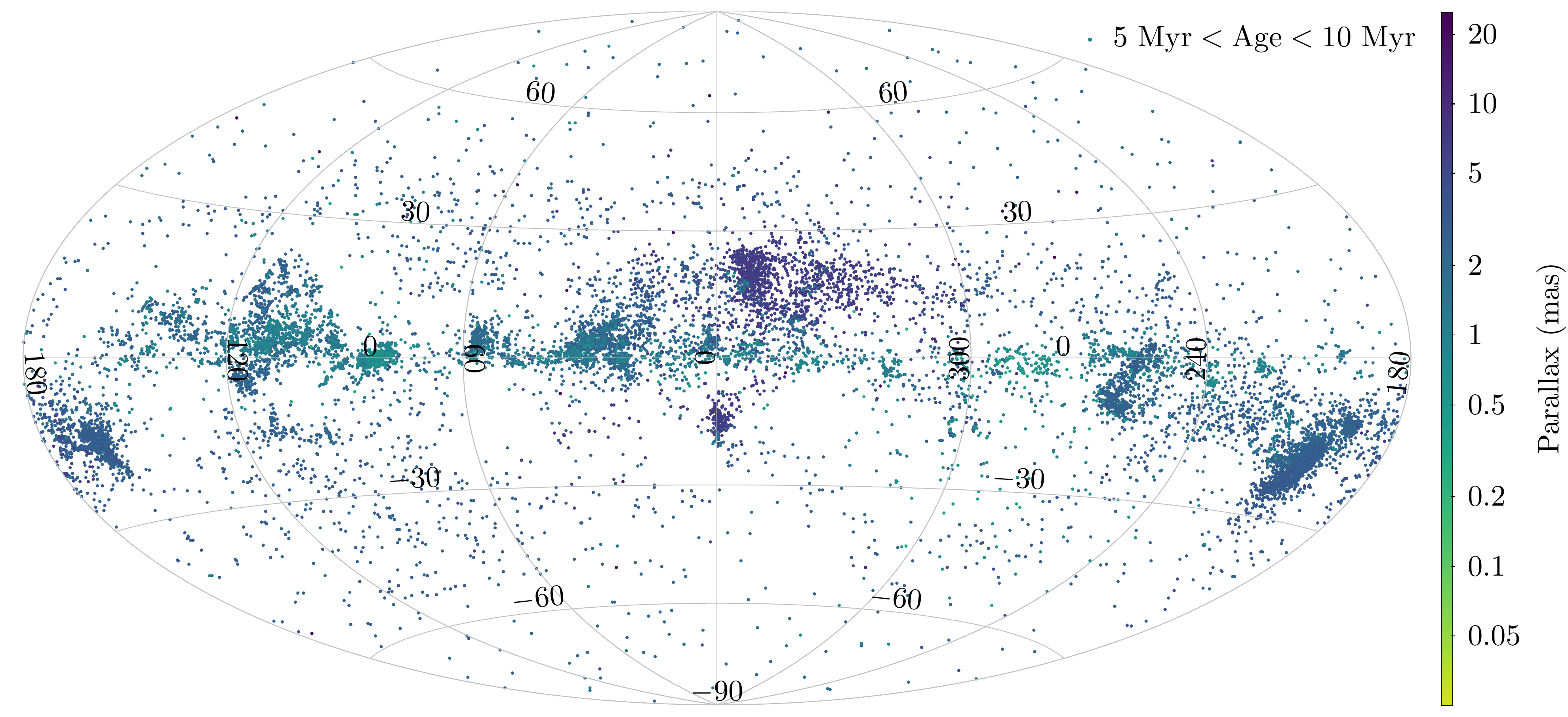}{.5\linewidth}{}
\fig{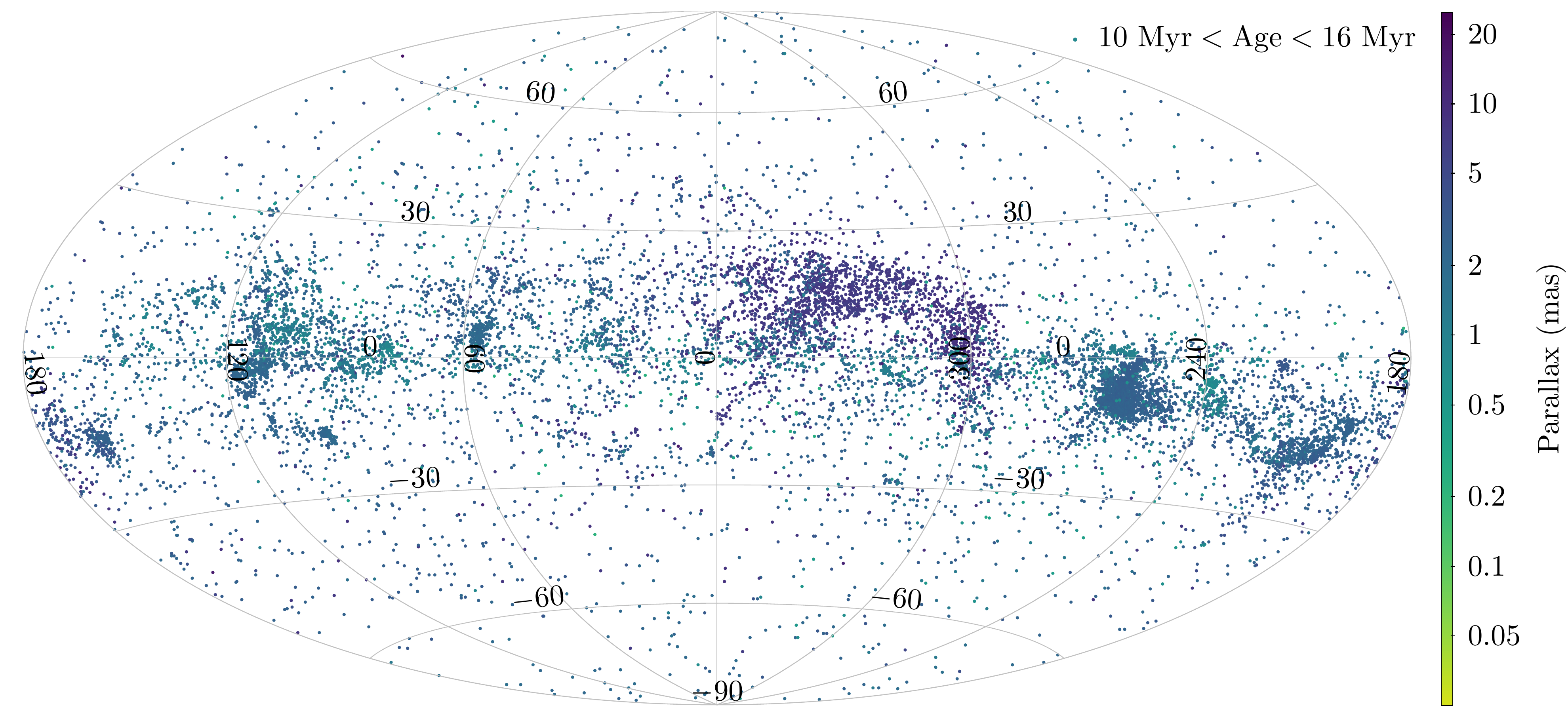}{.5\linewidth}{}}
\gridline{\fig{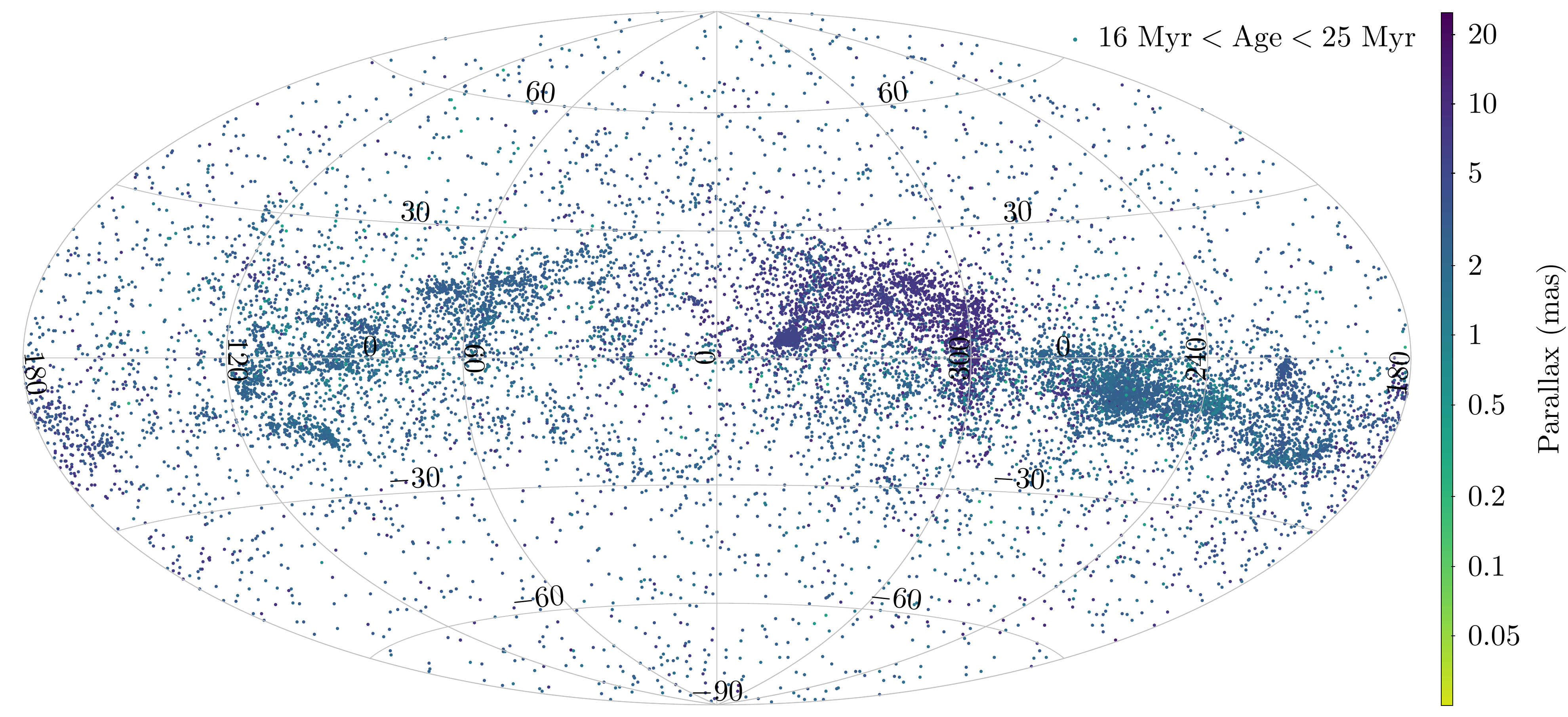}{.5\linewidth}{}
\fig{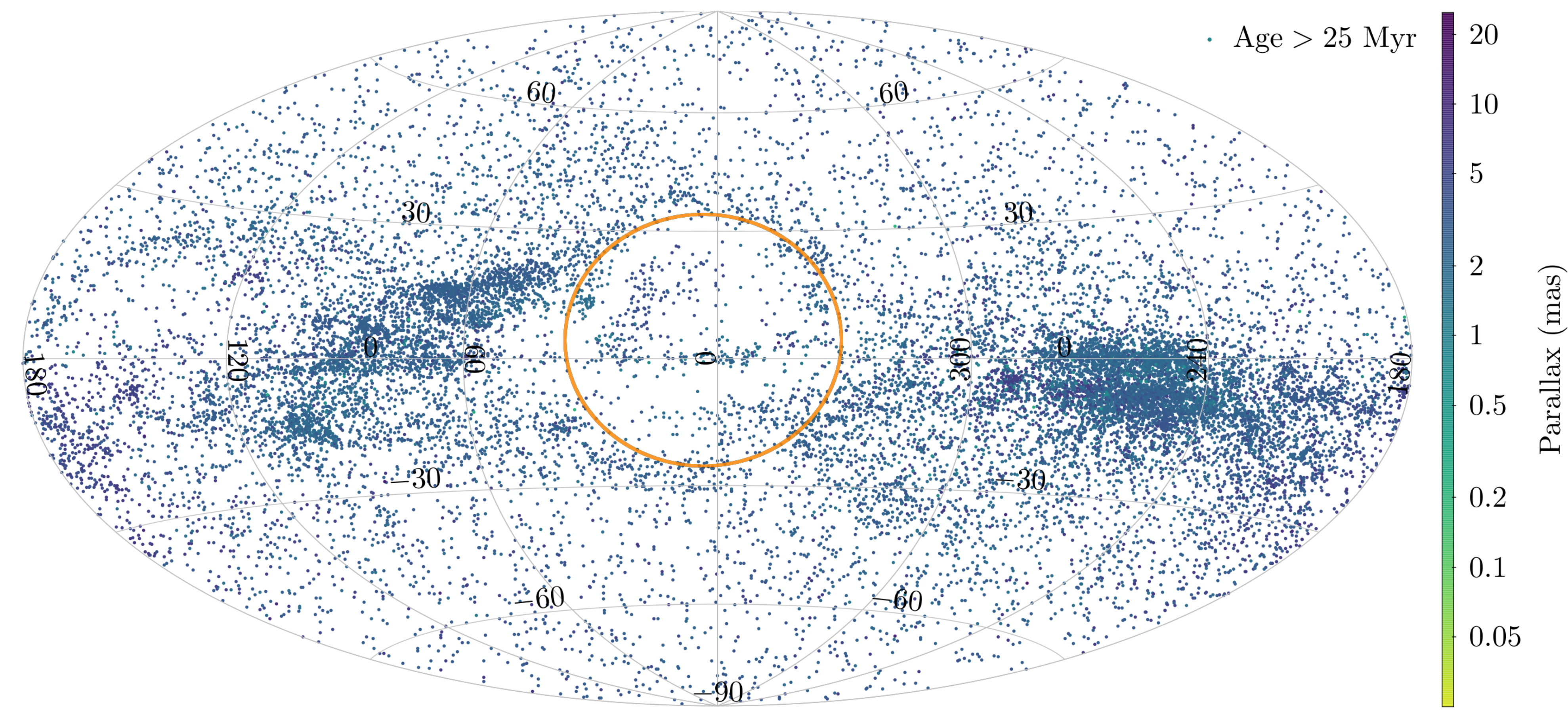}{.5\linewidth}{}}
\caption{Distribution of stars at various age ranges for sources with 85\% confidence threshold, color coded by parallax, to demonstrate the various features that emerge for different epochs of star forming history. Note the ring-shaped structure apparent in the last panel, discussed in section \ref{sec:30MyrBubble}. \label{fig:AgeRanges}}
\end{figure*}

Table \ref{tab:sources} contains the catalog of the sources in the evaluation catalog that can be identified as PMS sources with at least 70\% confidence. This catalog consists of 197,315 sources. Figure \ref{fig:outputs} shows the distribution of the identified sources along the sky, according to the different cuts in confidence levels, color-coded by their estimated ages. Figure \ref{fig:AgeRanges} shows the spatial distribution of stars at different age slices, to better highlight the star forming history of the solar neighborhood.

\begin{deluxetable*}{cccccc}
\tablecaption{Sample of output catalog based on DR2, with age, classification, and predicted extinction.\label{tab:sources}}
\tablehead{
\colhead{Gaia DR2} & \colhead{$l$} & \colhead{$b$} &  \colhead{Predicted Age} &  \colhead{PMS} &  \colhead{Predicted $A_V$\tablenotemark{a}}\\
\colhead{Source ID} & \colhead{(deg.)} & \colhead{(deg.)} &  \colhead{(dex)} &  \colhead{Probability} &  \colhead{(3d pos., mag.)}}
\startdata
25220048169856 & 176.382 & -48.739 & 7.369 $\pm$ 0.177 & 0.740 $\pm$ 0.100 & 0.382 $\pm$ 0.008 \\
2194034672517070720 & 97.309 & 9.949 & 6.817 $\pm$ 0.028 & 0.762 $\pm$ 0.131 & 0.822 $\pm$ 0.008 \\
5224626096939825024 & 298.281 & 14.145 & 6.106 $\pm$ 0.032 & 0.955 $\pm$ 0.009 & 0.378 $\pm$ 0.001 \\ 
\enddata
\tablenotetext{}{Only a portion shown here. Full table is available in an electronic form.}
\tablenotetext{a}{Spatially averaged $A_V$ estimate at a given $l$, $b$, and $\pi$, not to be confused with the true $A_V$ of a star, which can be significantly higher, particularly in the young stars still associated with dusty envelopes and/or disks.}
\end{deluxetable*}

The evaluation catalog extends up to 5 kpc in parallax, however, at larger distances, increasingly fewer low mass stars can be detected. Therefore, 90\% of all sources classified as PMS sources are located within 1 kpc, and 70\% are located within 500 pc. The difference becomes more extreme at particular age ranges. Only lower mass stars can be still be identified as PMS at older ages (e.g., $>$30 Myr), thus, the ability to identify them at larger distances is suppressed compared to younger (e.g., $<5$ Myr) stars. Similarly, the confidence with which PMS stars can be identified tends to be lower both for sources that are older as well as for sources that are more distant (compared to younger nearby sources), as both of them are located preferentially closer to the main sequence and/or the red giant branch.

We note that only $\sim$30,000 out of $\sim$200,000 candidate PMS stars presented here were also used in the training and testing sample from \citetalias{kounkel2020}. Furthermore, while some of them may have also been previously included in other studies of pre-main sequence stars \citep[most notably in][see Section \ref{sec:zari}]{zari2018}, the bulk of the catalog are new identifications. 

\begin{figure*} 
\epsscale{1}
\plottwo{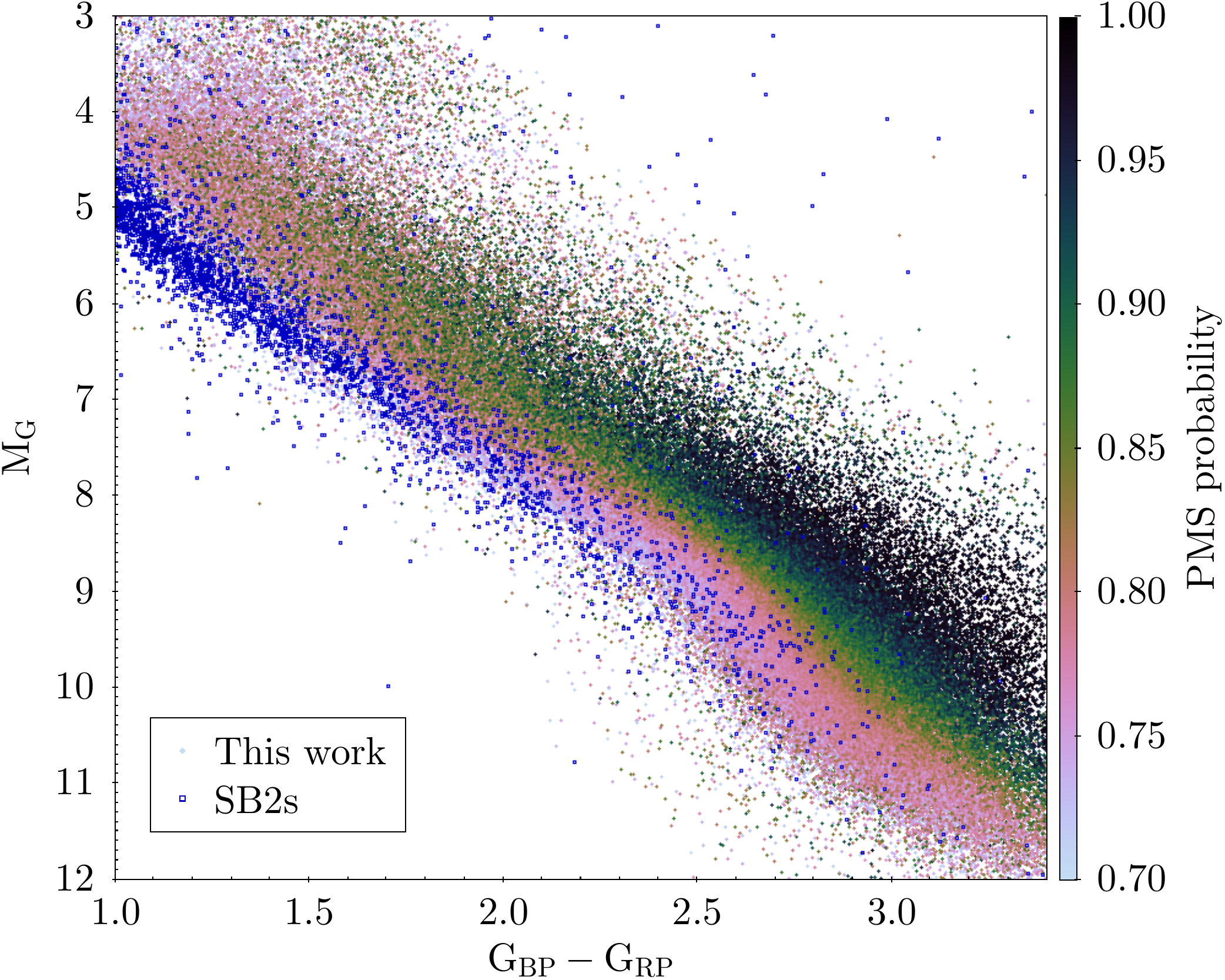}{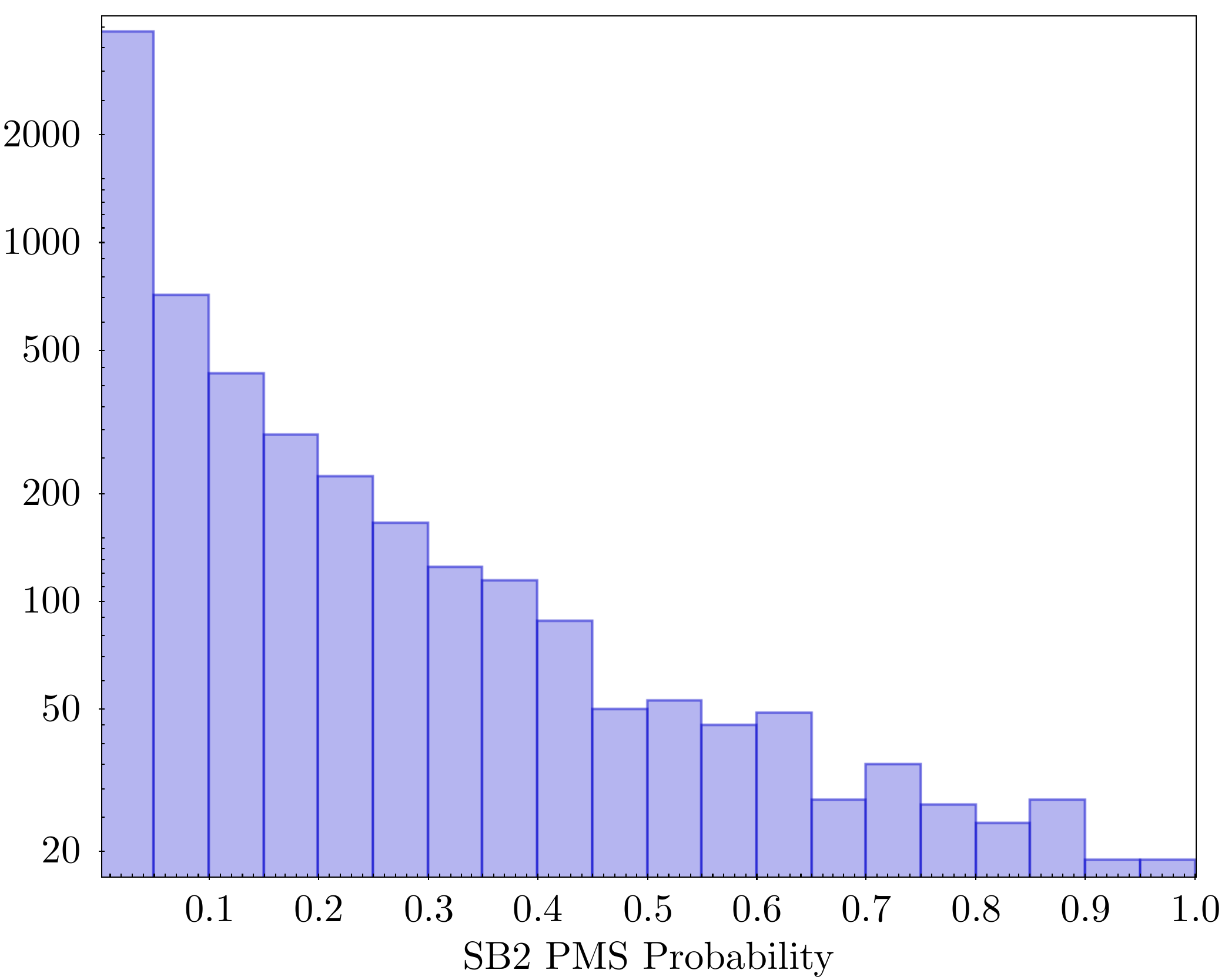}
\caption{Left: HR diagram of the evaluation sample, color-coded by the average probability of a source being PMS in each 2-dimensional bin. On top of it, in blue, are the stars identified as double lined spectroscopic binaries in the APOGEE data (M. Kounkel, et al, in prep). These sources preferentially trace the photometric binary sequence. Right: Distribution of probabilities of classifying these SB2s as pre-main sequence. Note, although fields associated with notable star forming regions (e.g, Orion, Upper Sco) have been excluded form the sample, some bona-fide PMS stars may have still persist in other fields. \label{fig:binarysequence}}
\end{figure*}

Few notes of caution should be given to the unresolved binaries. The classifier largely avoids the binary sequence of evolved stars, particularly at higher probabilities (Figure \ref{fig:binarysequence}). No confusion occurs in the sources younger than $\sim$40--60 Myr in the sources, as they are located above the binary sequence. However some of contamination from binaries can be present at lower probabilities (e.g., at thresholds $<$70--80\%). For example, at 70\% threshold there is a hint of overdensity from the binary stars in Praesepe, which is a 600 Myr cluster that does not contain PMS stars. At 80\% threshold, this overdensity disappears.

\begin{deluxetable*}{cccccc}
\tablecaption{Sample of output catalog based on EDR3, with age, classification, and predicted extinction.\label{tab:edr3}}
\tablehead{
\colhead{Gaia EDR3} & \colhead{$l$} & \colhead{$b$} &  \colhead{Predicted Age} &  \colhead{PMS} & \colhead{Predicted $A_V$\tablenotemark{a}}\\
\colhead{Source ID} & \colhead{(deg.)} & \colhead{(deg.)} &  \colhead{(dex)} &  \colhead{Probability} & \colhead{(3d pos., mag.)}}
\startdata
110951890948096 & 176.087 & -48.388 & 0.871$\pm$0.019 & 7.421$\pm$0.035 & 0.415$\pm$0.008 \\
265532058523264 & 175.988 & -47.522 & 0.767$\pm$0.054 & 7.435$\pm$0.018 & 0.436$\pm$ 0.027\\
313670051618560 & 174.933 & -48.774 & 0.831$\pm$0.102 & 7.872$\pm$0.215 & 0.646 $\pm$ 0.050 \\
\enddata
\tablenotetext{}{Only a portion shown here. Full table is available in an electronic form.}
\tablenotetext{a}{Spatially averaged $A_V$ estimate at a given $l$, $b$, and $\pi$, not to be confused with the true $A_V$ of a star, which can be significantly higher, particularly in the young stars still associated with dusty envelopes and/or disks.}
\end{deluxetable*}

The reason for this is that the sources that are on the binary sequence have been included in the training sample, thus the classifier has learned that the colors that correspond to these main sequence binaries are likely false positives. It cannot effectively separate true single PMS stars older than 60 Myr that overlap with the binary sequence. However, as such young stars are rare in comparison to main sequence stars, the classifier downweights them both equally to minimize the loss. Thus, minimizing contamination from main sequence binaries results in a lack of stars $>$60 Myr included in Table \ref{tab:sources}. Nonetheless, with independently derived membership (such as from analyzing the distribution of stars in the phase space in older regions regions with bona fide PMS stars, like in $\alpha$ Per), it is nonetheless possible to estimate their ages without relying on the classifier.

Unfortunately, however, Sagitta is unable to separate pre-main sequence unresolved binaries from single PMS stars. In young populations where there is a clearly defined binary sequence for the cluster, the stars on that binary sequence get assigned preferentially younger ages by $\sim$0.1-0.15 dex than the age of the single stars in the same cluster. This is consistent with the relative ages for single and binary stars that could be estimated with traditional isochrone fitting. Thus, the ages of sources that could be suspected to be unresolved binaries or tertiaries \citep[such as in the cases where binary sequence for a cluster is apparent, which can be seen in regions as young as 8 Myr with a mono-age population e.g.,][and especially in the younger regions without a clearly defined binary sequence]{bouma2020} should be treated with care.

Recently released \textit{Gaia} EDR3 \citep{gaia-collaboration2020} has changed the definition of the bandpasses compared to DR2. The parallax has also been improved by $\sim$30\%. There are no strong systematics in the performance of Sagitta when applied to EDR3, and the measurement of age and classification compared to DR2 is generally consistent with each other within 1$\sigma$ according to the reported errors. Applying the pipeline to the EDR3 sources that meet the same quality checks as described in Section \ref{sec:testsample} results in a larger catalog of stars that can be identified as likely PMS, but that is primarily driven by these sources previously not meeting the required precision in parallax and/or fluxes. The sources that are newly identified as PMS in EDR3 tend to be fainter and be located at preferentially larger distances, extending the sensitivity limits of the survey. The resulting catalog is included in Table \ref{tab:edr3}. We note that while the analysis in this paper, including the subsequent sections, is limited to the DR2 data, overall conclusions are consistent in the sample derived from EDR3.

\subsection{Spectroscopic validation}

\subsubsection{Li I}

Outside of known star forming regions, currently, only a few sources have existing spectra. LAMOST has coverage only of the northern hemisphere, furthermore, it avoids large parts of the galactic plane. Despite that, LAMOST DR5 has $\sim$5,900 stars coincident with our catalog. While most of them are concentrated in the Orion Complex, Taurus, and Perseus, there are some are more distributed across the galactic plane, and the sky in general. One of the ways through which it is possible to confirm a star to be pre-main sequence is the presence of Li I absorption line, which approaches equivalent width $Eqw_{Li}\sim0.5$ \AA\ in low mass stars with the age of a few Myr. However, it depletes rapidly in a color-dependent fashion \citep[e.g.,][]{baraffe2010}. Selecting LAMOST spectra to have signal-to-noise ratio in $r$ band $>$30 to ensure robust detection ($\sim$2000 sources), we measure $Eqw_{Li}$ for the stars overlapping with our sample.

\begin{figure*} 
\epsscale{1}
\plottwo{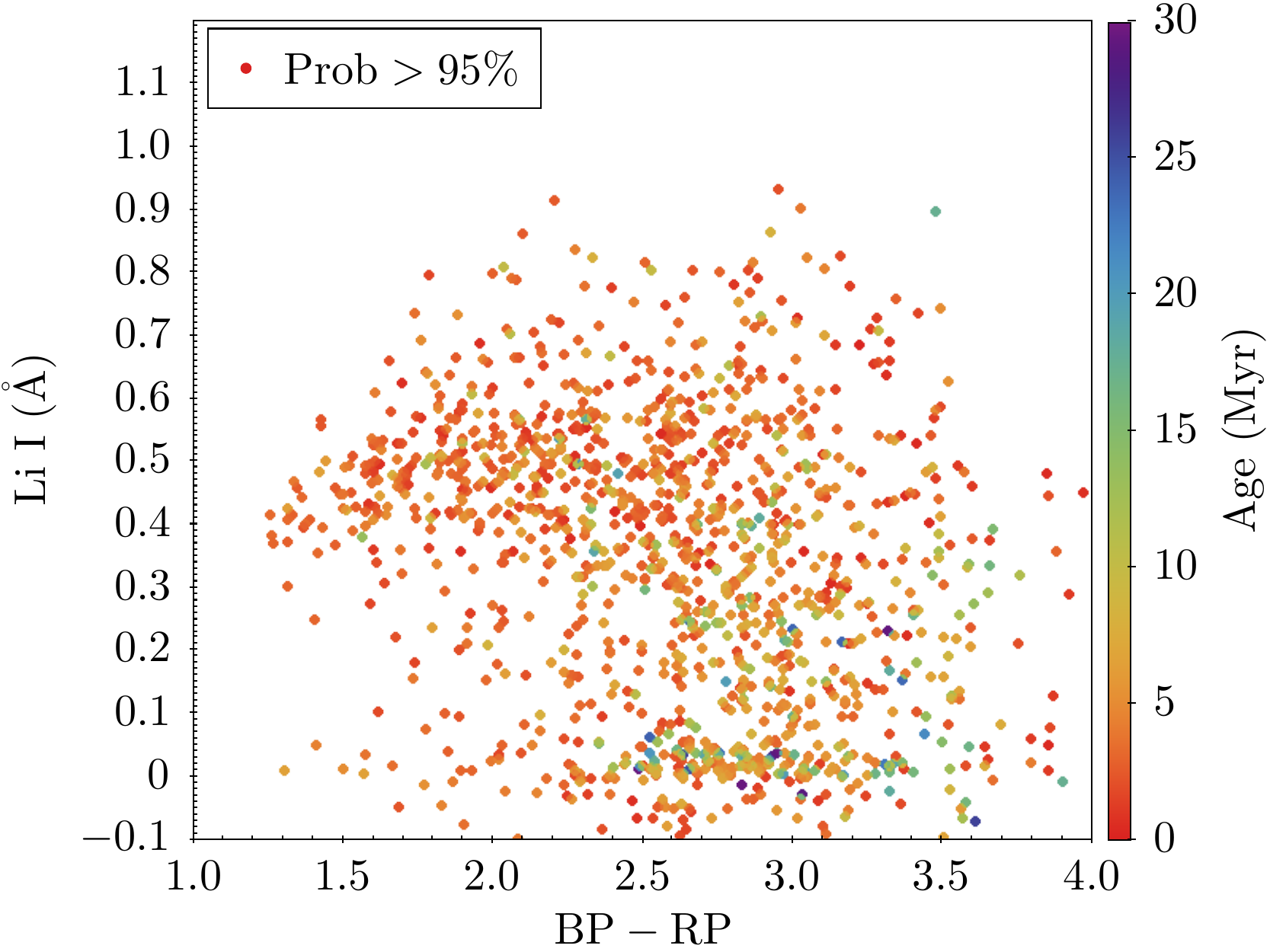}{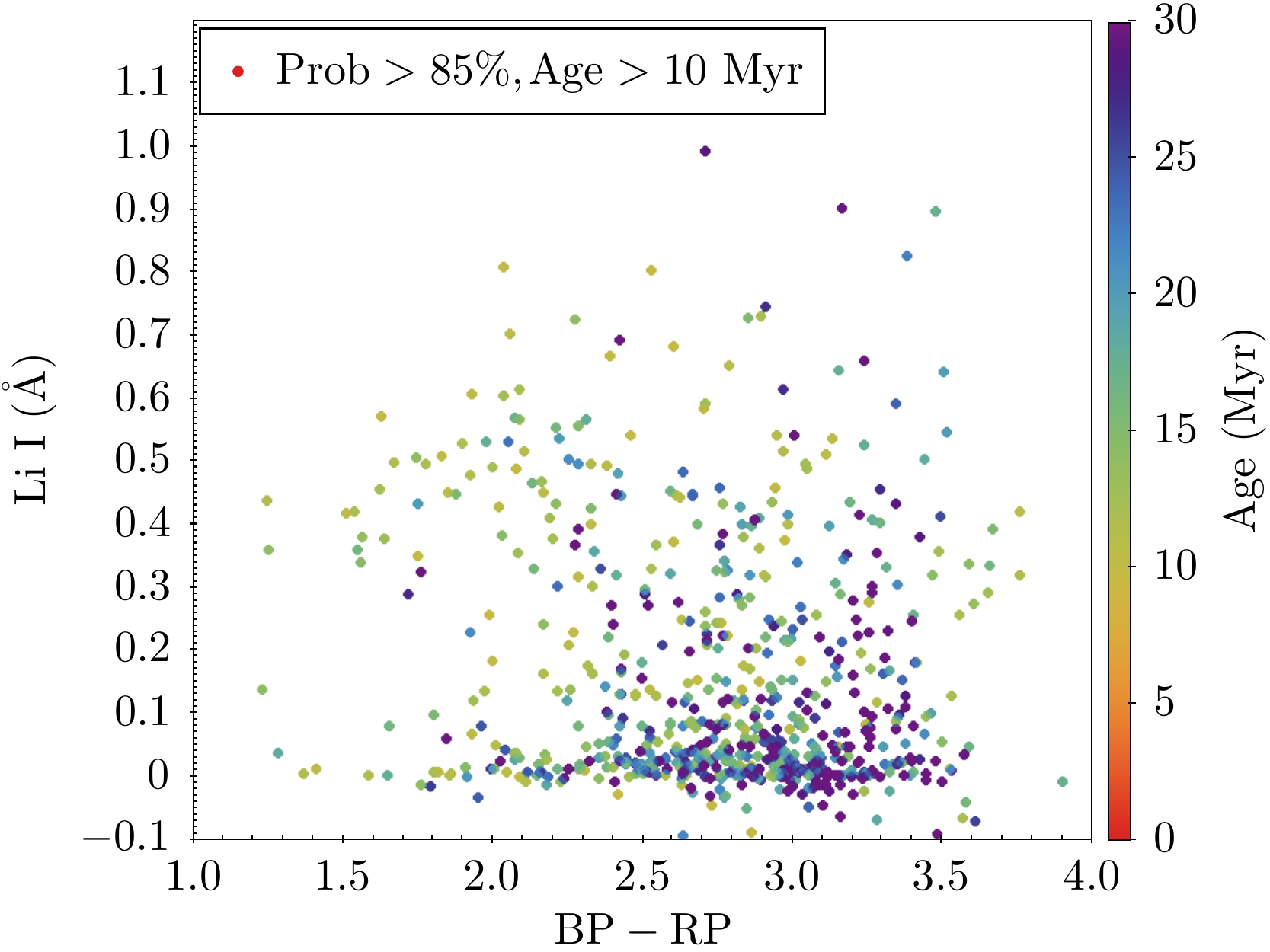}
\caption{Distribution of Li I equivalent widths in the highest confidence sample and in the medium confidence samples (restricted to only sources older than 10 Myr) as a function of color. Note that Li I begins depleting at BP-RP$\sim$3 even in the youngest and most rigorously selected sample; at the older ages this is the primary color range that we are sensitive to with the photometric selection. \label{fig:lamostli}}
\end{figure*}

We consider $Eqw_{Li}>0.1$ \AA\ to be a firm confirmation of youth in the stars. We note that stars as young as 20 Myr should deplete most of their Li I content in the color range we are most sensitive to with our selection at those ages (Figure \ref{fig:lamostli}). Indeed, spectra of many low mass stars appear to have strongly defined absorption lines with $Eqw_{Li}\sim0.05$ \AA, which this cut would not include. However, in the interest of being conservative in our estimates and avoiding possible confusion with other lines, in this exercise such sources would be considered to be more evolved. In the future, as more optical spectra are available, a more careful consideration of such systems should be possible.

\begin{figure} 
\epsscale{1}
\plotone{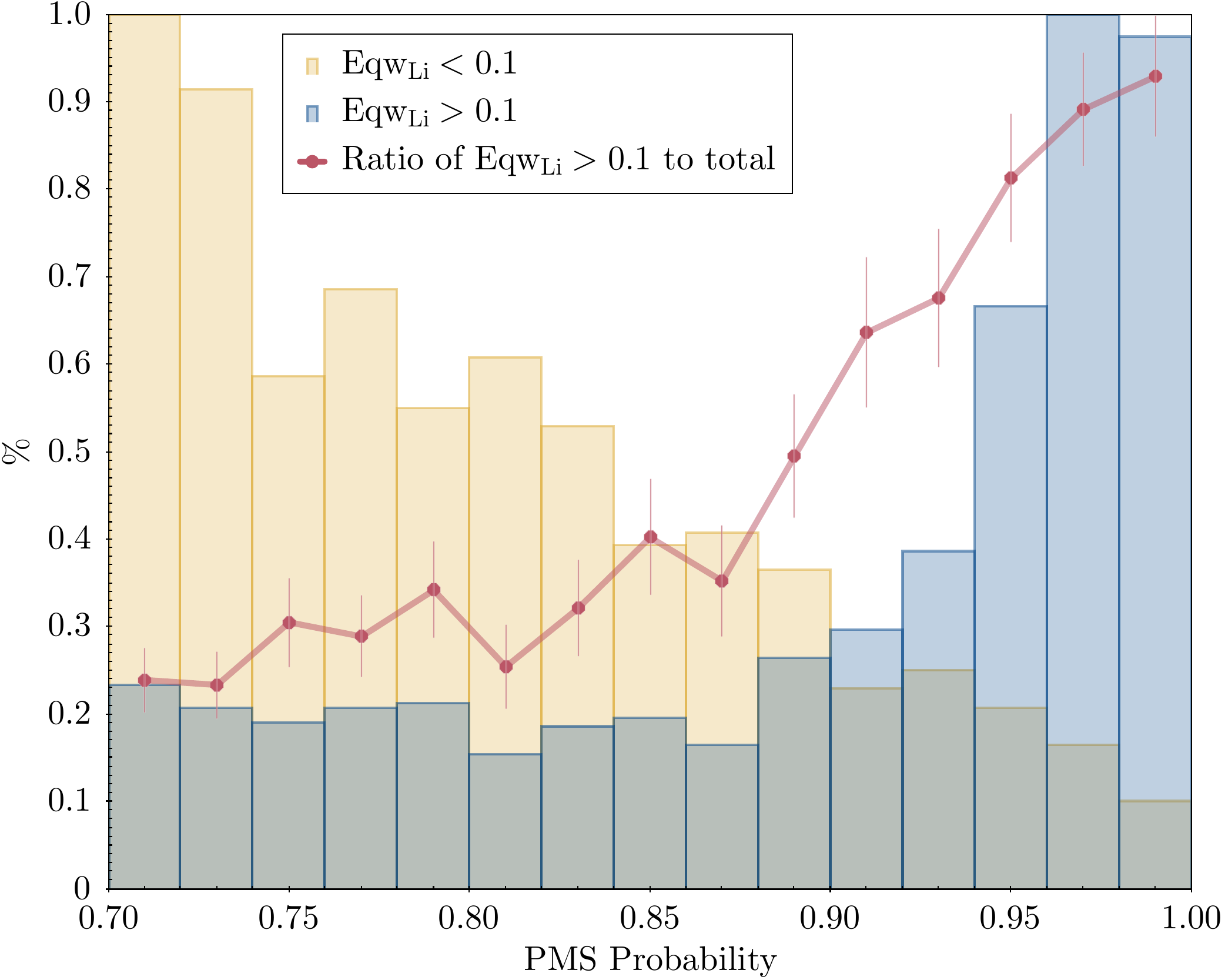}
\caption{Distribution of sources with confirmed signature of youth based on the LAMOST spectra. Blue histogram shows the distribution of sources with a clear detection of Li I with $Eqw_{Li}>0.1$ \AA. Yellow histogram shows sources in which Li I has been mostly depleted (which may include a number of PMS sources with age$>$20 Myr). The histograms are normalized to their peak. The red line shows the ratio of sources with $Eqw_{Li}>0.1$ \AA\ to the full LAMOST sample. \label{fig:lamostprob}}
\end{figure}

We evaluate the fraction of sources with $Eqw_{Li}>0.1$ \AA\ as a fraction of total sample with respect to their reported probabilities of being PMS in Figure \ref{fig:lamostprob}. Sources identified at high probabilities can almost uniformly be confirmed to be young. The parity with more evolved sources is reached at $\sim$85--90\% as the confirmed YSOs plateau at lower probabilities and the number of more evolved stars (including contaminants and older PMS stars that depleted their Li) increases.

Based on this result, we suggest two confidence thresholds for applying our model outputs depending on desired context: a ``highest confidence" threshold at $>$0.95 PMS probability, and a ``medium confidence" threshold at $>$0.85 PMS probability. In total, 24,626 sources meet the highest confidence threshold. Almost all of these sources have a clear Li I detection. This sample tends to be very young, with 97\% of its sources having predicted ages of $<$20 Myr. The medium confidence threshold contains 77,283 sources and roughly coincides with the aforementioned plateau of Li I detection. The sample defined by this threshold contains the bulk of YSOs with clear Li I detection while still retaining many YSOs which have undergone lithium depletion, and therefore contains a much larger proportion of older YSOs up to ages of roughly 45 Myr. To illustrate the differences in apparent source distribution within both thresholds, many of the plots within Sections \ref{sec:validation} and \ref{sec:discussion} have been constructed with these PMS probability cutoffs.

\subsubsection{Activity indicators}

Although a presence of Li I is the most direct method of confirming stellar youth, a number of other tracers can suggest it. In particular, young stars are magnetically active, this activity produces several prominent emission lines.

One such line is H$\alpha$. Classical T Tauri stars have a strong H$\alpha$ emission with a very wide profile in excess of -10 \AA\ for the late K-early M dwarfs \citep{white2003}. At this state, this emission is largely driven by accretion from the disk. After the disk depletes, weak lined T Tauri stars, H$\alpha$ emission weakens to $>$-10 \AA\ (for stars of a similar spectral type), but still remain strong due to the magnetic activity. Eventually, a star becomes inactive, and its H$\alpha$ equivalent width weakens to $\sim$0\AA. Such process is slow, it may take on an order of $500$ Myr to 1 Gyr, as such, it is not particularly robust in separating populations $<40$ Myr to that are older, e.g., 100 Myr. However, in the field, $\sim$80\% of M dwarfs are inactive, and only $\sim$20\% are still active \citep{newton2017}.

\begin{figure*}
\gridline{\fig{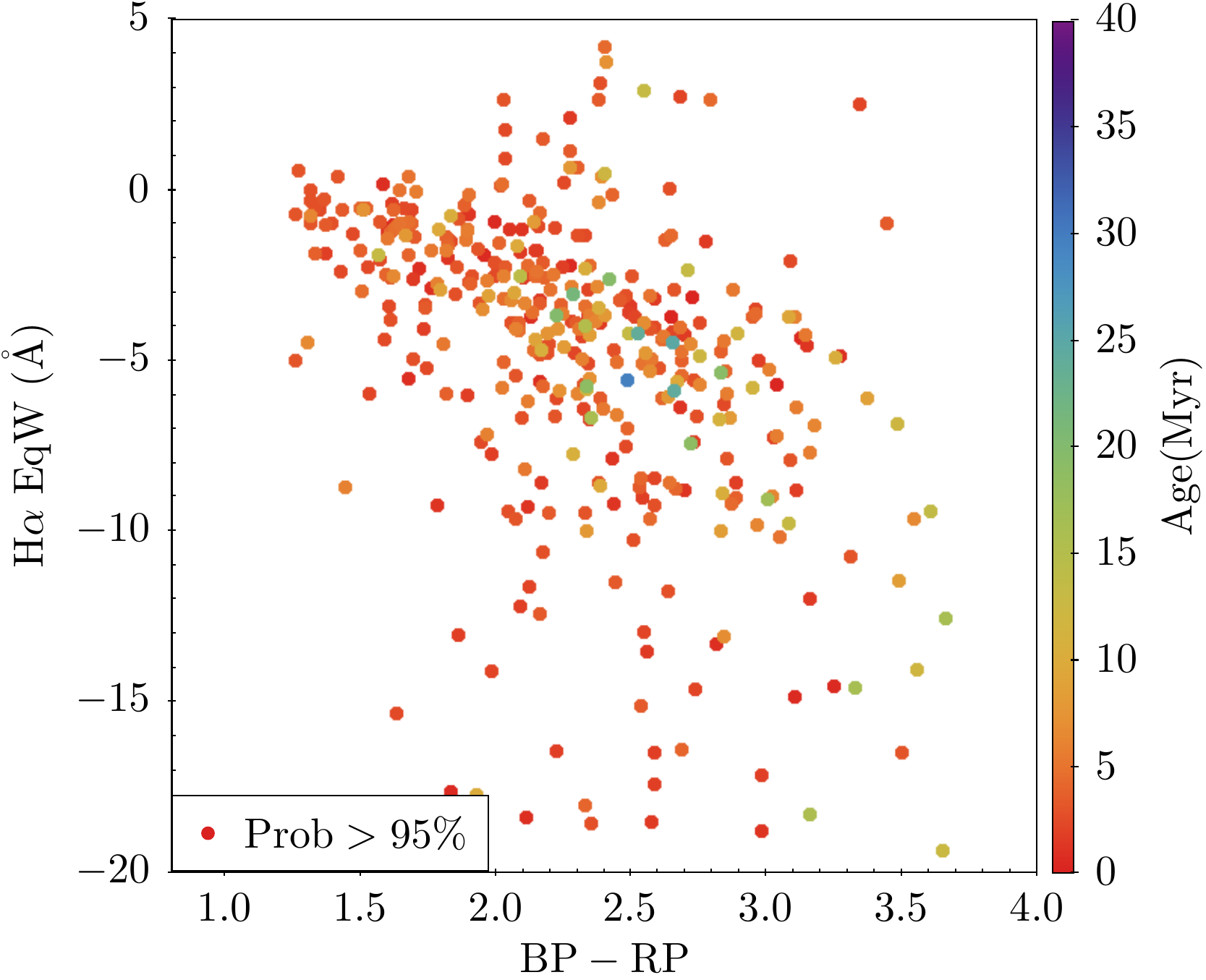}{.33\linewidth}{}
\fig{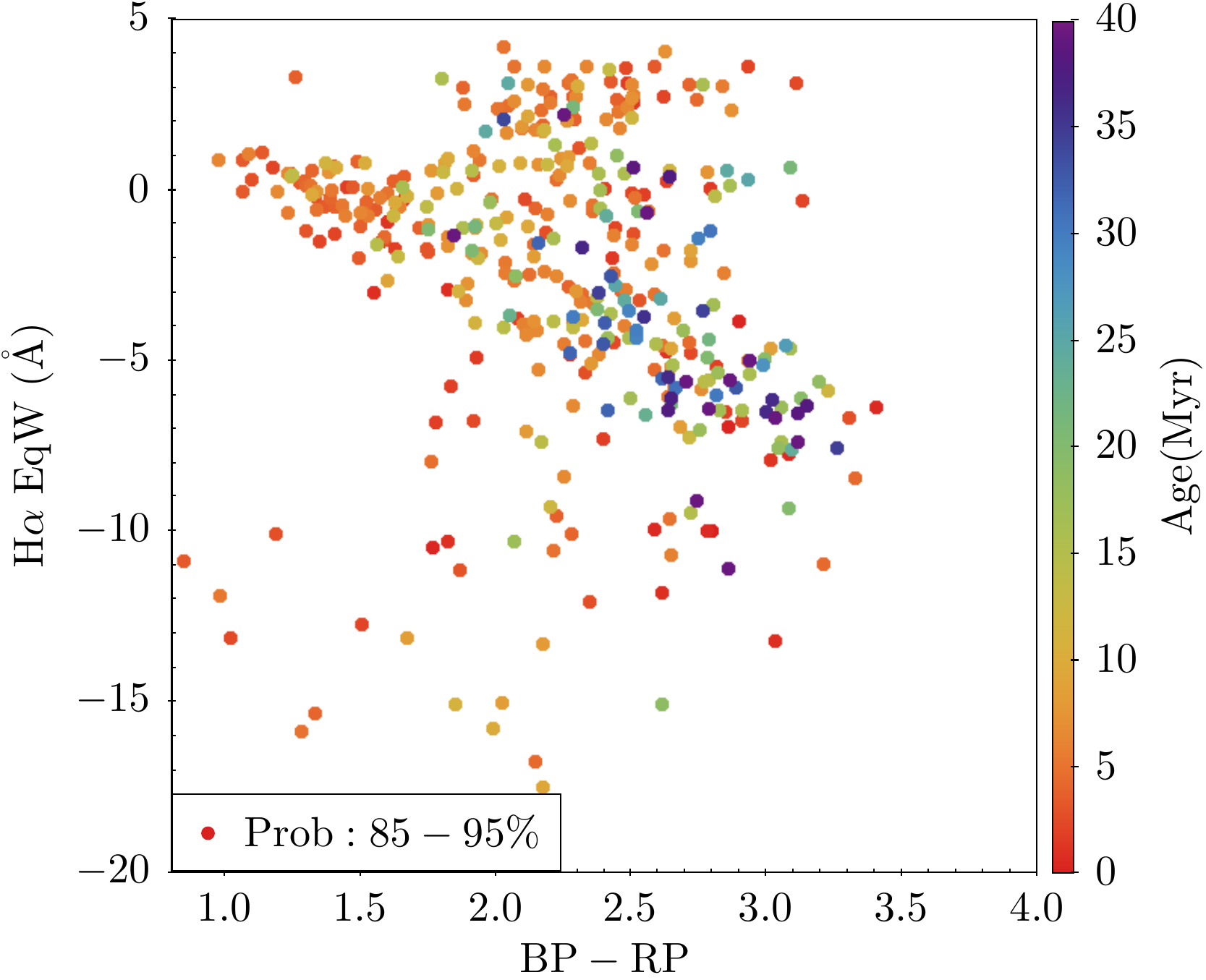}{.33\linewidth}{}
\fig{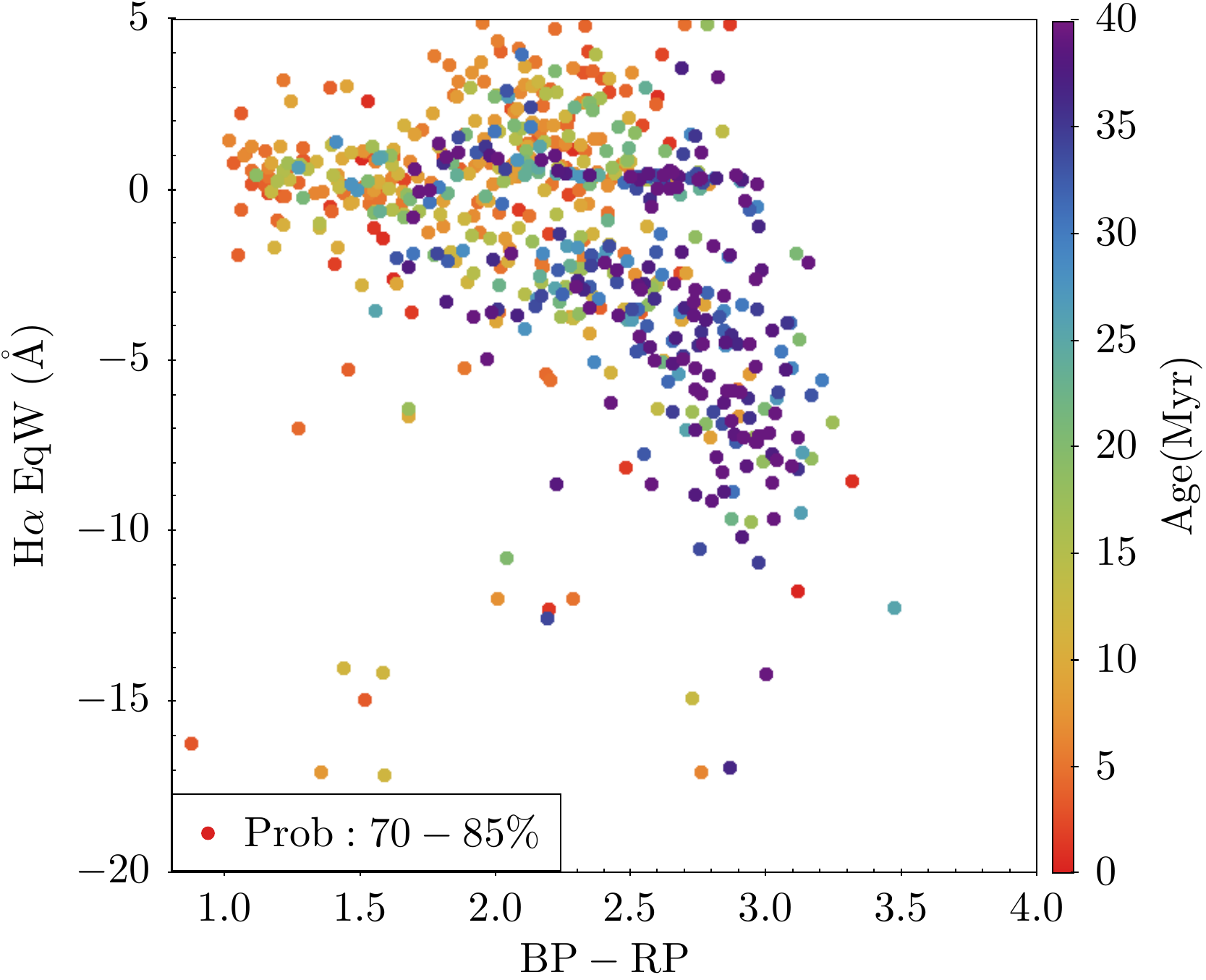}{.33\linewidth}{}
}
\gridline{\fig{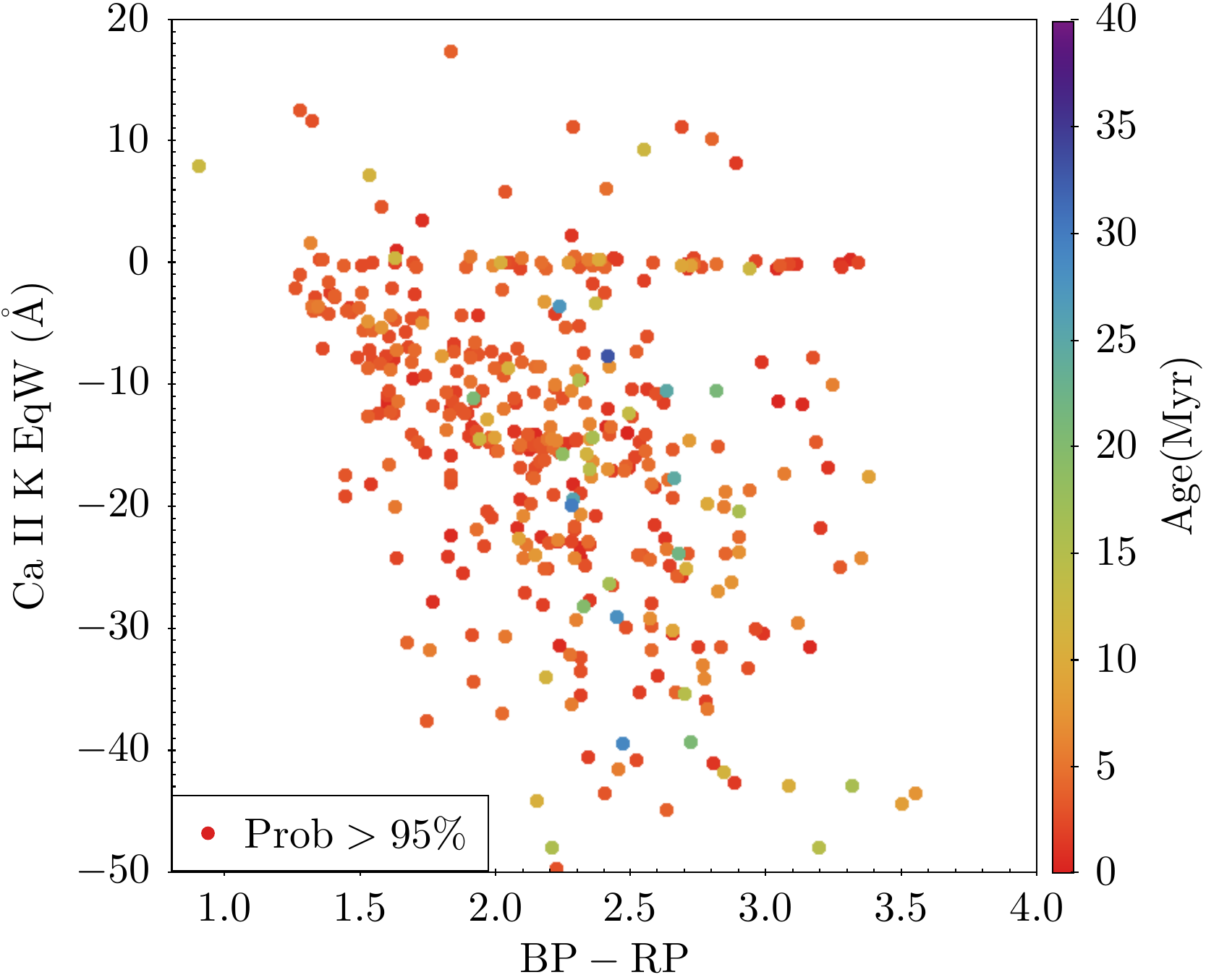}{.33\linewidth}{}
\fig{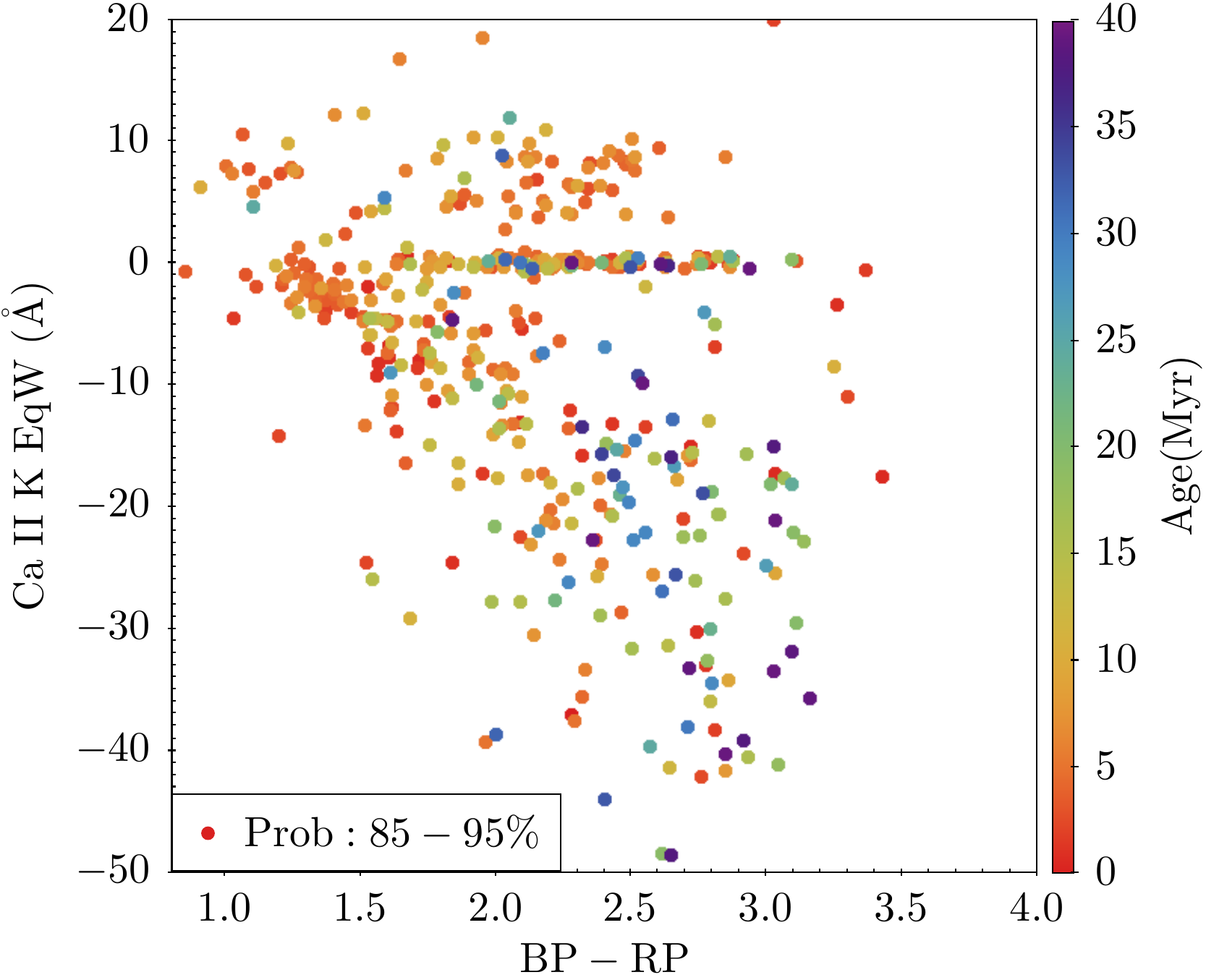}{.33\linewidth}{}
\fig{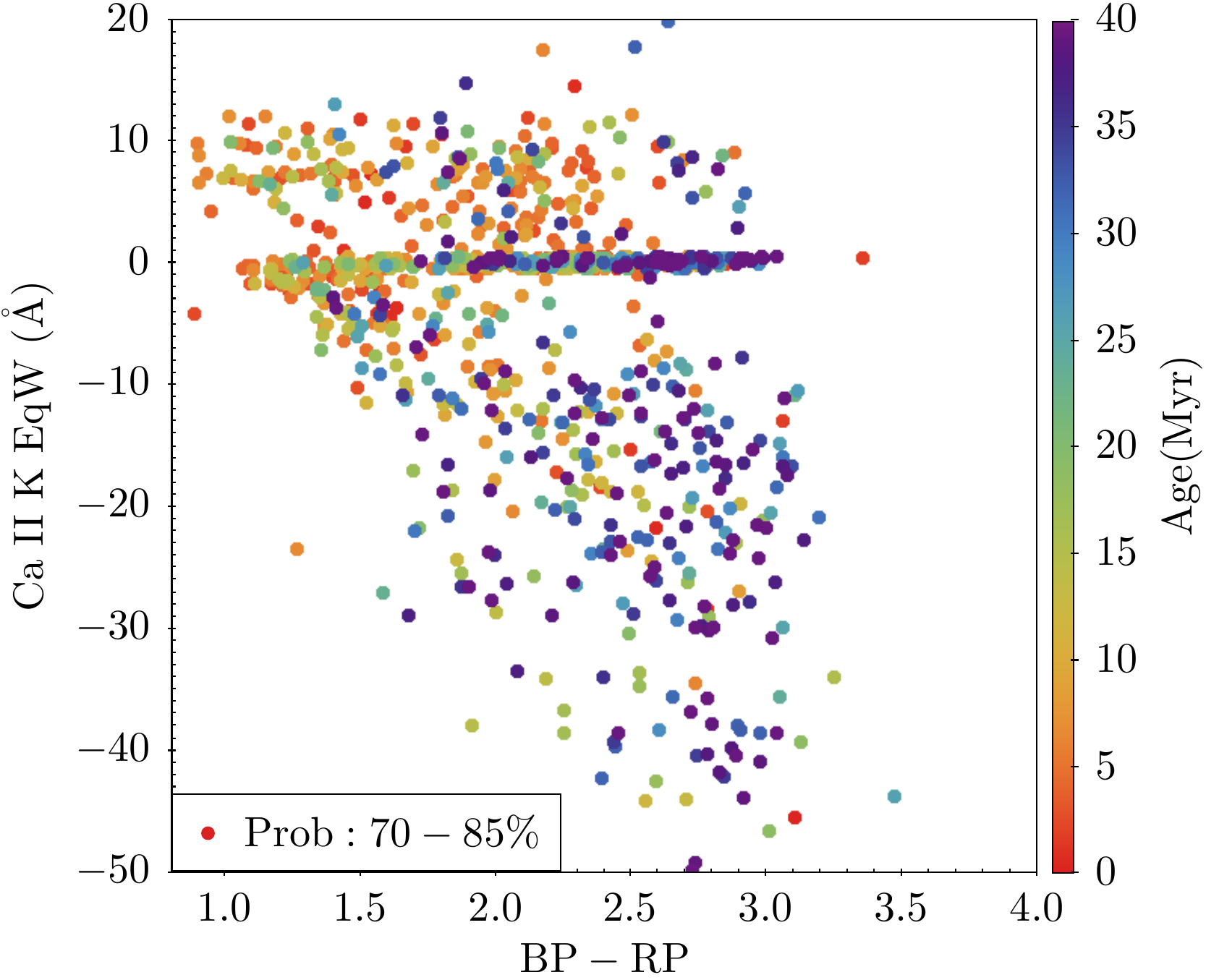}{.33\linewidth}{}
}
\gridline{\fig{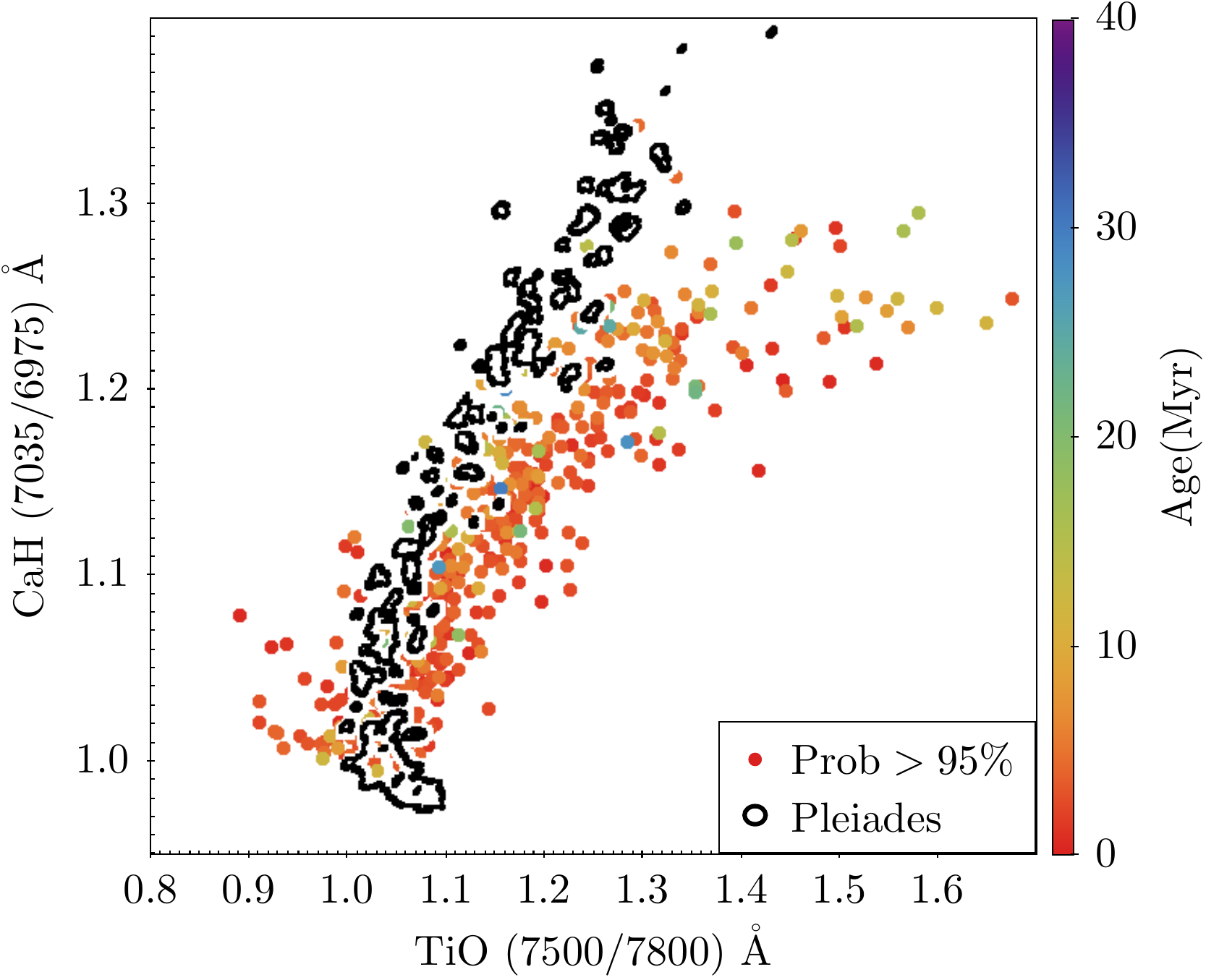}{.33\linewidth}{}
\fig{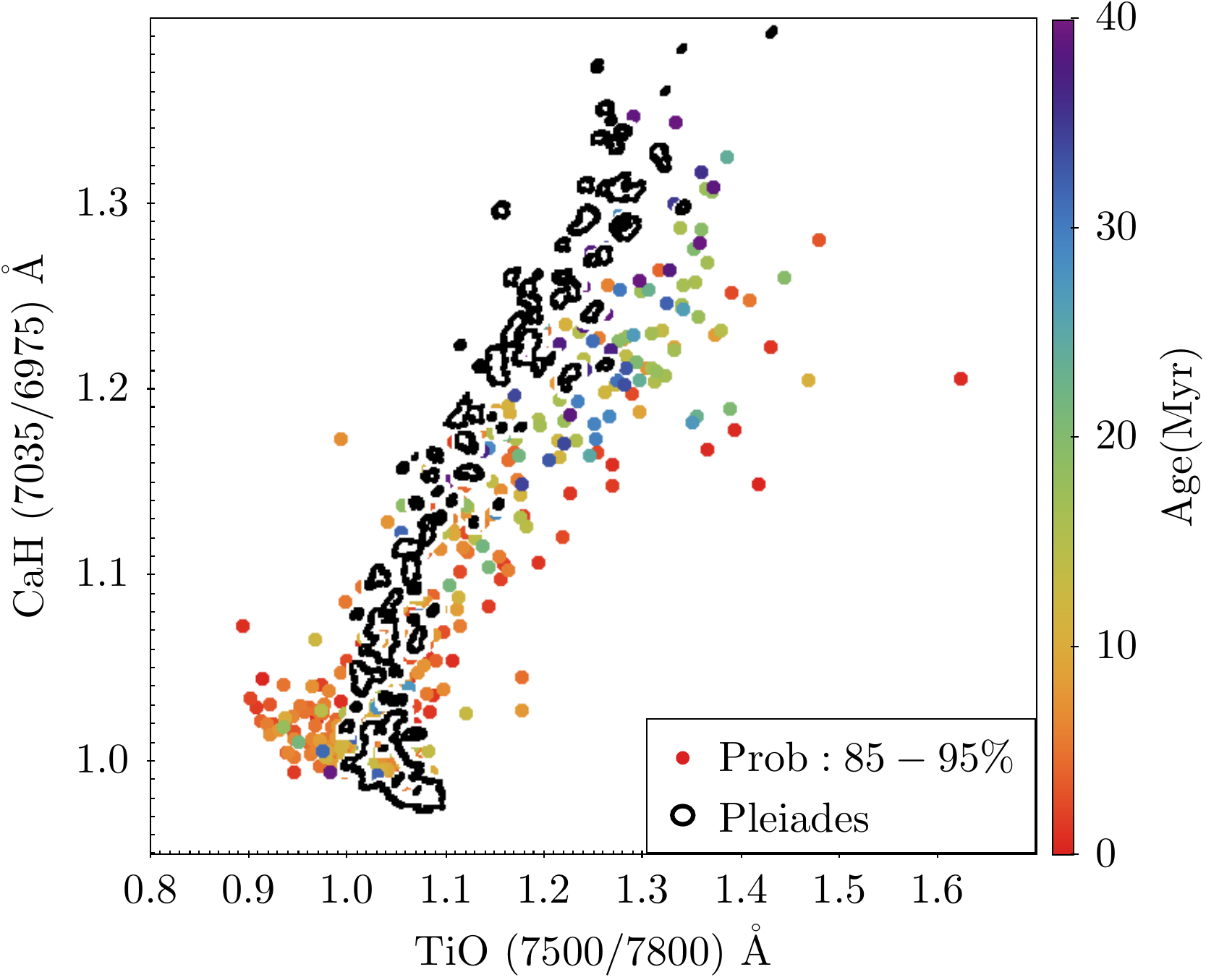}{.33\linewidth}{}
\fig{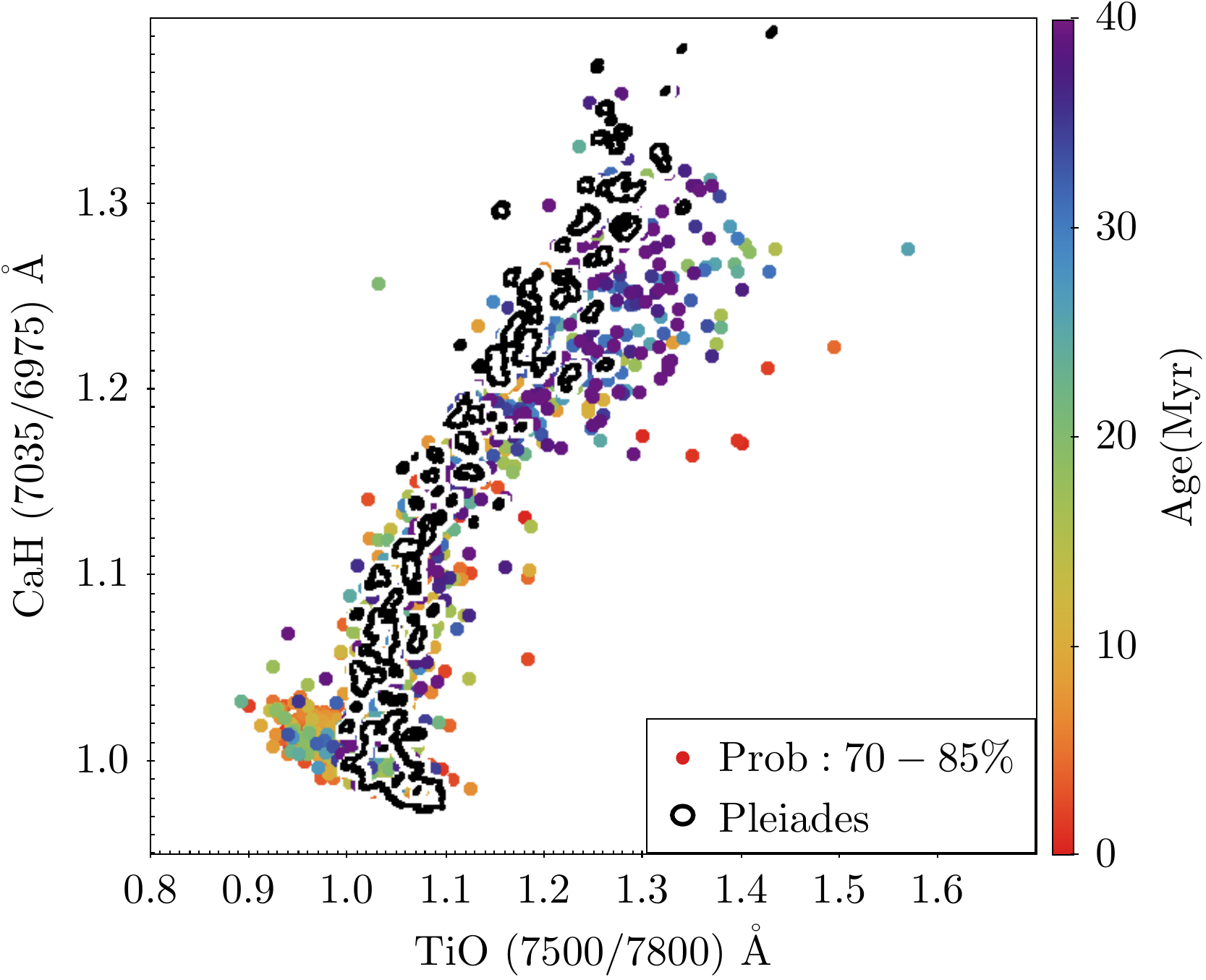}{.33\linewidth}{}
}

\caption{Age sensitive features in LAMOST spectra for the sources in common with our catalog. The sources are color-coded by the derived age of the stars. Three columns separate the sample into different classification confidence intervals. Left: sample with highest confidence with probability $>$95\%. Middle: medium confidence with probability between 85 to 95\%. Right: low confidence with probability of 70 to 85\% (this sample is not used in the subsequent analysis). Top row shows the H$\alpha$ equivalent width as a function of color. The middle row  shows Ca II K equivalent width. Both of these lines are used as activity indicators, with strong emission signifying their youth. Bottom row shows the surface gravity sensitive spectral indices CaH and TiO \citep{wilking2005}. The black contours show the observed distribution of these stars among the members of the Pleiades, to demonstrate the typical distribution among the main sequence dwarfs. Grey contours outline the distribution in our sample. Note that most of the highest and medium confidence sources can be separated from the dwarfs, as is expected of the pre-main sequence subgiants. \label{fig:actlogg}}
\end{figure*}

In the medium confidence sample ($>85$\% probability), 84\% have H$\alpha$ emission consistent with being CTTS or WTTS, with only 16\% being inactive (Figure \ref{fig:actlogg}, top row). Among the sources with low confidence (70--85\%), 60\% are active and $\sim$40\% are inactive. Curiously, if we consider only the sources we identify as older, with the age of $>$7.5 dex, in the medium confidence sample, the fraction of the active stars increases from 84 to 88\%, and in the low confidence sample, it increases from 60 to 73\%.

Other activity tracers may also be present in the spectra, such as Ca II H \& K lines. They deplete faster than H$\alpha$, generally persisting for $<$300 Myr \citep{clark-cunningham2020}, after which the typical emission strength in field stars is $>$-2 \AA. However, unlike H$\alpha$, which is ubiquitous, Ca II H \& K emission appears to be somewhat more stochastic. Pleiades is one of the clusters observed by LAMOST, and it can be used as a benchmark for the young stars we identify, as all of the stars in the Pleiades are fully on the main sequence, but it is only somewhat older than the pre-main sequence stars we identify.

In total, we find that only 25\% of stars in the Pleiades still have Ca II K emission in excess of -$5$\AA. In contrast in our medium confidence sample sample, we find 60--70\% have strong Ca II K emission, depending on their age (both for the stars younger than 10 Myr, and the stars older than 20 Myr). This strongly suggests that, as expected, the population of stars we identify is consistent with being younger than the Pleiades. In the highest confidence sample, the fraction of strong Ca II K emitters is comparable, $\sim$70\%, suggesting that this is the limit due to stochasticity. This fraction decreases to 33\% in the low confidence sample (Figure \ref{fig:actlogg}, middle row row).

We note that in the subsequent analysis, restricting the sample only to the sources that do have Ca II emission does not skew the spatial distribution of the sample.

\subsubsection{Surface gravity sensitive features}

Pre-main sequence stars have not yet completed their process of contracting onto the main sequence, as such, they have a somewhat lower \logg\ than the main sequence dwarfs. Unfortunately, although efforts have been made to measure calibrated \logg\ values from the spectra of young star observed as a part of large surveys \citep{olney2020}, this is not yet widely available across optical spectra, including LAMOST.

Instead of using \logg\ directly, however, it is possible to examine known surface gravity sensitive features. \citet{wilking2005} have developed several spectral indices that can be used as a proxy, these indices include CaH 6975 \AA ~and, to a lesser extent, TiO 7140 and 7800 \AA ~features, which are most effective when used in a combination with one another.

We measure these indices in the LAMOST spectra both for the Pleiades and for the stars selected by Sagitta (Figure \ref{fig:actlogg}, bottom row row). We find a clear separation between them, this separation persists both in the highest confidence and in the medium confidence samples, regardless of the age. This separation is consistent with what is expected due to the surface gravity difference between the main sequence dwarfs and pre-main sequence subgiants. The separation is most apparent at larger indices. Indices close to $\sim$1 in both bands correspond to hotter stars with less pronounced TiO or CaH bands.

On the other hand, the low confidence sample, particularly at the lowest thresholds ($<$75--80\%) is starting to have a considerable overlap with the Pleiades. Similarly to the activity indicators, this may be attributed to both sampling older pre-main sequence stars and to a higher degree of contamination. We exclude low confidence sources with $<$85\% probability from the subsequent analysis.

\subsection{Properties of known star forming regions} \label{sec:sfrs}

\begin{figure*}
\gridline{\fig{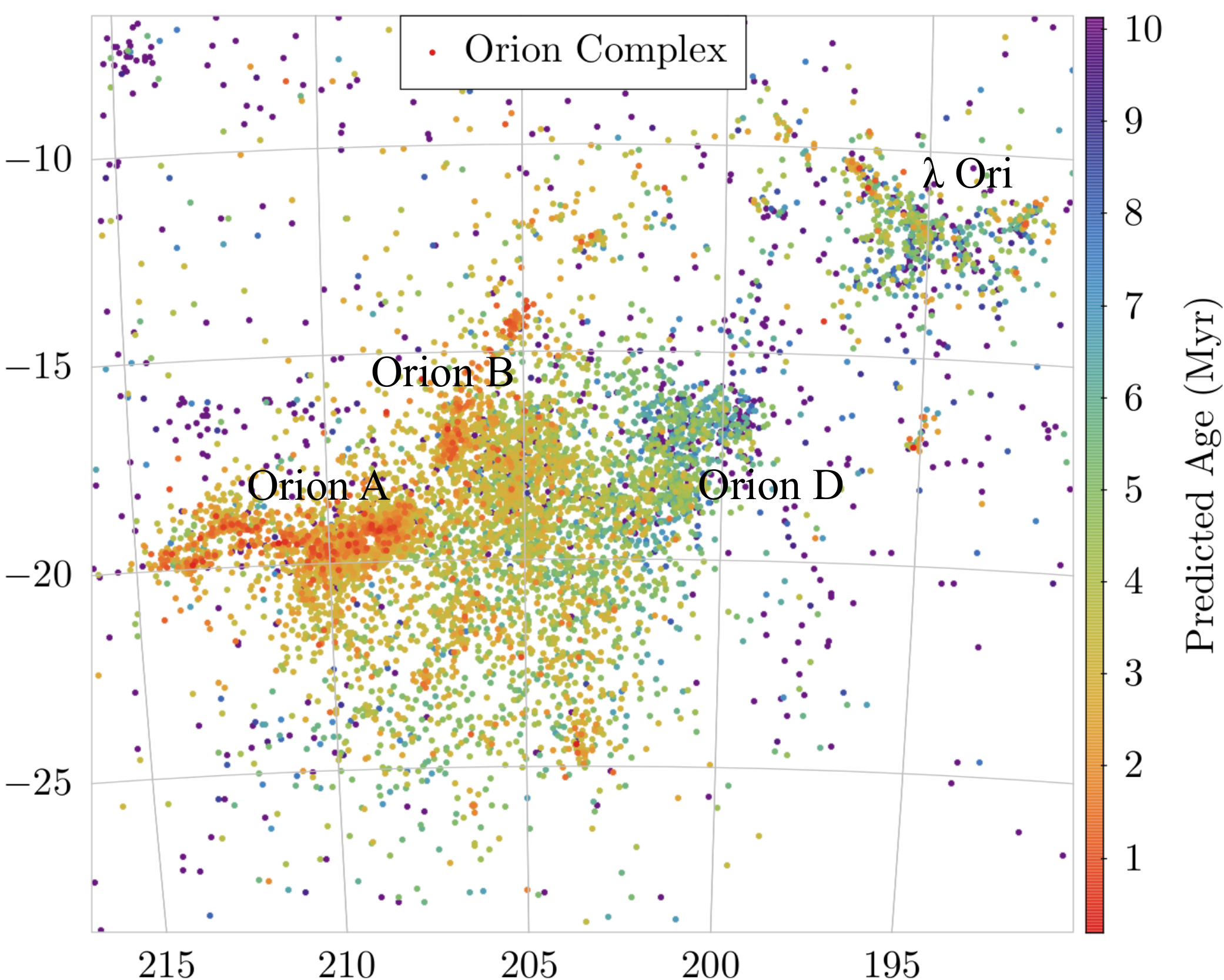}{.45\linewidth}{} 
\fig{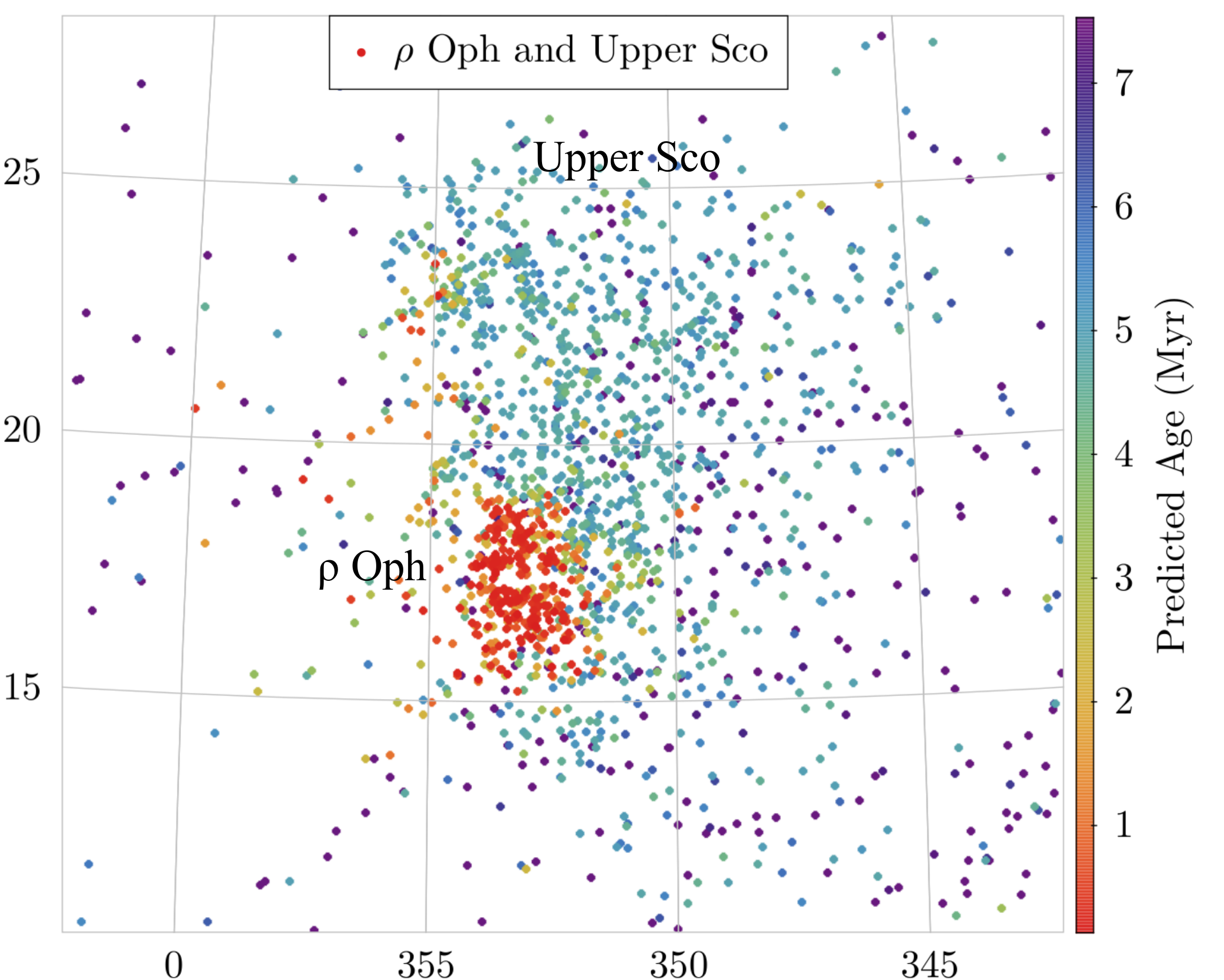}{.45\linewidth}{}}
\gridline{\fig{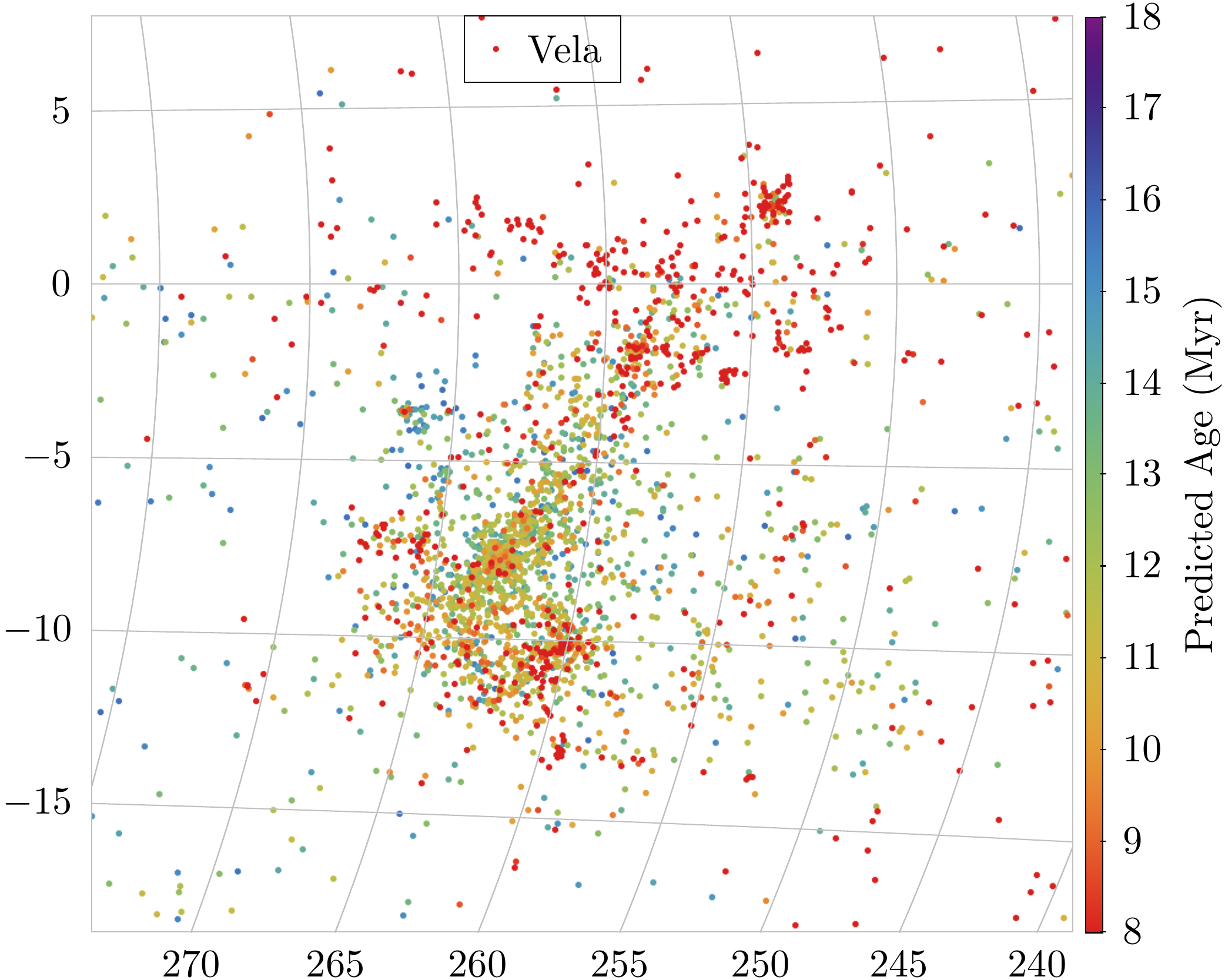}{.45\linewidth}{}
\fig{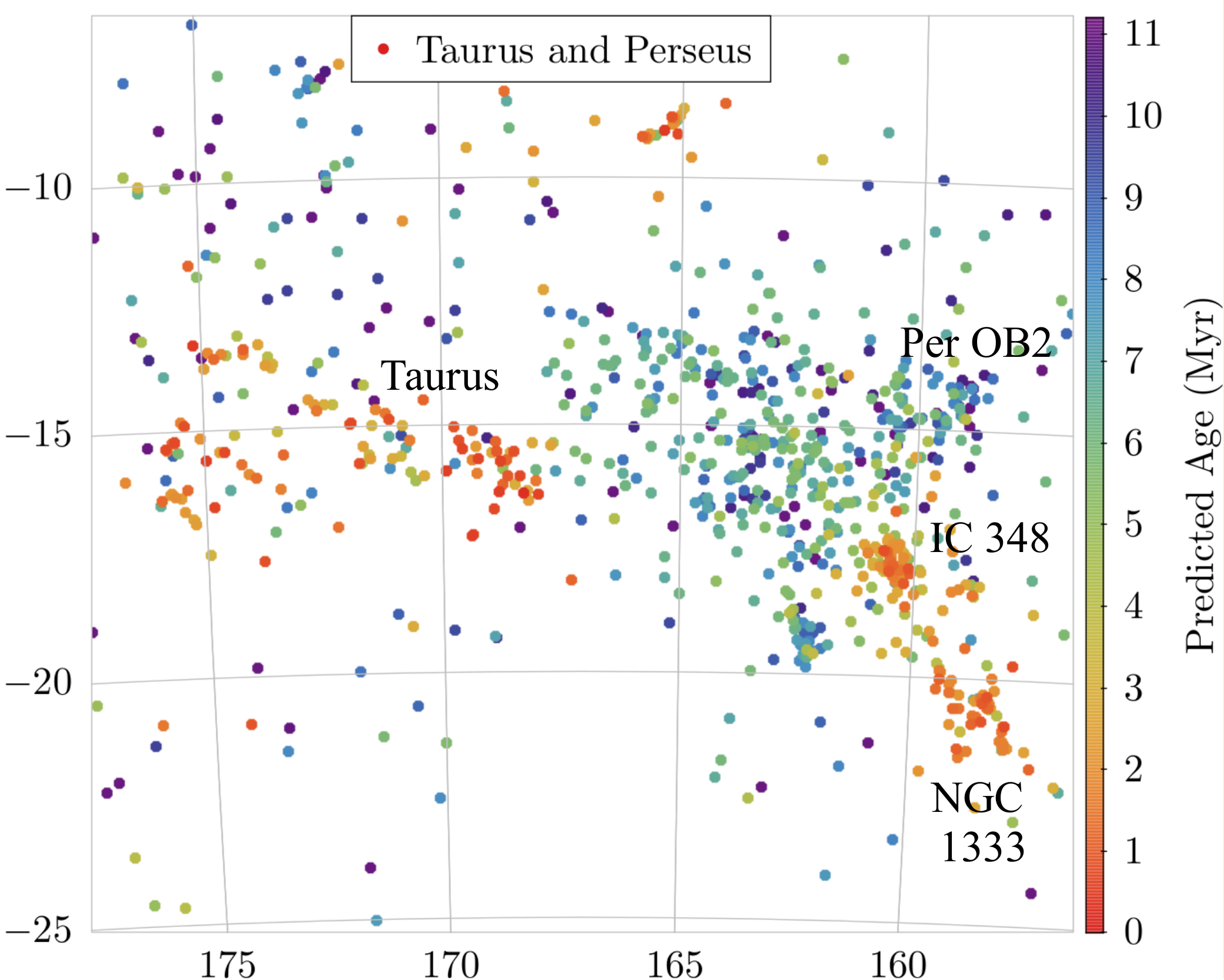}{.45\linewidth}{}}
\gridline{\fig{Serpens_SFR_detail_labeled.pdf}{.45\linewidth}{}
\fig{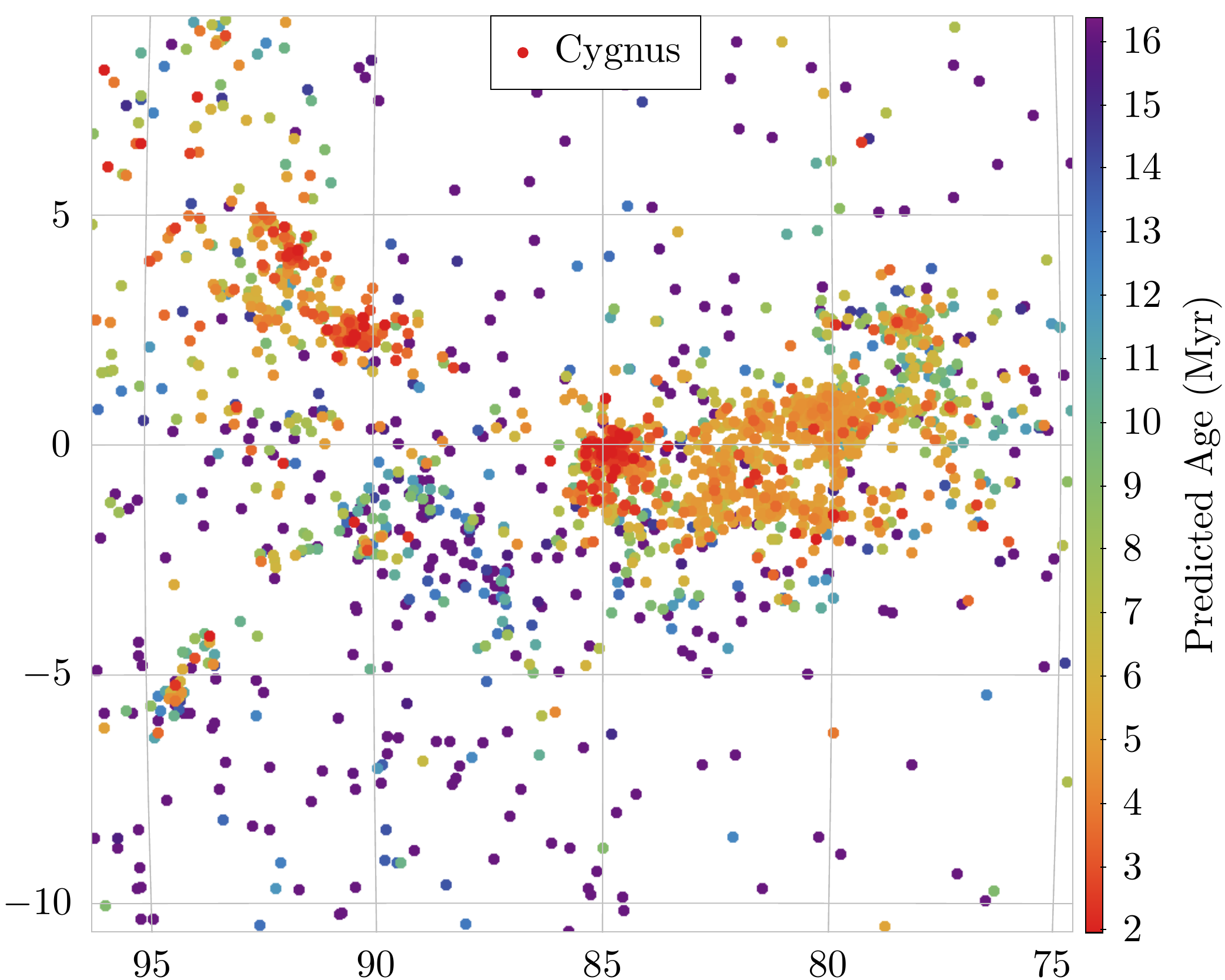}{.45\linewidth}{}}
\caption{Distribution of stars identified by \textit{Sagitta} towards various star-forming regions, color-coded by stellar ages. Probabilities have been selected between 85\% and 95\% to best represent each region, based on the age and distance to each. To remove contamination from other nearby regions, the ages for Taurus and Vela have been restricted to $<$ 15 Myr, and to highlight the more distant region of Cygnus, distance was restricted to $> 500$ pc. \label{fig:SFRs}}
\end{figure*}

\begin{figure*}
\gridline{\fig{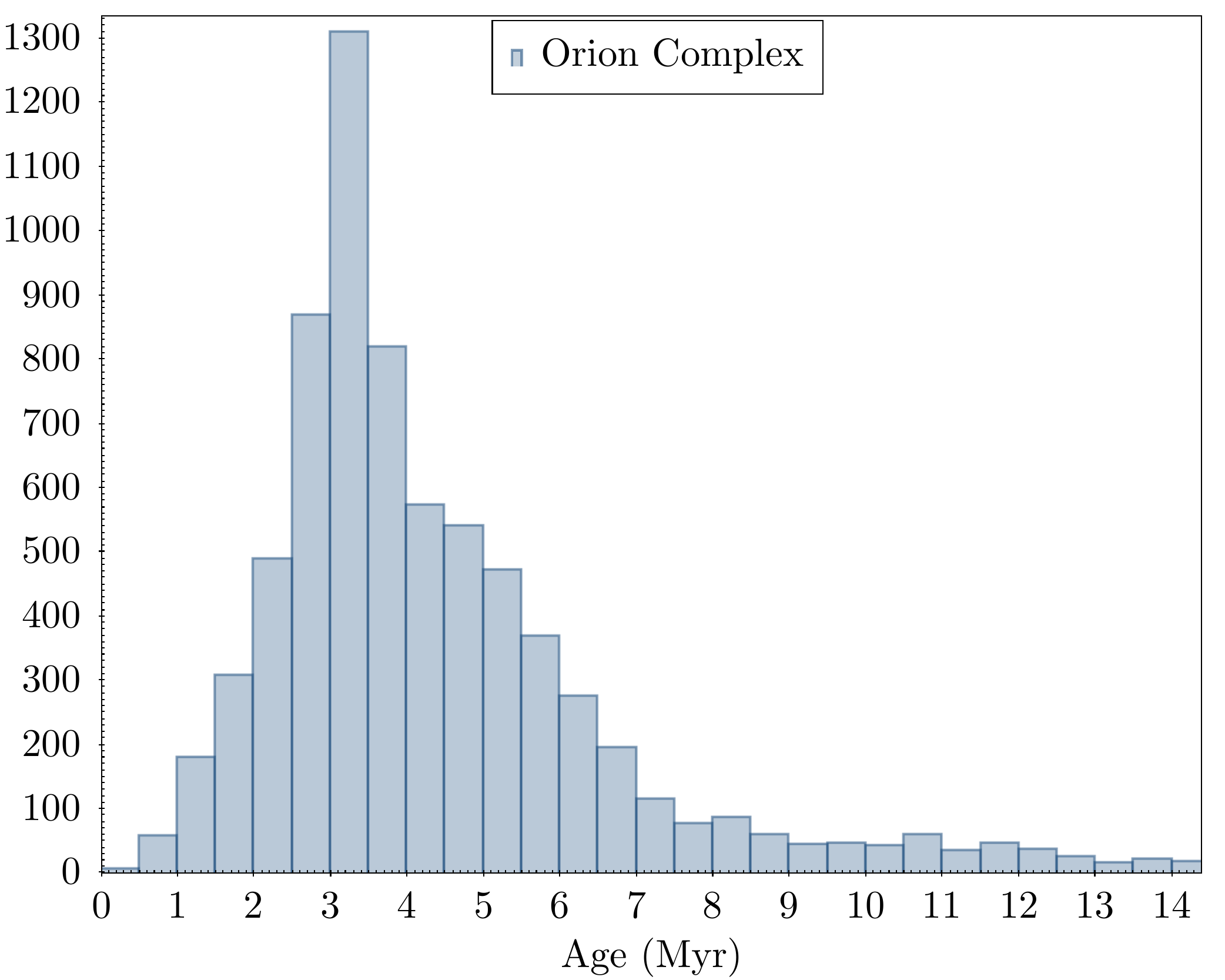}{.45\linewidth}{} 
\fig{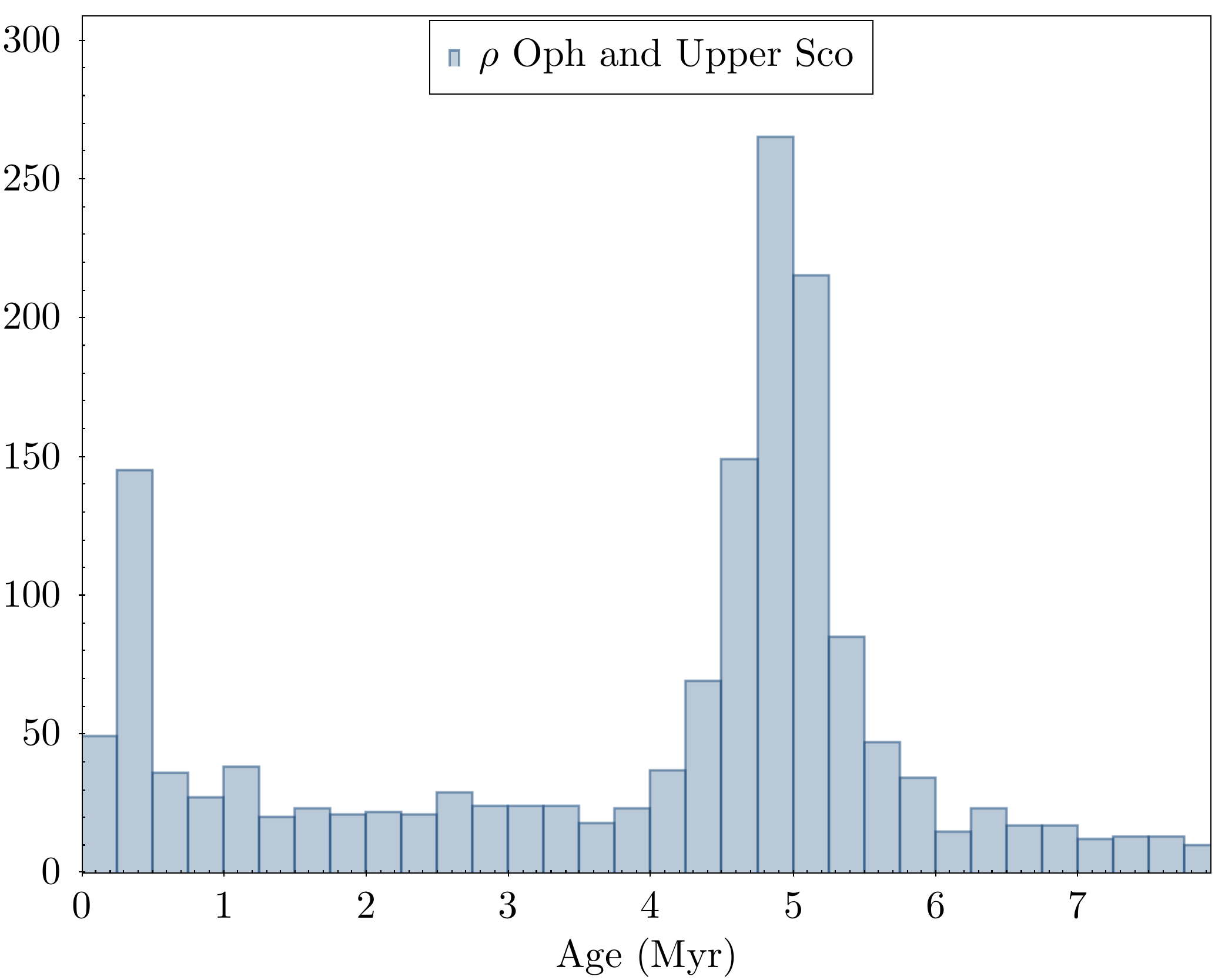}{.45\linewidth}{}}
\gridline{\fig{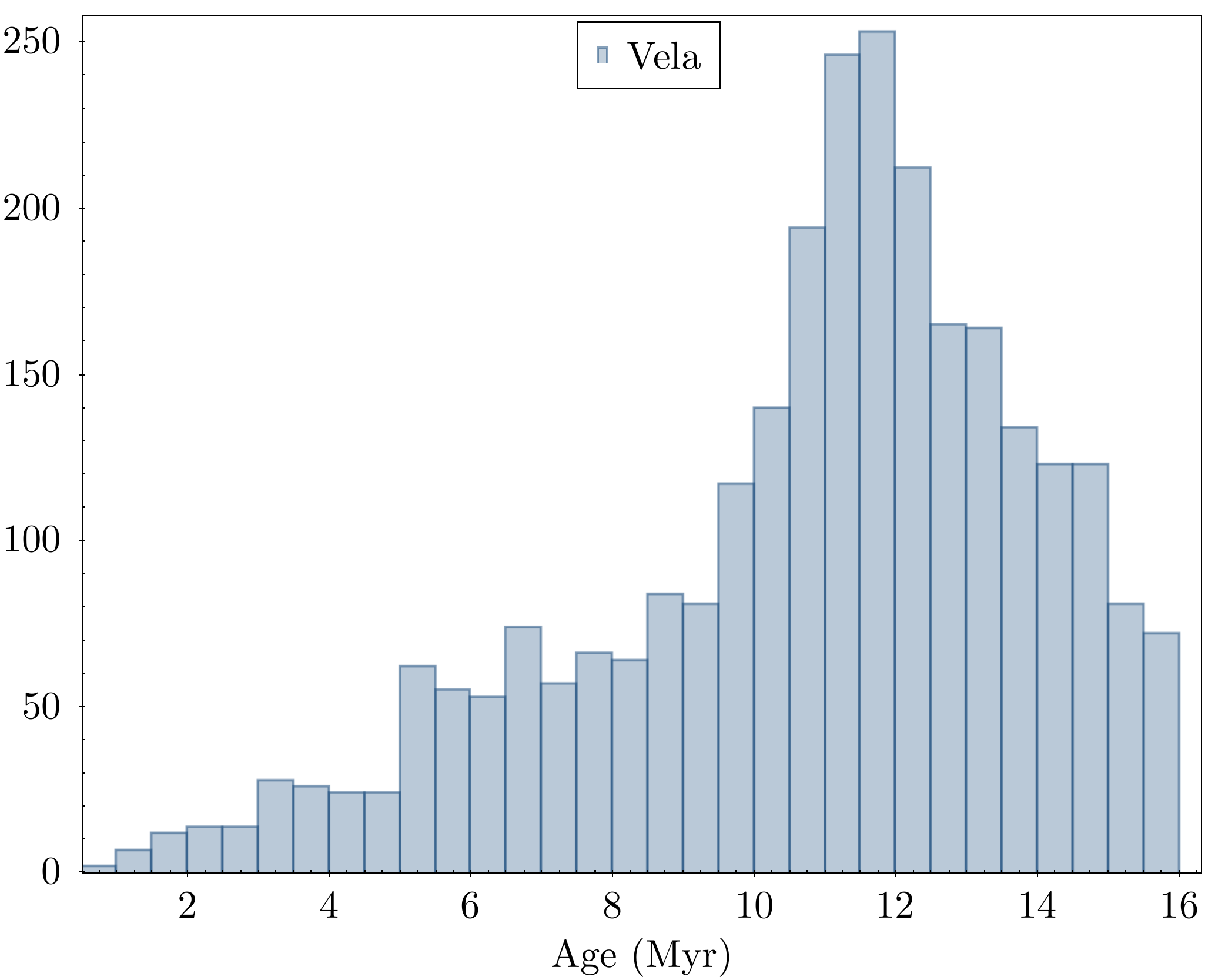}{.45\linewidth}{}
\fig{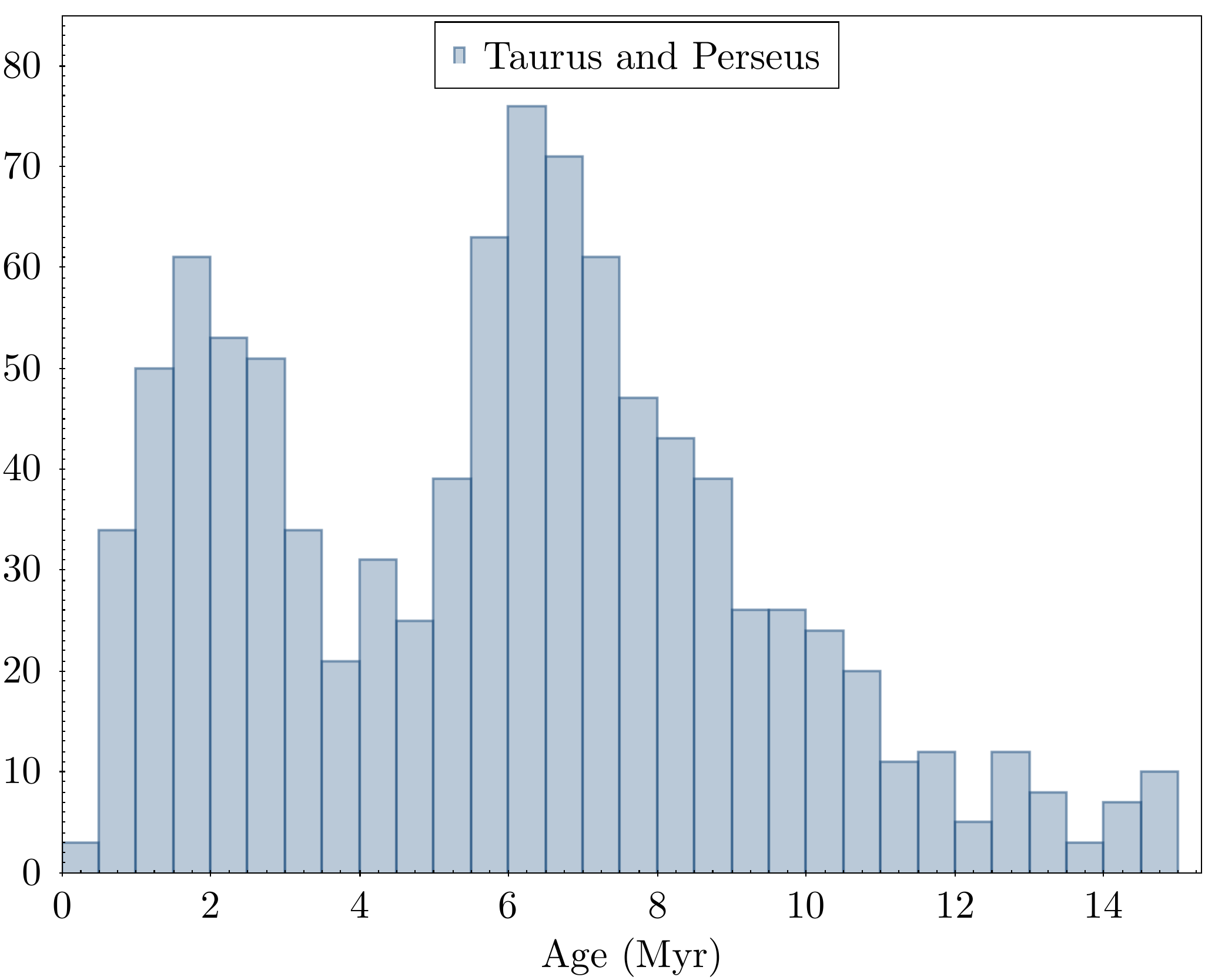}{.45\linewidth}{}}
\gridline{\fig{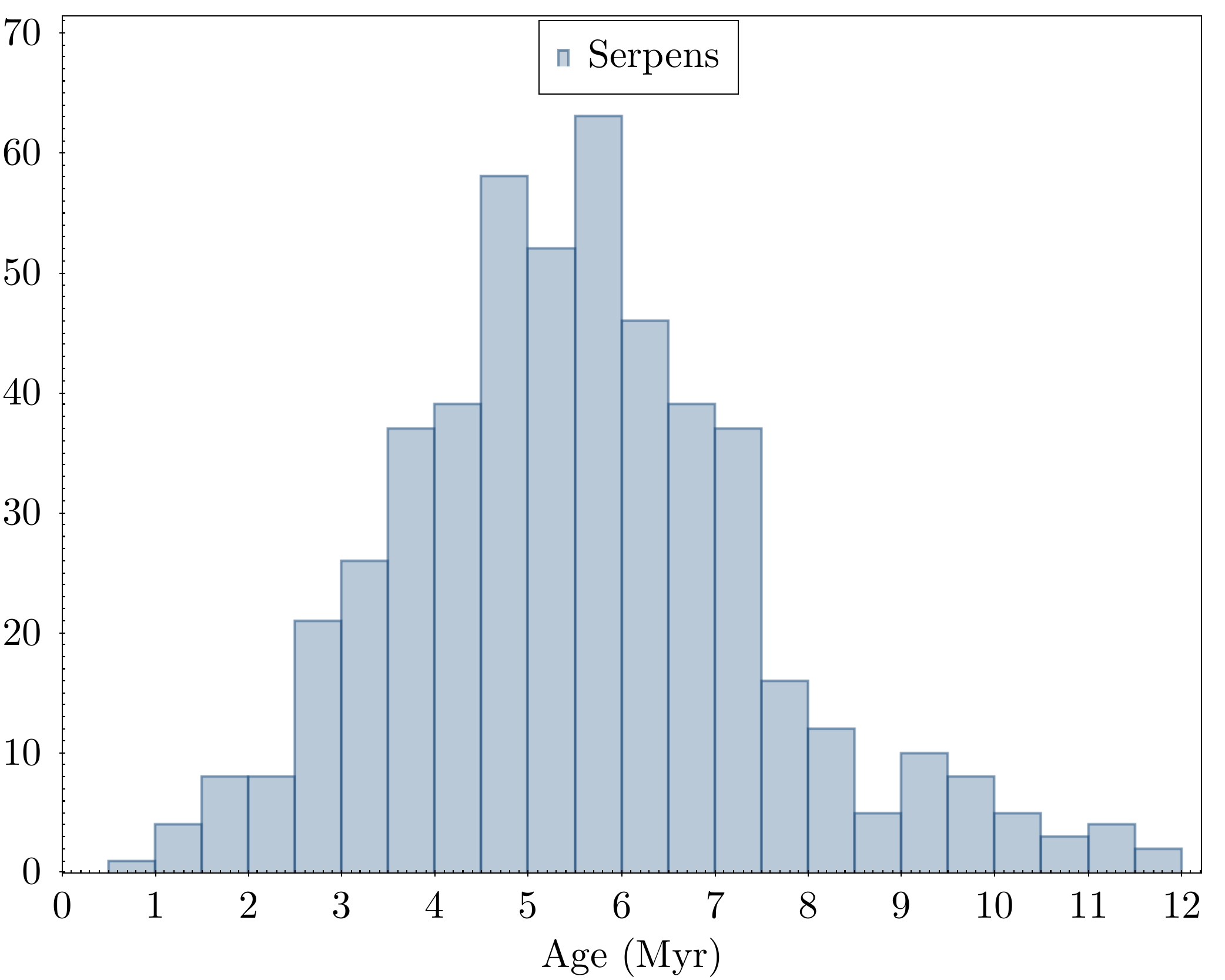}{.45\linewidth}{}
\fig{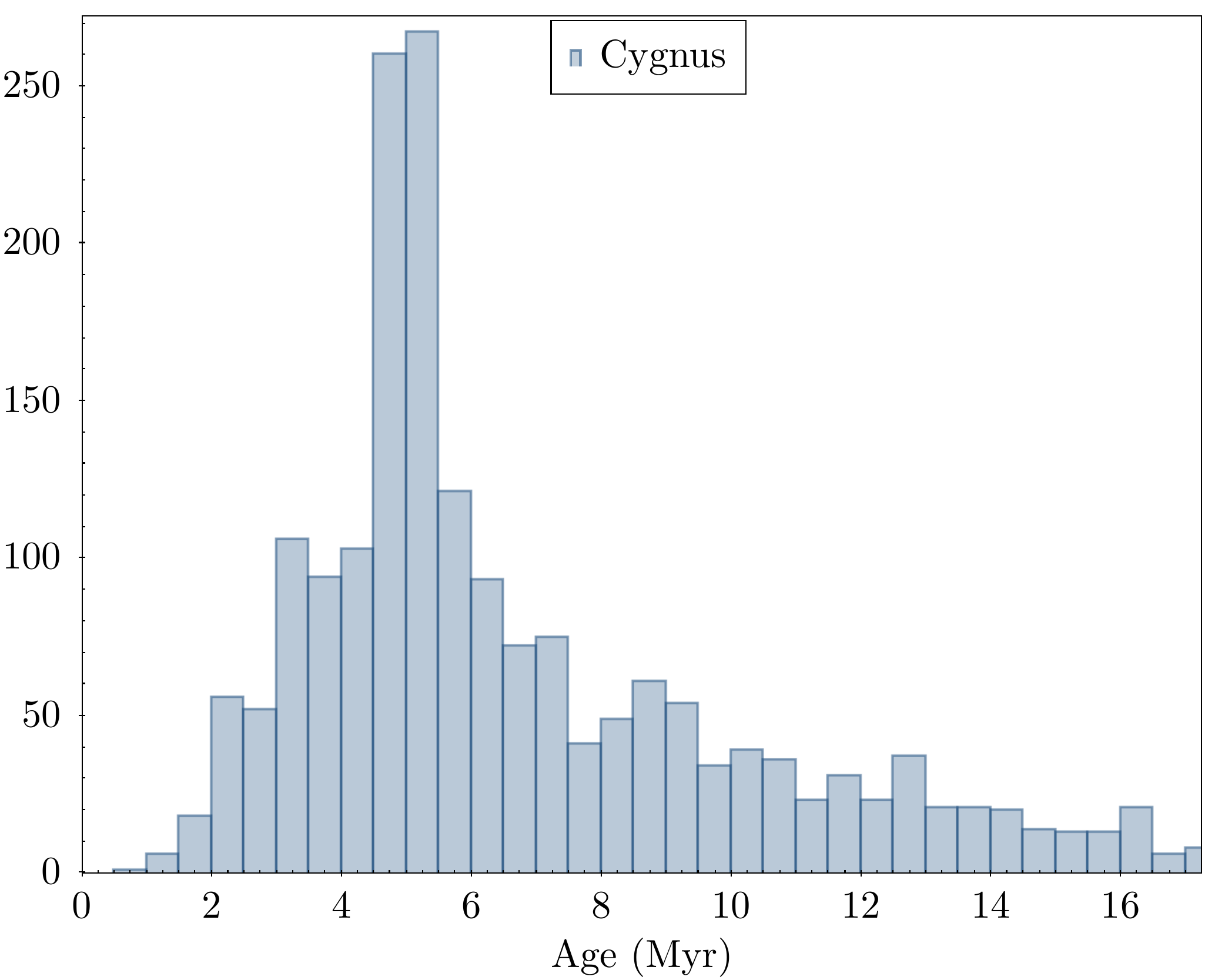}{.45\linewidth}{}}
\caption{Distribution of ages for star-forming regions in Figure \ref{fig:SFRs}. Note that the membership of each region was not assessed beyond the position on the sky and some distance and/or age ranges, therefore contamination from unrelated stars is present in each field of view. \label{fig:SFRHist}}
\end{figure*}

Figure \ref{fig:SFRs} shows the zoom-in view of various star forming regions that were used as a benchmark in evaluating the measured stellar ages. Figure \ref{fig:SFRHist} shows the corresponding distribution of ages for each individual SFR view. 

\subsubsection{Orion Complex}
The Orion Complex is the closest region of ongoing massive clustered star formation, containing $>10,000$ stars with ages of $<1$ Myr to $>10$ Myr. The youngest stars are found in the Orion A and B molecular clouds and older stars are found in Orion D and $\lambda$ Ori \citep{kounkel2018a}. Sources from Orion represent a significant fraction of the input training set, and provide a valuable evaluation for the performance of both the classifier and the regression model. 

The model recovers $\sim$7,500 sources with $>$95\% threshold, and  $\sim$9,000 within $>$90\% threshold. The age gradients and the typical ages that have previously been observed by \citet{kounkel2018a} are well recovered by the model.

We note that the evaluation catalog has a hole near Trapezium in the Orion Nebula, as nebulosity degrades the photometric quality. Thus, few sources met the cuts specified in Section \ref{sec:testsample}. Applying Sagitta on a separate catalog that has not been as constrained by the quality of inputs, it is able to recover both the members and the appropriate ages.

\subsubsection{$\rho$ Oph and Sco--Cen}

The $\rho$ Oph star-forming region has already been discussed as a means of evaluating the performance of the classifier, but both $\rho$ Oph and the surrounding, slightly older Upper Scorpius region are also notable for their peculiar star forming history. 

We recover the typical age of $<1$ Myr for $\rho$ Oph, and $\sim$5 Myr for Upper Sco \citep{preibisch2008,pecaut2016}. Although the transition between the two region is rather sharp, without a significant age spread larger than $\pm$1 Myr, there is some overlap between the two populations, furthermore, the eastern part of Upper Sco contains somewhat younger stars. From the Upper Sco, along the rest of Sco--Cen, we recover a relatively smooth gradient in age 15--20 Myr towards the Lower Centaurus Crux that has previously been observed in other works. Members of the younger Lupus clouds (such as III and IV) get recovered with the characteristic age of $\sim$3 Myr. The southern portion of CrA that is still associated with the molecular gas has a typical age of $\sim$4 Myr; the northern portion that has since cleared its gas has an age of $\sim$6--8 Myr. Nearby, there is also $\sim$20 Myr population that also appears to be related to Sco--Cen.

While the average ages we measure for these regions are consistent to the literature values (as the literature values were originally used for training), we note that are able to go from discrete region-specific estimates to a more homogeneous map of star forming history.

\subsubsection{Vela}

Towards Vela there are two unrelated populations found in a similar volume of space. One is Vela OB2, which is associated associated with $\gamma$ Velorum, and has a typical age of $\sim$10 Myr. The other populations has an age of $\sim$30--35 Myr, and it contains an open cluster NGC 2547 \citep{jeffries2009, jeffries2014, beccari2020}. 

Due to its youth, we recover Vela OB2 at higher confidence -- the bulk of the members can be identified at the threshold of $>$95\%, containing $\sim$2,700 stars. On the other hand, NGC 2547 becomes apparent only with the threshold lowered to $\sim$85\%, containing $\sim$10,000 stars. We recover the average ages of both these populations.

Vela OB2 in particular shows curious star forming history. The stars that are located towards the northern group H \citep{cantat-gaudin2019} are preferentially younger than those near the central part.

\subsubsection{Taurus and Perseus}

Due their proximity and youth, the Taurus molecular clouds contain some of the best studied young stars. This region does not contain any clusters, rather, it is a collection of several diffuse clouds, some of which are up to 30 pc apart. The most up-to-date membership of this region is presented in \citet{luhman2018}. We are able to recover most of these members within our evaluation catalog, as well as add a number of new candidates.

We are able to recover the typical age of $\sim$1--3 Myr for much of the previously known members. There have also been a suggestion of an older nearby $\sim$16 Myr population \citep{kraus2017}, which we are also able to recover. As is the case with the younger stars, this population is an assembly of diffuse clumps of stars, resembling evolving cirrus clouds.

Along a similar sight line as Taurus (but at a somewhat larger distance) lies Perseus. We are able to recover the age of $\sim$6--8 Myr for Per OB2 (with some substructure), as well as ages of 1--3 Myr for younger clusters IC 348 and NGC 1333 \citep{azimlu2015}.

\subsubsection{Serpens}
Serpens contains several young clusters located at a distance of $\sim$450 pc. Similarly to the work of \citet{herczeg2019}, we are able to recover stars towards Serpens Main (with the age of $\sim2$ Myr), as well as Serpens Northeast, and Serpens far-South (with the age of $\sim$3--5 Myr). There also appears to be substantial diffuse population 5--8 Myr population surrounding them. Unfortunately, W40, likely the youngest region in this star forming complex, is too deeply embedded to be seen in the optical regime.

We recover $\sim$500 sources towards Serpens up to the threshold of $>$95\%, and $\sim$1500 the threshold of $>$90\%. 

\subsubsection{Cygnus}
The star-forming regions in Cygnus, (particularly Cygnus OB2) are located at much larger distance than other regions discussed in this work. Furthermore, it is located behind a considerable layer of extinction \citep{wright2016}. Because of this, it can only be recovered in full at lower thresholds from the classifier. Nonetheless, we are able to recover ages of $\sim$4--10 Myr for Cyg OB2.

\subsection{Comparison to Other Catalogs}
\subsubsection{Zari et al. (2018)}\label{sec:zari}
\cite{zari2018} identified pre-main sequence stars younger than 20 Myr within 500 pc using \textit{Gaia} DR2 data. The available catalog provides three confidence intervals for pre-main sequence sources based on their kinematic distributions. Their total sample contains 43,719 sources, with 23,686 satisfying the strictest kinematic threshold.

\begin{figure*}
\gridline{
\fig{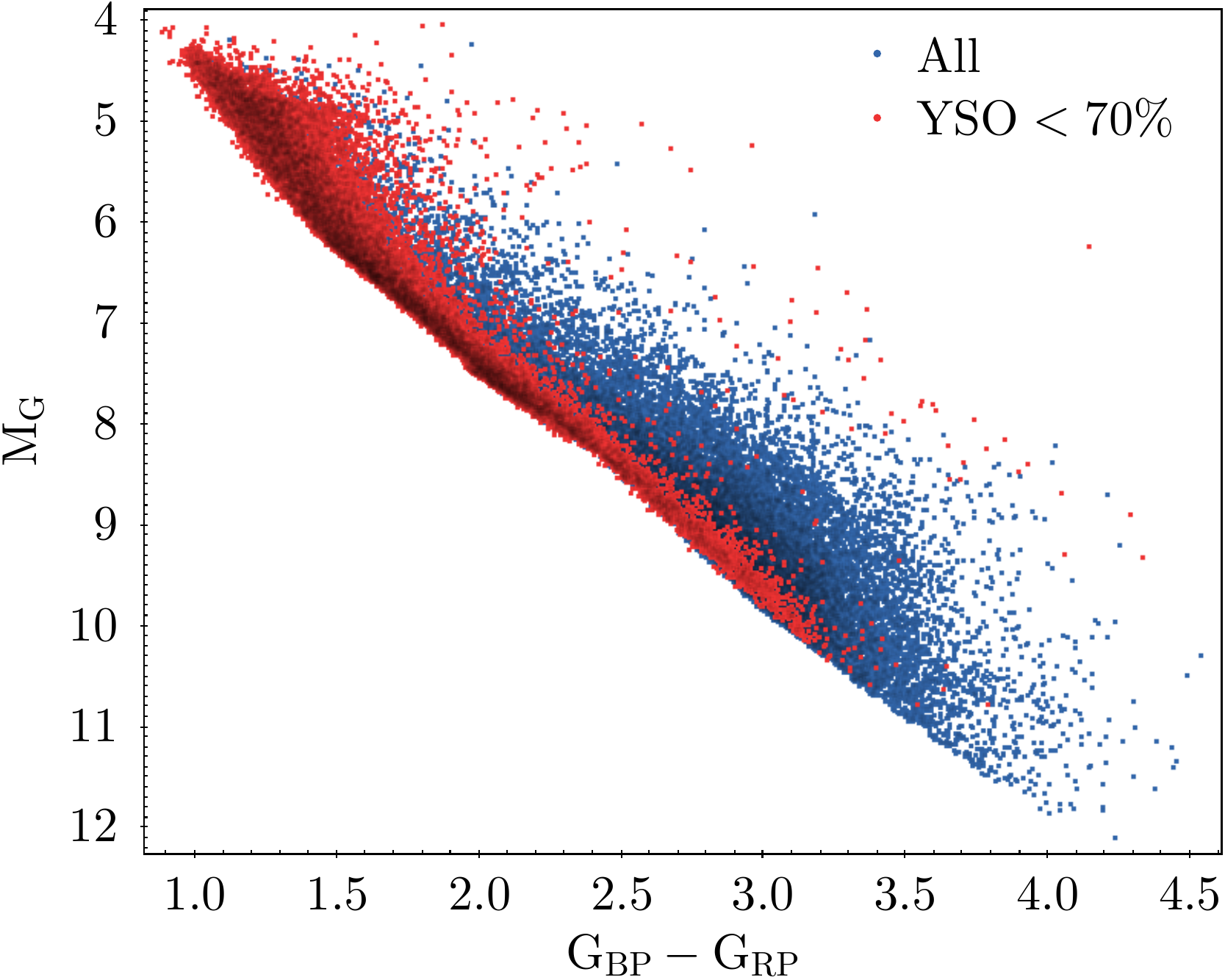}{0.35\linewidth}{}
\fig{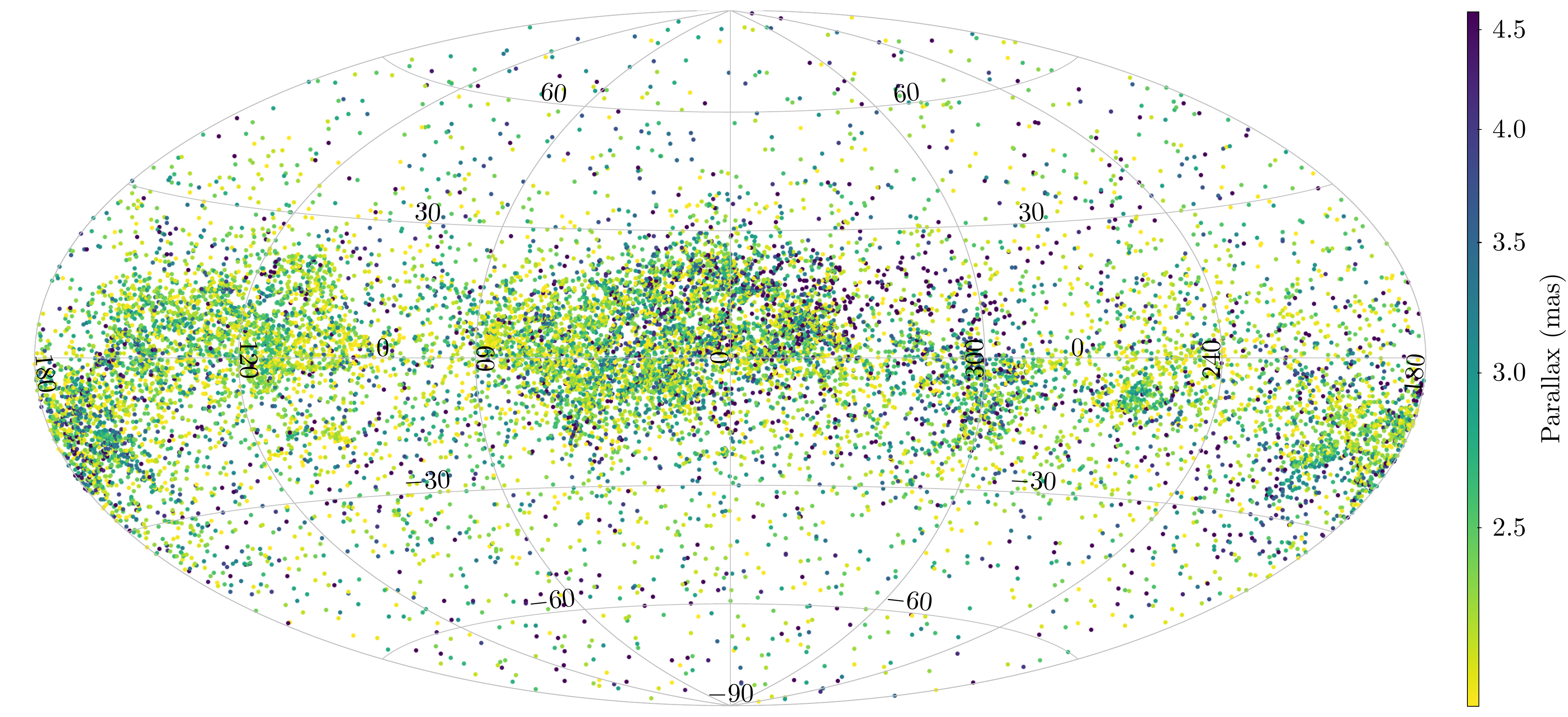}{0.65\linewidth}{}
}
\caption{The top panel shows a Hertzsprung-Russel Diagram constructed from young stars in \cite{zari2018}, analyzed with \textit{Sagitta}. Red points represent sources that Sagitta assigns $<70$\% likelihood of being pre-main sequence. The bottom panel shows the spatial distribution for the aforementioned sources with $<70$\% likelihood of being young, color coded by parallax.  \label{fig:Zari}}
\end{figure*}

In examining full catalog from \citet{zari2018}, \textit{Sagitta} confirms 18,488 of the objects they identify as pre-main sequence with $>$90\% probability, and 23,115 with $>$80\% (we note that in the full evaluation sample there are 30,689 stars younger than 20 Myr at 90\% probability within 500 pc, and 44,225 at 80\% probability). There is a particularly good agreement in the identified stars located within $\sim$200 pc within the appropriate age bounds. The sources to which we assign lower confidence in classifications with Sagitta ($<$70\%) are preferentially located at the distance of $>$400 pc (close to the maximum distance of the \citet{zari2018} study), and they tend to be bluer ($G_{BP}-G_{RP} < 2$). Most of these sources do not appear to trace known star forming regions, rather they trace the extinction patterns. Similarly, they do not have much coherence in their radial velocities, as would be expected in young populations. It is likely that these sources are contamination from the main sequence in their sample (single or unresolved visual binary) due to extinction (Figure \ref{fig:Zari}). However, it is possible that some of them do trace some patterns in the star forming history through massive stars (to which we lose sensitivity) that we cannot trace with lower mass counterparts.

\subsubsection{Marton et al. (2016, 2019)}
\cite{marton2019} used Random Forest to classify pre-main sequence sources using \textit{Gaia} DR2 and WISE photometry. Their catalog includes classifications for 101,838,724 \textit{Gaia} sources, with 1,509,781 located within 5 kpc and having greater than 90\% likelihood in being classified as YSOs. The classification was done only on the areas of the sky above a given opacity threshold using the Planck dust opacity map. On the surface level, this catalog recovers the underlying shape of various star forming regions (such as Orion A \& B, $\rho$ Oph, Taurus-Auriga, and others). However, this is somewhat misleading, as the opacity threshold pre-selects molecular clouds, and masks out young populations that are no longer associated with the molecular gas. Within that mask, however, the catalog is prone to contamination, even at a very high level of reported certainty.

\begin{figure*}\gridline{
    \fig{marton_histogram.pdf}{0.33\linewidth}{}
    \fig{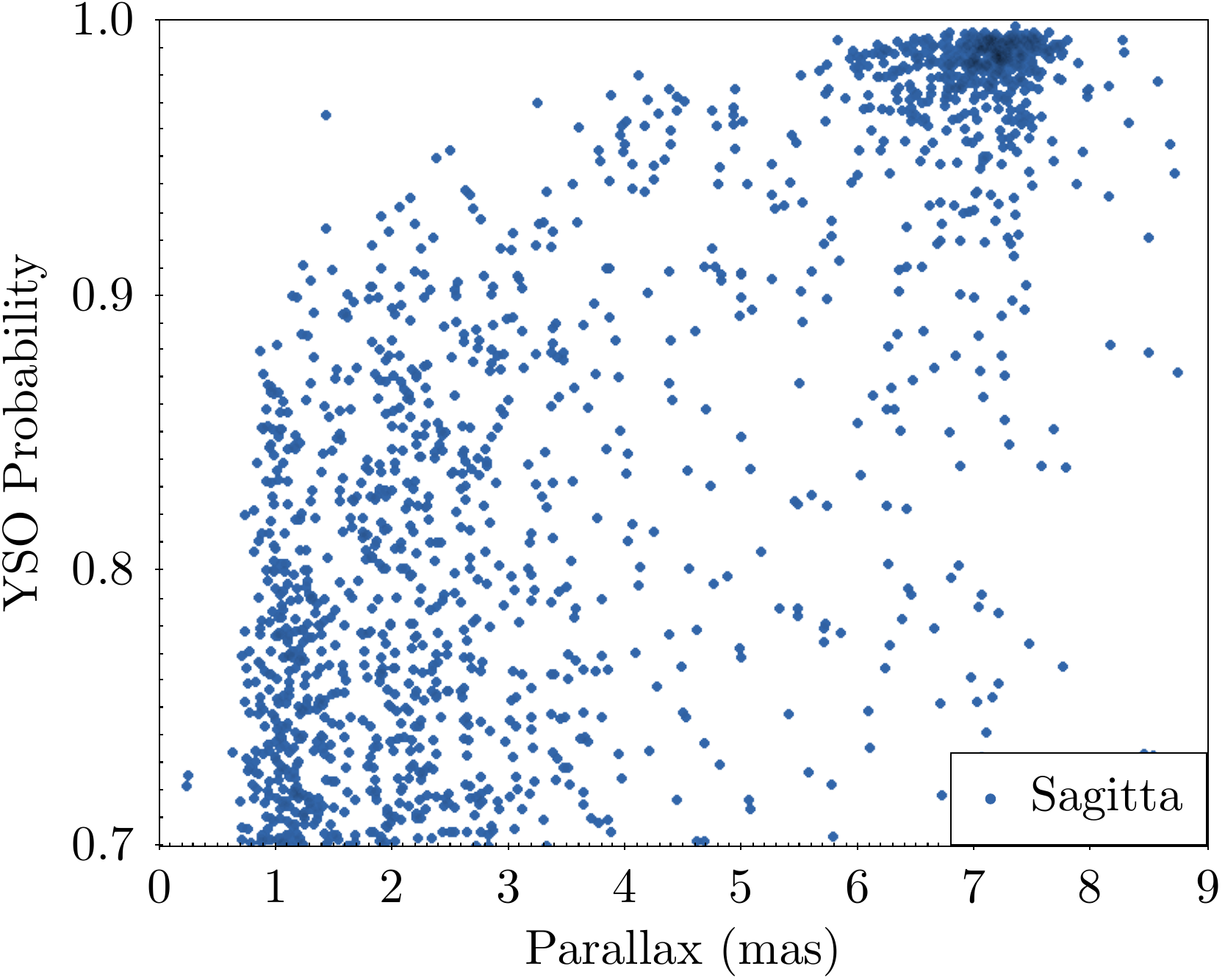}{0.33\linewidth}{} 
    \fig{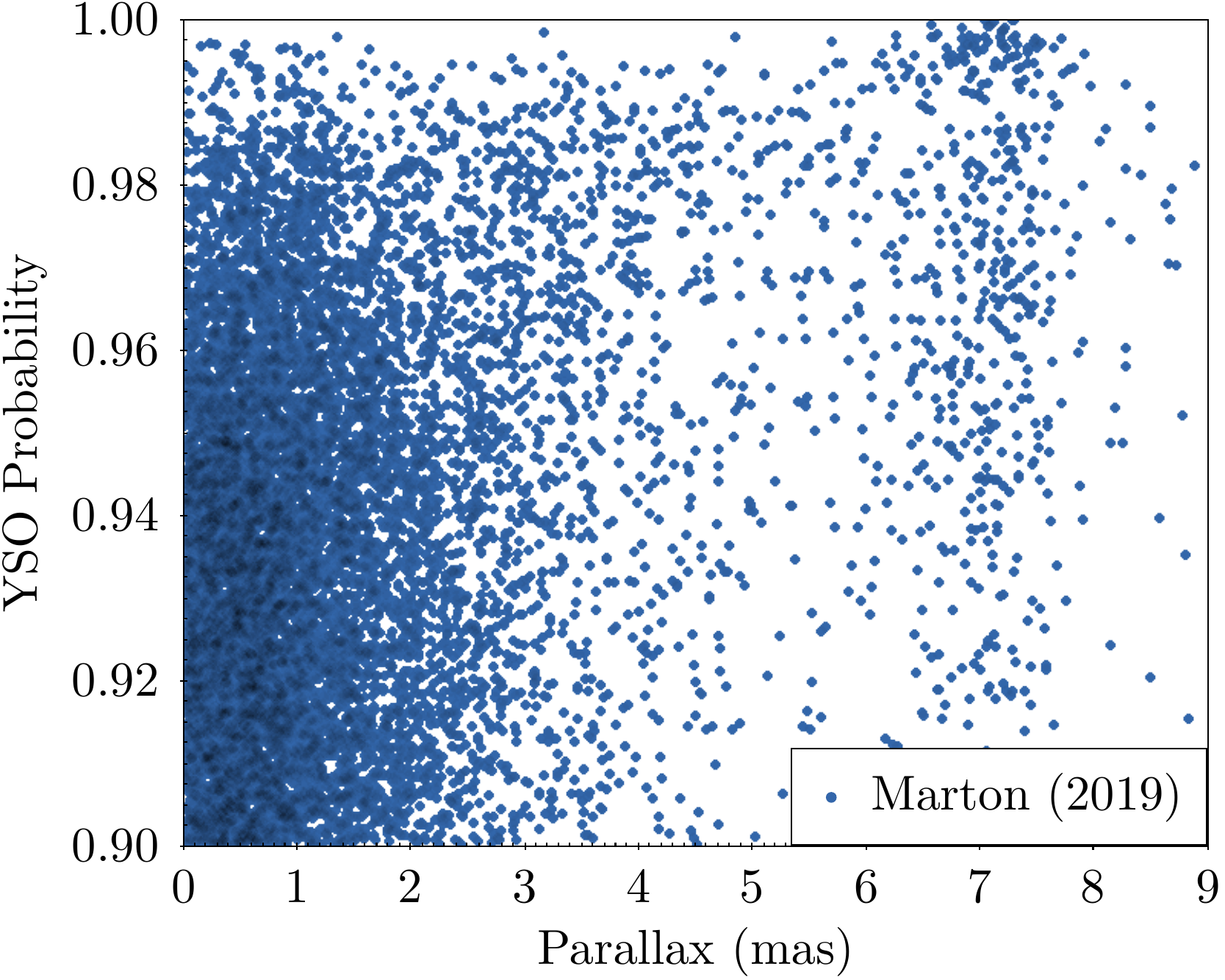}{0.33\linewidth}{}
    }
    \caption{Comparison of distribution of parallaxes for sources towards $\rho$ Oph and Upper Sco star-forming region as identified by \cite{marton2019} and Sagitta (this work) at different thresholds. Sources with small parallaxes are likely contamination due to extinction.
    \label{fig:martonhist}}
\end{figure*}

As mentioned in Section \ref{sec:classifiervalidation}, $\rho$ Oph is a particularly useful region for evaluating contamination. We compare the performance of \textit{Sagitta} versus the \citet{marton2019} catalog in Figure \ref{fig:martonhist}. The left panel shows the distribution of parallaxes of both models identified with likelihood $>95$\% towards that star forming region. Notably, while the classification from \cite{marton2019} does recover some sources at the appropriate parallax for the region ($\sim$7.5 mas), the vast majority of the stars classified with high confidence in their Gaia-ALLWISE model as YSOs have distances that are more consistent with reddened background stars. In contrast, with the same confidence threshold, our classification identifies a larger number of bona-fide YSOs in the appropriate parallax range overall, with only a small degree of contamination. Examining the Sagitta classifications using different confidence thresholds, and using the parallax information to assess the reliability of the classification, we see the number of likely contaminants decrease and the number of bona fide PMS candidates increase as the threshold increases. On the other hand, in the Gaia-ALLWISE catalog, the overall fraction of contaminants to bona fide YSO candidates remains mostly flat throughout the entire probability distribution, and a marginal degree of confidence is not achieved until $>99$\%.

A similar situation persists in the other star forming regions. The Gaia-ALLWISE catalog recovers many more candidate members of these regions than prior censuses have found, spread mostly uniformly within the outline of the clouds, regardless of the intrinsic underlying density distribution of those populations. As such, it is likely that their machine learning approach shows overreliance to extinction as a proxy for youth.

In total, our evaluation catalog recovers only 11,654 sources at $>$70\% confidence threshold in common with the catalog of \citet{marton2019} for the sources they classify as YSOs at 90\% confidence threshold. 

We note that in large part, the contamination in the Gaia-ALLWISE catalog is driven by noisy (and possibly partially mislabeled) data in the training sample, which then propagates to the noise in the predictions \citep[to a lesser degree, this is also the issue in the work of][for distant sources]{chiu2020}. The issue is further compounded in applying a trained model to very uncertain data. When we analyze the full catalog of YSOs from \citet{marton2019} that they classify at 90\% confidence, out of 1.7 million stars, Sagitta would assign 1/4 of them $>70$\% confidence, and 1/10 of them $>90$\% confidence, and the outputs from Sagitta would also be strongly contaminated by red giants with a large parallax error. As various data-driven classifiers are becoming more commonplace, this highlights the importance of rigorous vetting of the data that are processed by machine learning algorithms (both in training and in evaluating), to fully understand the limitations and ensure that these algorithms are not applied indiscriminately. 

Similarly, we examine the catalog from \cite{marton2016}, where they used Support Vector Machine methods to identify YSO candidates from WISE data alone, as that work predates the release of \textit{Gaia} DR2. This catalog contains 133,980 Class I \& II and 608,606 Class III YSOs, most of which are too reddened to be detected in the optical regime, making them difficult to evaluate. For the sources for which parallaxes are available, a significant fraction of them do appear to be reddened supergiants. Although, unsurprisingly, even in the subset of sources with optical emission, contamination in Class I \& II sources is somewhat less prominent than it is in Class III, as the former tend to have peculiar colors from the protoplanetary disks, compared to the latter which are just naked photospheres. In total we only recover 3,722 sources from this catalog compared to our evaluation catalog with the threshold of $>$70\%, or 17,472 without any quality checks on the data within the same threshold.

\subsubsection{Vioque et al. (2020)} 
\cite{vioque2020} used machine learning techniques to search Herbig Ae/Be stars using \textit{Gaia}, 2MASS, \textit{Sloan}, IPHAS, VPHAS+, and WISE photometry. They identify 8,470 candidate PMS stars, 693 classical main sequence Be stars, as well as providing a list of 1,309 sources that could belong to either type with above 50\% probability.

Their selection criteria is preferentially sensitive to the stars that are more massive than the sources we are able to identify as PMS candidates in this work. Furthermore, based on the availability of IPHAS and VPHAS+ data, they are restricted to only $\sim1^\circ$ within the Galactic plane. Because of this, our classifier identifies only a few sources in common with this catalog. From their catalog, we classify 3500 as PMS with Sagitta within 80\% threshold. These sources tend to be very young, with an average predicted age of approximately 5 Myr.

\subsubsection{Kuhn et al. (2020)}

Recently, \citet{kuhn2020a} have performed a data-driven selection of dusty YSOs from the Spitzer data across the Galactic plane. As their catalog is primarily focused on very reddened sources that do not necessarily have reliable \textit{Gaia} astrometry, the overlap of their selection with our evaluation catalog is minimal, only 456 stars. Similarly, of 36,423 sources that do have optical counterparts, regardless of data quality, our classifier would flag only 37\% of these sources as likely pre-main sequence with $>$70\% confidence. Nonetheless, the catalog does appear to be robust and complementary to our selection, identifying preferentially younger stars and providing a more complete selection, particularly in the more distant star forming regions, including some that we only barely recover (e.g., Sco OB1).

Nonetheless, the age estimator part of \textit{Sagitta} does appear to work well on this catalog, resulting in an average age of $\sim$4 Myr. Furthermore, it does appear to reveal some coherent age gradients in these star forming regions.

\section{Discussion}\label{sec:discussion}

\subsection{Local Bubble \& Gould's belt}

\begin{figure*} 
\plottwo{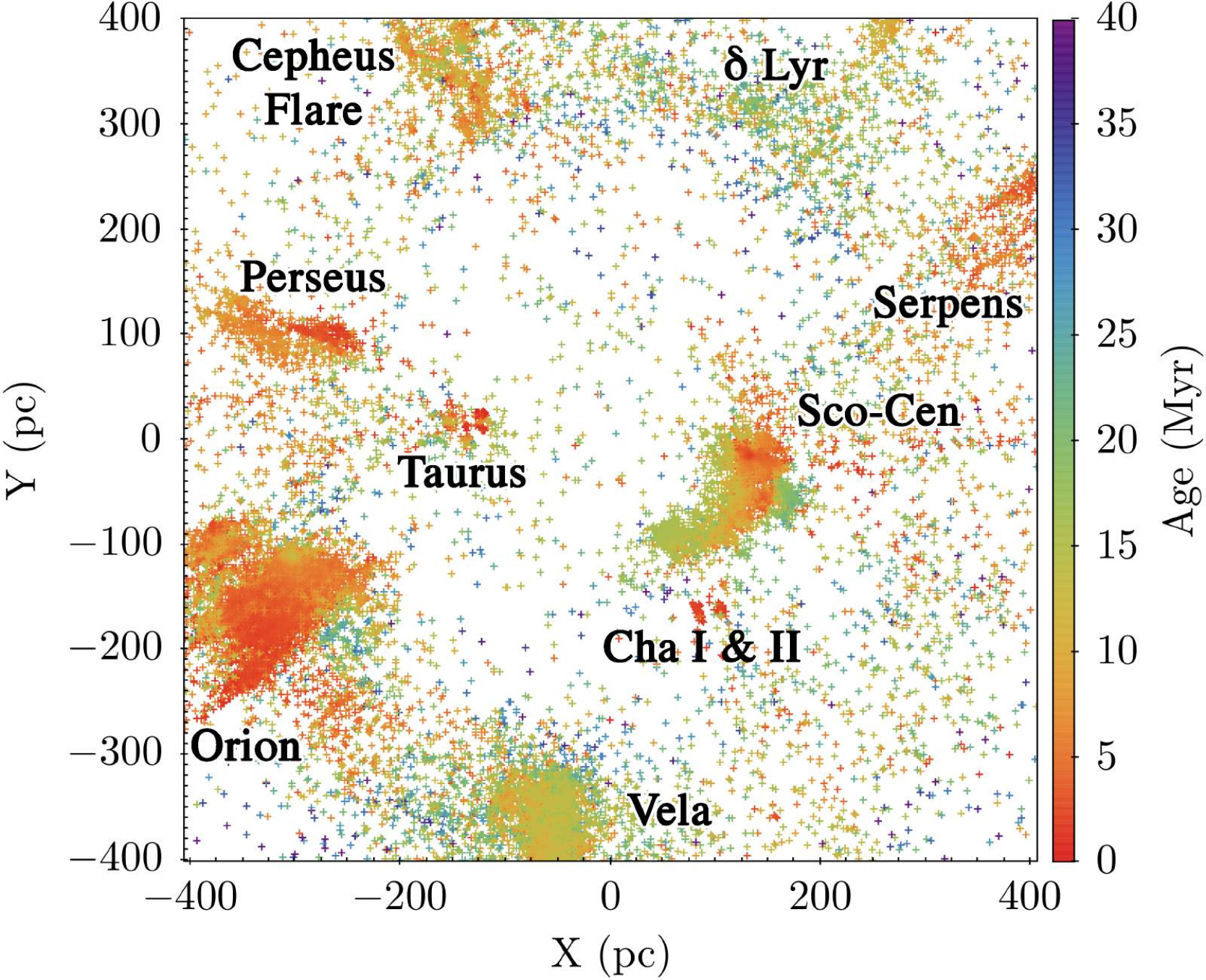}{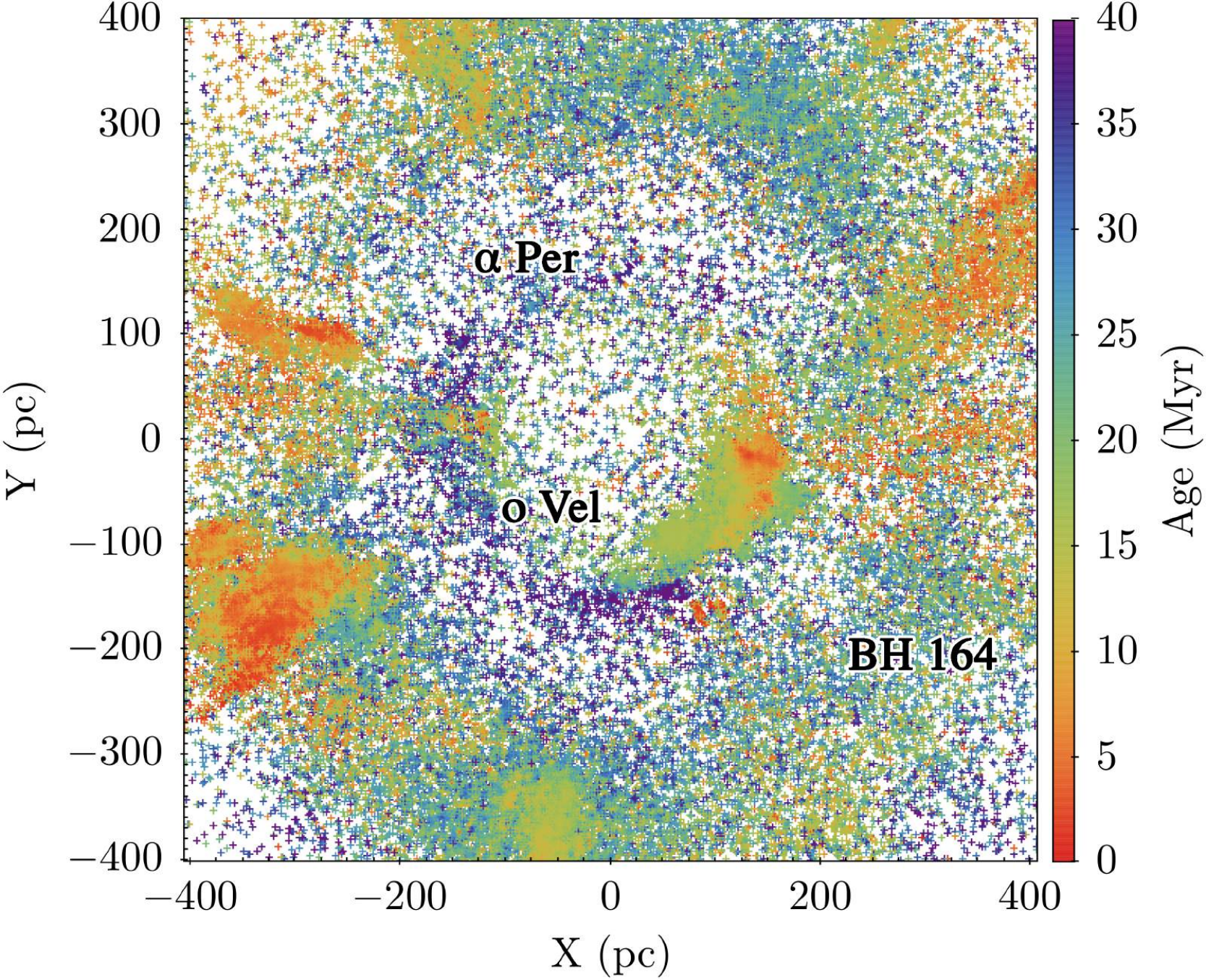}
\plottwo{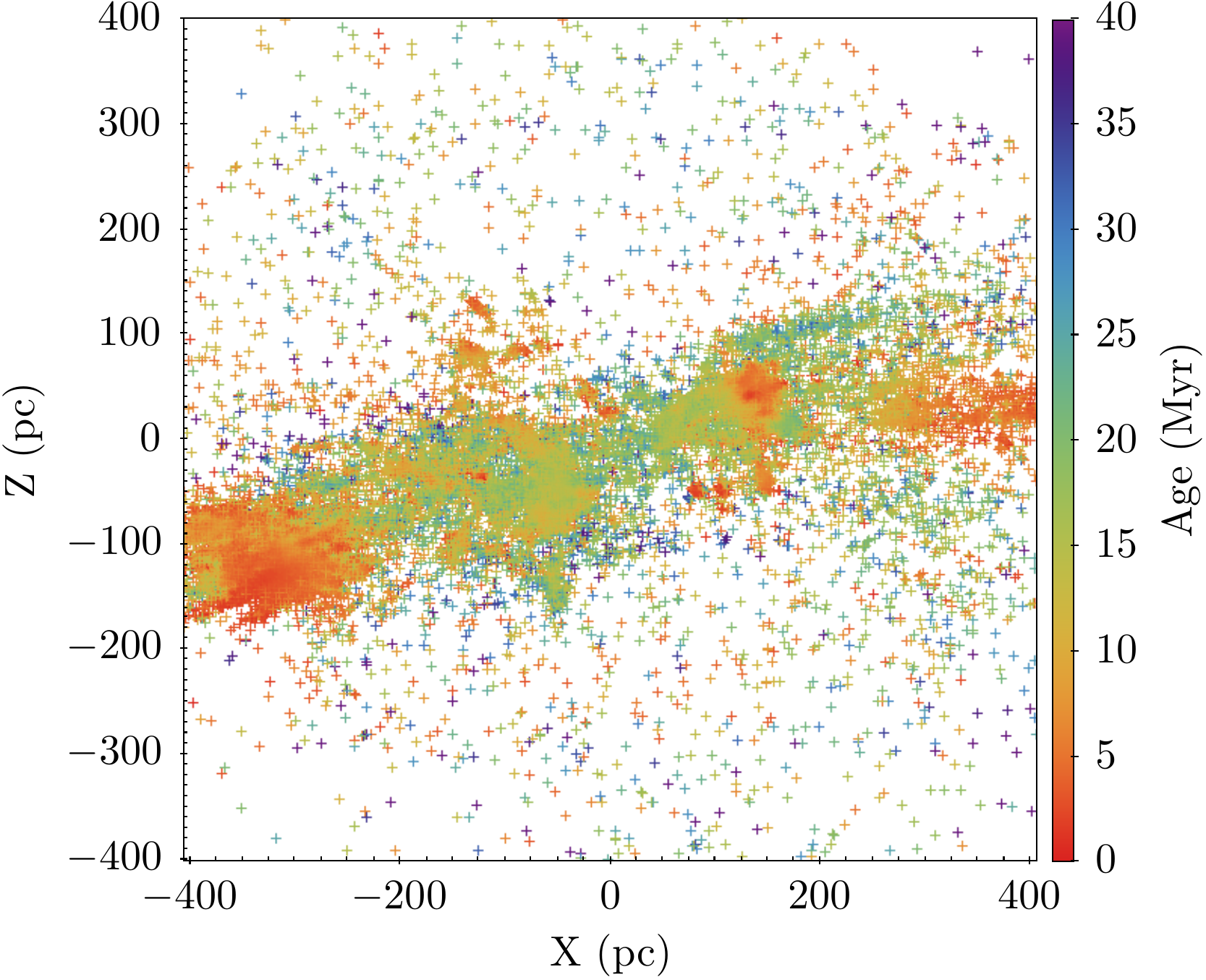}{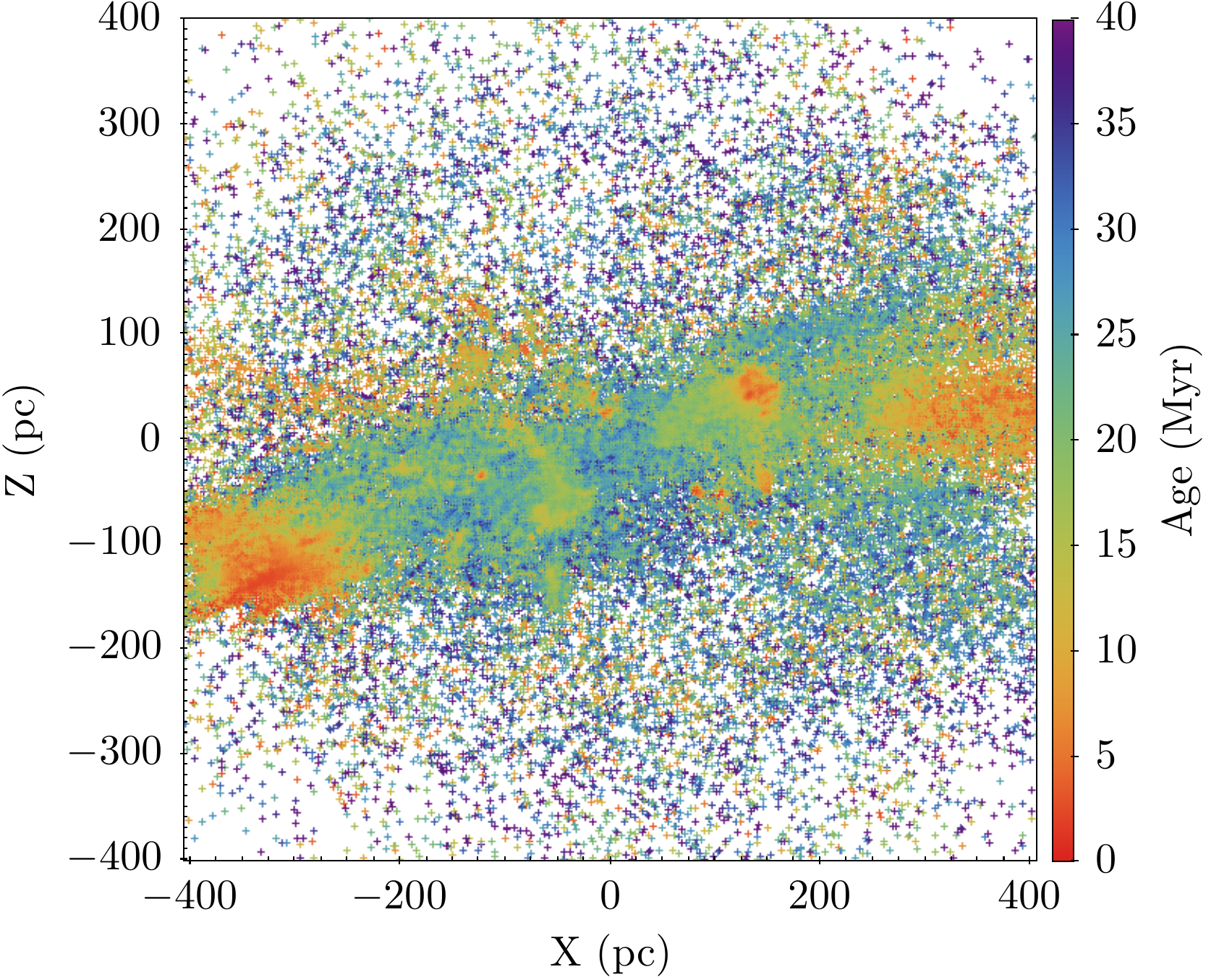}
\caption{Distribution of PMS stars (up to a confidence thresholds of 95\% - left, and 85\% - right) in the heliocentric rectangular reference frame, color coded by age, as a demonstration of the young stars tracing the outline of the Local Bubble. In X-Z and Y-Z projections, the distribution of PMS stars is largely planar, following the tilt of the Gould's Belt (See also Figure \ref{fig:circ}).\label{fig:localbubble}}
\end{figure*}

Gould's belt has been a long recognizable feature of the solar neighborhood, showing the apparent tilt of star forming regions, such as Sco-Cen, Orion, Taurus, and Perseus relative to the Galactic plane. Over the years, there have been a number of interpretations to the causes of this tilt, whether it is caused by a series of supernovae eruptions \citep{poppel2000}, or a collision of some sort with the disk \citep{comeron1994,bekki2009}.

\begin{figure*} 
\plottwo{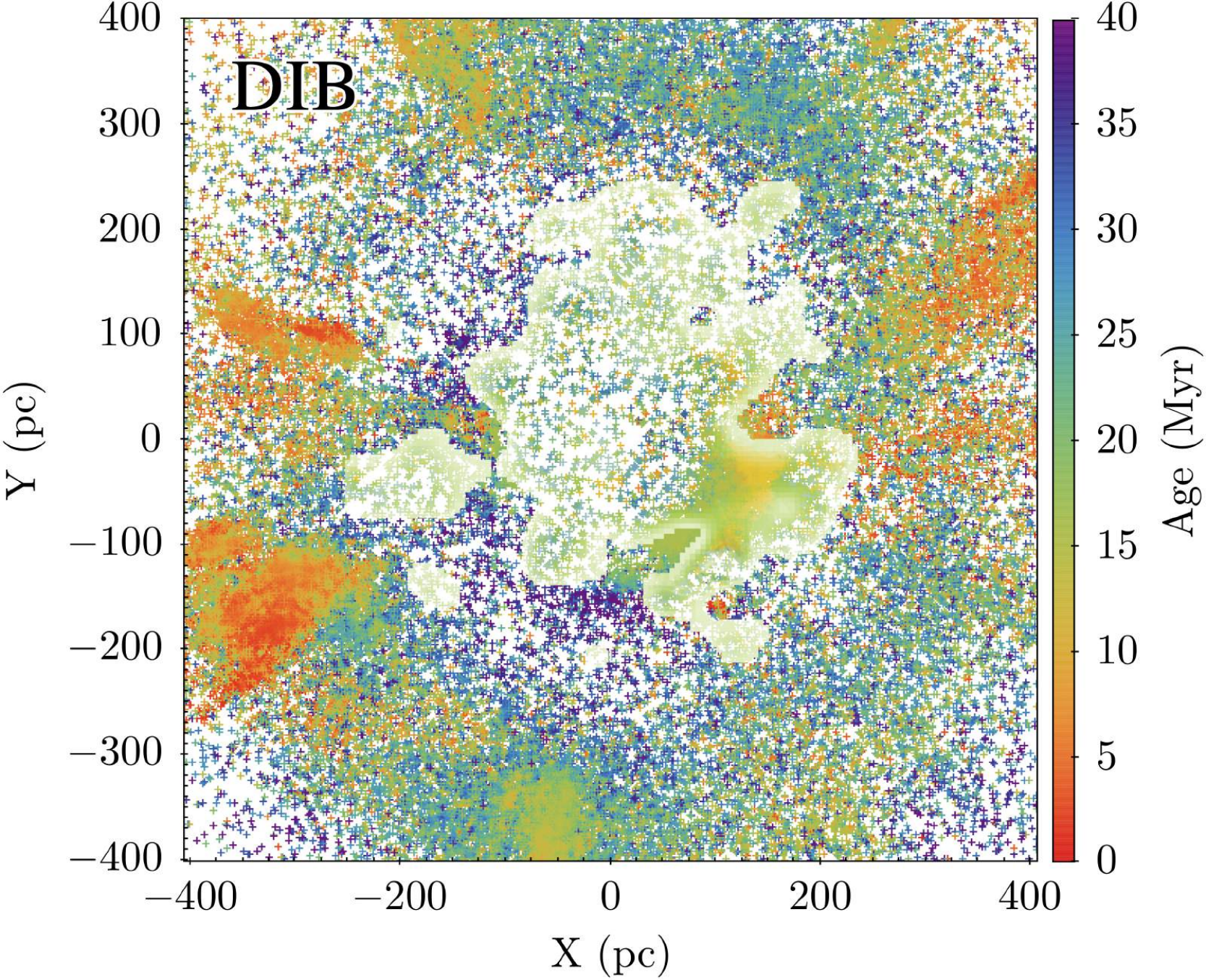}{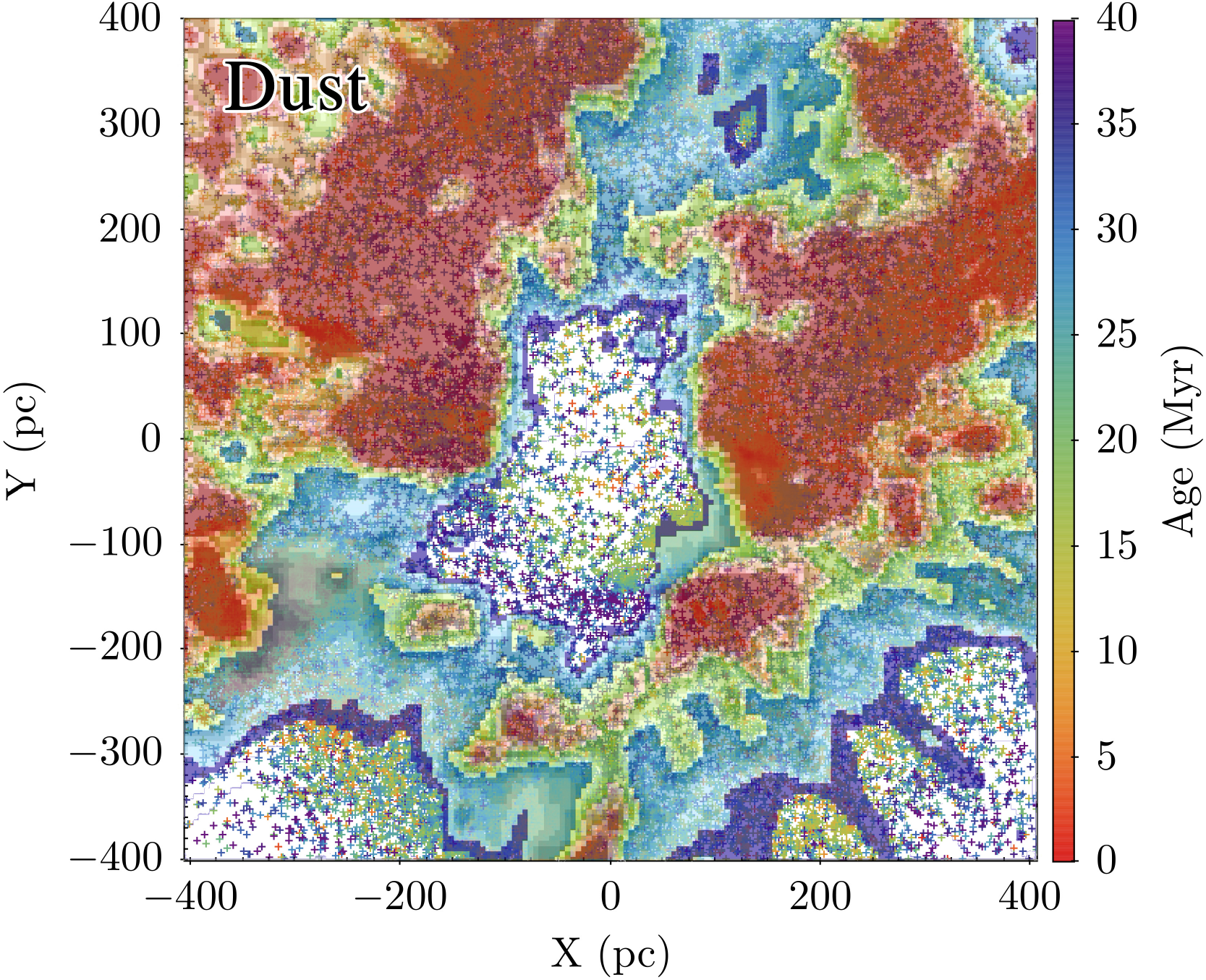}
\caption{Tracers of the Local Bubble superimposed at 85\% confidence map of the young stars in the solar neighborhood. Left: $\lambda$5780 diffuse interstellar band, adapted from \citet{farhang2019}. Right: dust clouds map from Lallement, R. in prep.} \label{fig:localbubble2}
\end{figure*}

Recently, some of the populations on one side of the Gould's belt (such as Orion) have been associated with the Radcliffe Wave \citep{alves2020} - a 2.7 kpc long structure that has a number of ripples protruding from the Galactic plane, up to 160 pc in amplitude. However, no interpretation has been given as a cause for these ripples. In \citetalias{kounkel2020} we have recovered a number of populations along the Radcliffe Wave. Although it appears to form a backbone to the Local Arm, it is made up only of the young populations - those older than 10-20 Myr no longer trace it. As such, it appears to be neither a standing nor a travelling wave, the formation of the ripples had to have been a recent phenomenon.

\citet{zari2018} have analyzed the distribution of stars younger than 20 Myr and found no evidence for a fully connected Gould's belt, rather that all of the individual populations appeared to be unrelated. Similarly \citet{bouy2015} have used a census of nearby OB stars and arrived to similar conclusions. With an improved census of PMS stars that extend towards the older ages we seek reexamine this.

Figure \ref{fig:localbubble} shows the top down map of PMS stars. There is a clear ring-like structure with the radius of $\sim$100--150 pc that connects Sco--Cen and Taurus as well as some of the older populations, such as $\alpha$ Per/Cas Tau OB association. Although hints of a complete ring are seen in stars $<20$ Myr old (particularly tracing the inner rim), it is the older stars that define this ring most clearly, which may be part of the reason why \citet{zari2018} have not identified it in their data.

Furthermore, beyond this ring, there appears to be a gap in the 3-dimensional distribution of PMS stars at the distances of $\sim$200--250 pc, with other populations becoming more prominent at distances of $>$300 pc.

This structure does not appear to be artificial, persisting even at higher confidence levels. Furthermore, as discussed in Section \ref{sec:classifiervalidation}, it cannot be attributed to the classifier favoring a particular set of distances in a truly uniform distribution of stars as the classifier does not necessarily intrinsically favor any specific distance in either the recovery fraction or in contamination (Figure \ref{fig:dist}). Although the coincidence of the two rings with the distance of massive populations, such as Sco Cen and the Orion Complex is suspect, these two rings do trace a number of open clusters and moving groups from \citep{kounkel2020} and \citep{cantat-gaudin2020}. Only a few of the previously known populations with the appropriate age to be recovered by Sagitta fall into the gaps of those rings. Furthermore, excluding the sources younger than 7.5 dex (corresponding to the most well established populations) disjoints the inner ring, leaving a considerable gap where Sco-Cen is located. If the overdensity of sources forming the ring was solely due to contamination from the older field stars due to an extra sensitivity at a given distance, it would be expected that this contamination would persist along all $l$.

Alternatively, rather than an enhancement in recovery, it is worth considering whether the two rings can be caused by a suppression in detection at distances of $<100$ and 200--300 pc. Indeed, Figure \ref{fig:upScoDistProb} does show a slight dip in the probabilities of YSOs if artificially placed at those distances. The suppression is only slight, however, generally most of these sources would be recovered within the same threshold.
Indeed, for the rest of the sample, artificially shifting the sources we identify in intermediate or high confidence thresholds to the location of these gaps still allows us to identify almost all of these sources.
As such, if the two rings were caused by this suppression, the gaps would be expected to fill up if we consider the sources from the lowest quality sample. This, however, is not the case -- the gaps do persist in the entire sample, regardless of the chosen threshold. Furthermore, as previously mentioned, no known young populations or clusters with ages $<$40 Myr breach these gaps.

To determine the effects of parallax on classifier recovery of the two ring structures, we selected all sources within the inner (radius $\sim 150$ pc) and outer ($\sim 300$ pc) rings as subsets of our overall catalog and synthetically shifted their parallaxes to distances roughly within the gap between the rings by respectively adding and subtracting 100 pc from their observed positions, then recalculated their apparent magnitudes as would be observed at these distances retaining their prior extinctions. We then recalculated these sources' Sagitta classifier certainties with these synthetic inputs. 

\begin{figure} 
\plotone{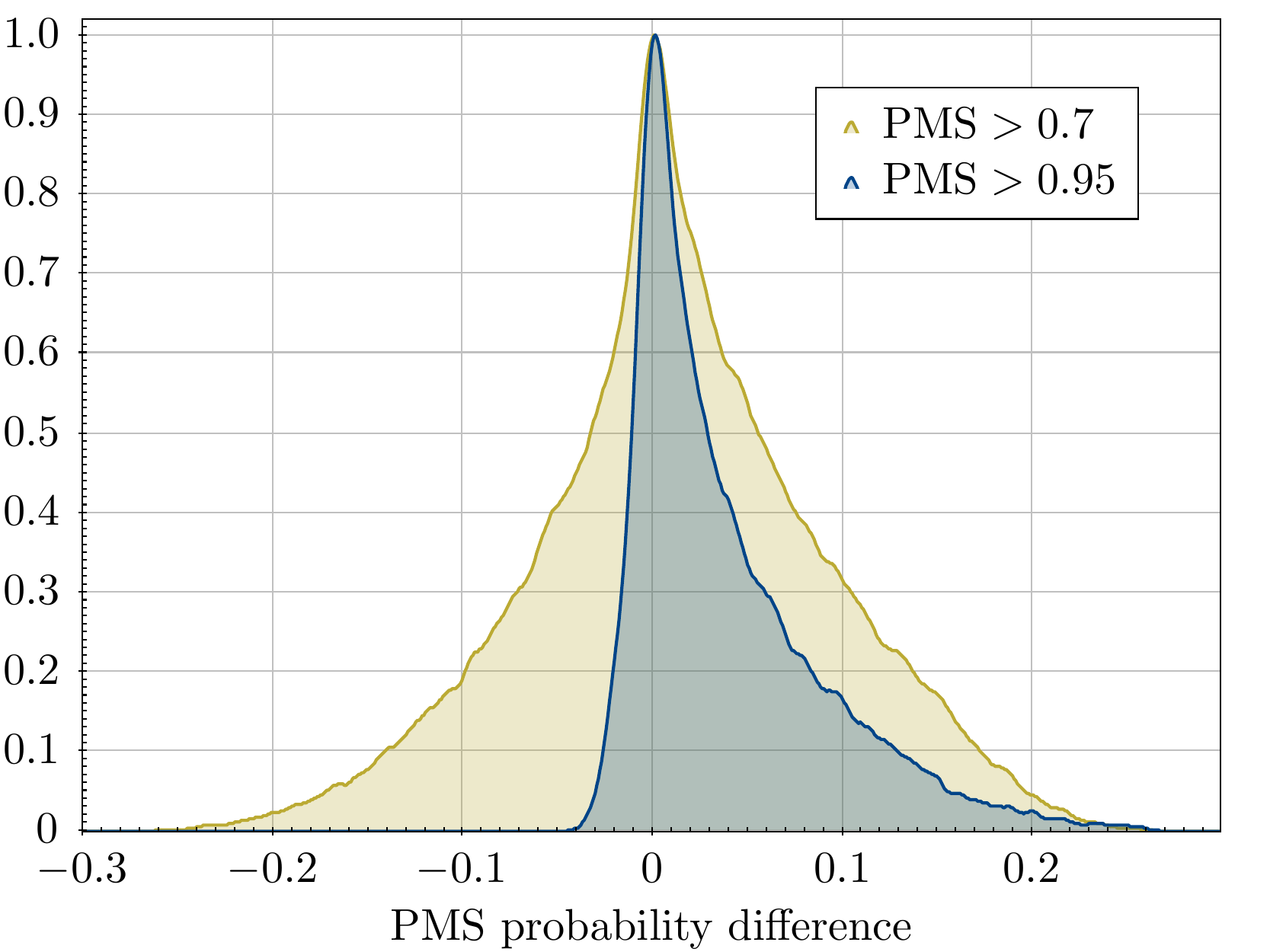}
\caption{Kernel density estimation showing the difference in the classification probability between the sample at the original distances, and the sample shifted at the distance of the gap between the two rings. Yellow curve shows the distribution for the full sample above 70\% threshold. Blue curve shows only the high confidence ($>$95\%) sample. The distribution is asymmetric as it is impossible to achieve confidence $>$100\%.} \label{fig:diff}
\end{figure}

Even with shifted distances, the recalculated YSO certainties reproduced the two ring structures fairly well. While some scatter is introduced into the overall distribution, the makeup of both rings are largely maintained, including the stars with older ages between known younger populations. Of the selected sources in the inner and outer rings, 70\% of sources which initially had classifier certainty of 95\% or higher and 80\% of sources which initially had certainty of 85\% or higher were recovered in the same threshold when shifted. Moreover, for stars initially in the $>95$\% threshold, $\sim 90$\% were recovered within the shifted $>85$\% threshold. (Figure \ref{fig:diff}) Based on these results, these two ring structures appear to be an accurate representation of the distribution of SFRs in the solar neighborhood rather than systematic pattern.

Outside of very low mass populations such as TW Hya (which we do recover), a lack of significant young star forming regions within 100 pc has long since been known. Indeed, the Sun is located near the center of the Local Bubble, a cavity of significantly lower density of neutral hydrogen compared to what is typically found in the interstellar medium. Using various tracers, such as diffuse interstellar bands \citep[that can trace the X-ray dissociation within the Bubble][]{farhang2019} or extinction \citep{lallement2018,leike2019} to trace the morphology of the local bubble as in Figure \ref{fig:localbubble2}, it fits well within the identified ring of stars. 

It has been noted Ophiuchus and Taurus have velocities that are comparable in magnitude, but opposite in direction, both moving with local standard of rest radial velocity (LSR RV) of +5 \kms\ away from us. They would have been in proximity of one another $\sim$20-25 Myr ago \citep{rivera2015}. This trace back age is comparable to the average age of the stars in the ring, although there are also a number of $\sim$40 Myr old stars that compose it.

We examine the available LAMOST radial velocities for the sources in our catalog. To exclude various well-characterized regions (which may systematically skew the distribution due to their density of stars), we examine the sources that are located at high galactic latitudes, associated with the less populated spherical shells near these rings. Specifically, we limit the catalog to the sources with $b<-30^\circ$ or $b>10^\circ$, as well as $\pi>2$ mas.

\begin{figure} 
\plotone{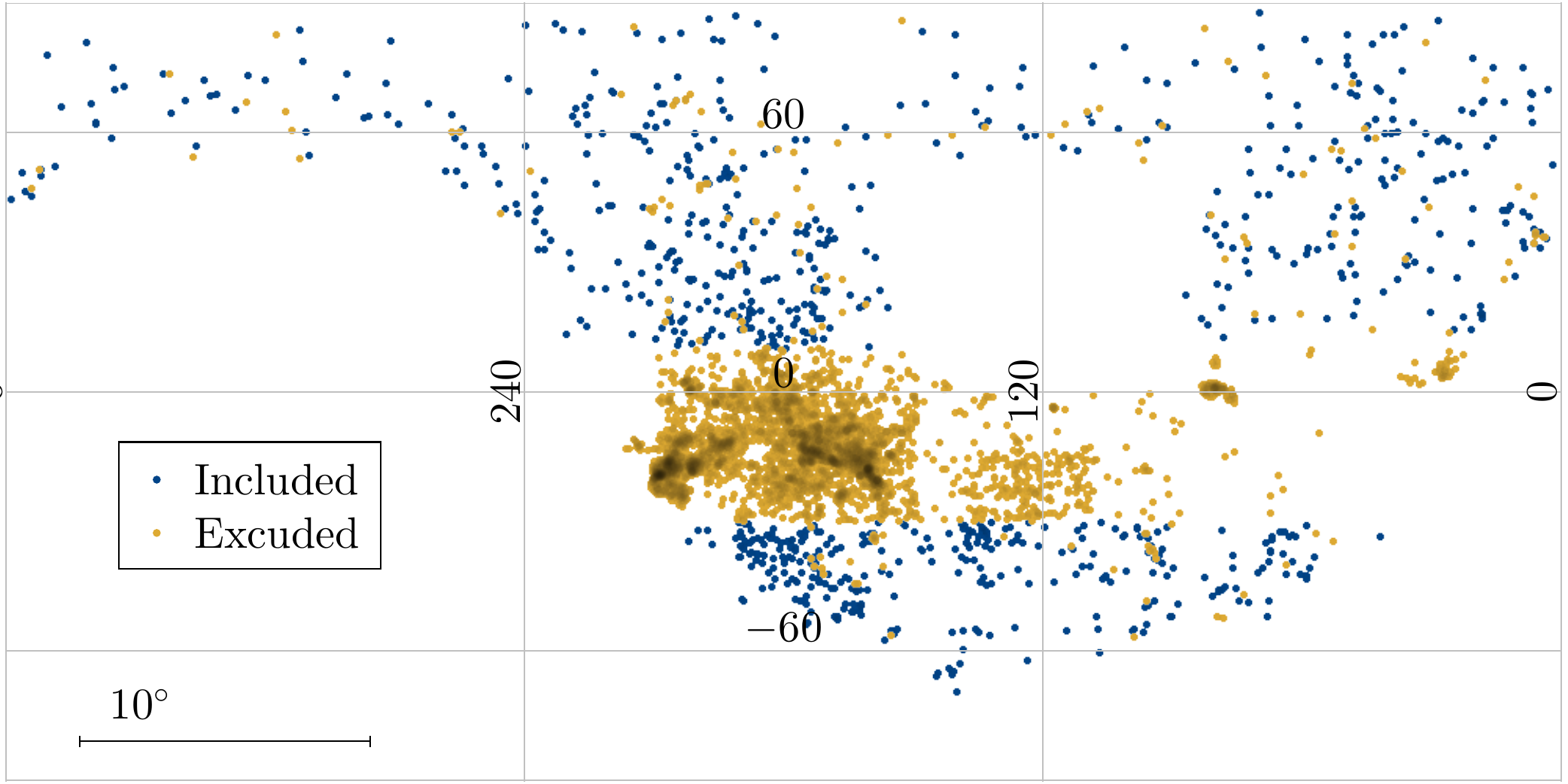}
\plotone{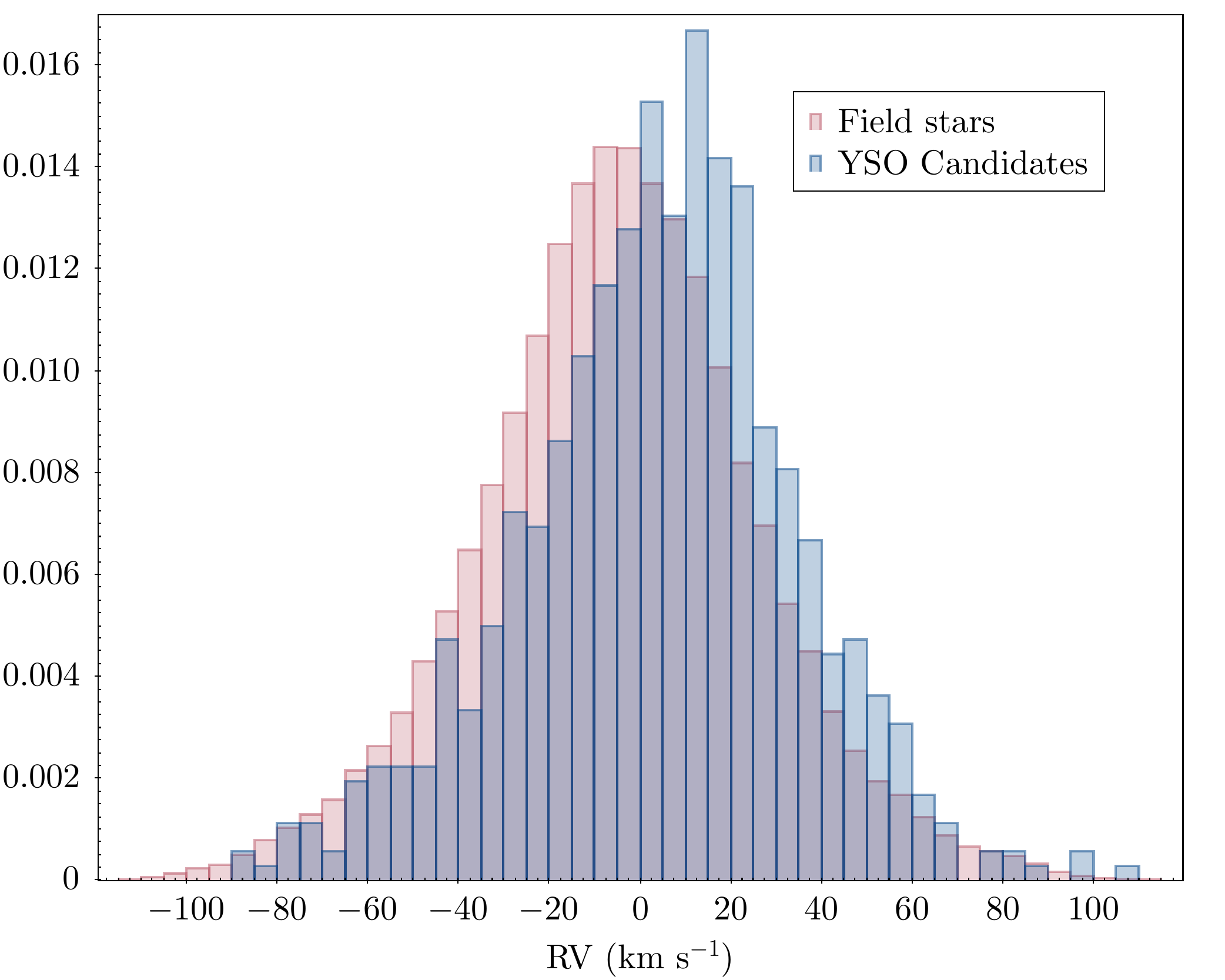}
\caption{Top: Spatial distribution of sources (in galactic coordinates) selected by Sagitta in this work that have been observed by LAMOST DR5. The sources were selected using criteria of $b<-30^\circ$ or $b>10^\circ$, as well as $\pi>2$ mas, to exclude the overdensities in Taurus, Perseus, and Orion, which could skew the RV distribution. Bottom: Distribution of local standard of rest RVs of the selected nearby high galactic latitude sources in comparison to the random field stars (which tend to be old and relaxed) satisfying the same spacial cuts. The histograms are scaled by the area. Note that the evolved field stars at these high galactic latitudes tend to show a small blueshift as they oscilate back towards the Galactic plane, and the PMS candidates are preferentially redshifted, consistent with a signature of an expanding bubble.} \label{fig:lamost}
\end{figure}

The typical LSR RVs of the field stars observed by LAMOST withing the same footprint are $-4.5\pm30$ \kms, i.e., there is a slight preference for the stars to be moving towards us - as they are falling back towards the midplane from a larger height above it. On the other hand, examining RVs of the YSO candidates have mean LSR RVs of $+5$ \kms (Figure \ref{fig:lamost}), which is similar to the typical LSR RVs of Taurus and Ophiuchus. Although there is likely some contamination in the catalog, the RVs appear to be dominated by a clear signature of an expanding bubble. Applying a Kolmogorov-Smirnov test, the probability that the two populations come from the same parental distribution is $\sim10^{-16}$, which is sufficient to reject the null hypothesis at $\sim8\sigma$ level. Separating the sample into the two rings, the inner one ($\pi>5$ mas) alone can reject the null hypothesis with $P\sim10^{-7}$, and the outer one ($\pi<5$ mas) with $P\sim10^{-8}$. This spherical expansion can account for the bulk of the sources found at high Z in Figure \ref{fig:localbubble}.

As more and more optical and near-infrared spectra of the PMS candidates become available, it should be possible to not only more unequivocally separate the true YSOs from possible contaminants, but also fully characterize 3d spacial motions of these stars. This would allow a more in-depth analysis on the origin and the dynamical evolution of this structure.

Recently, \citet{kerr2021} have performed a detailed analysis of the star-forming history of the Sco-Cen association. Similarly to what we observe, they find that Sco-Cen has a semi-circular arc morphology, and that there is sequential star formation within the Complex, which may be indicative of the history of triggering, with the velocity of propagation of $\sim$4 \kms. This velocity is consistent with the velocity of the expansion of the Local Bubble.

At a current glance, this appears to be an effect of a supernova explosion. It should be noted that different populations do appear to have a somewhat different peculiar velocity relative to one another, at least in the proper motion space. Thus most likely, instead of shockwave clearing a gas of a particular population \citep[as has been the case of supernovae in young star forming regions such as Orion, and potentially Vela;][]{kounkel2020a,grossschedl2020,cantat-gaudin2019} the shock front associated with the expanding bubble may have rammed into the neighboring clouds, not dissimilar to a scenario described by \citet{inutsuka2015} in which molecular clouds trace interaction regions between even shorter lived bubbles. As such, the formation of the Local Bubble may be a cause of one of the ripples along the Radliffe Wave, and that other supernova driven superbubbles in general may be a mechanism for other such ripples.

The Local Bubble does not immediately explain the lack of star forming regions at the distances of $\sim$200--250 pc. Such a gap can be seen in the distribution of present day molecular clouds \citep{zucker2020}, and it is also present in the catalog from \citet{zari2018}. Although it is not impossible, it would be surprising for two unrelated events to occur in the vicinity of the Sun to form two separate rings at $\sim100$ pc and at $\sim$300 pc. Instead, it may be possible that they are produced by a related event, possibly caused by multiple shock fronts. This may suggest a common origin for the populations that are a part of the Gould's belt.

\subsection{30 Myr Bubble}\label{sec:30MyrBubble}

\begin{figure*} 
\plottwo{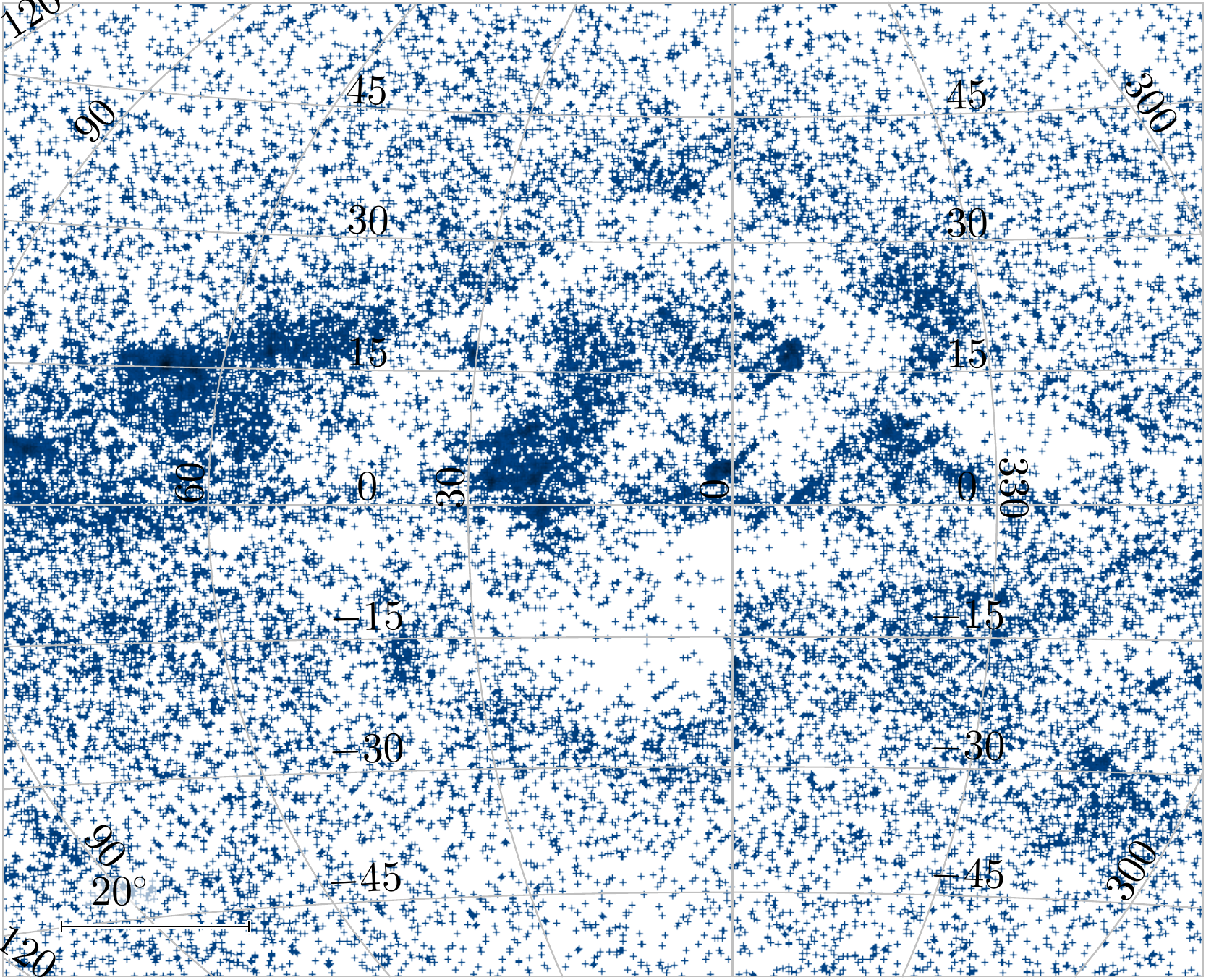}{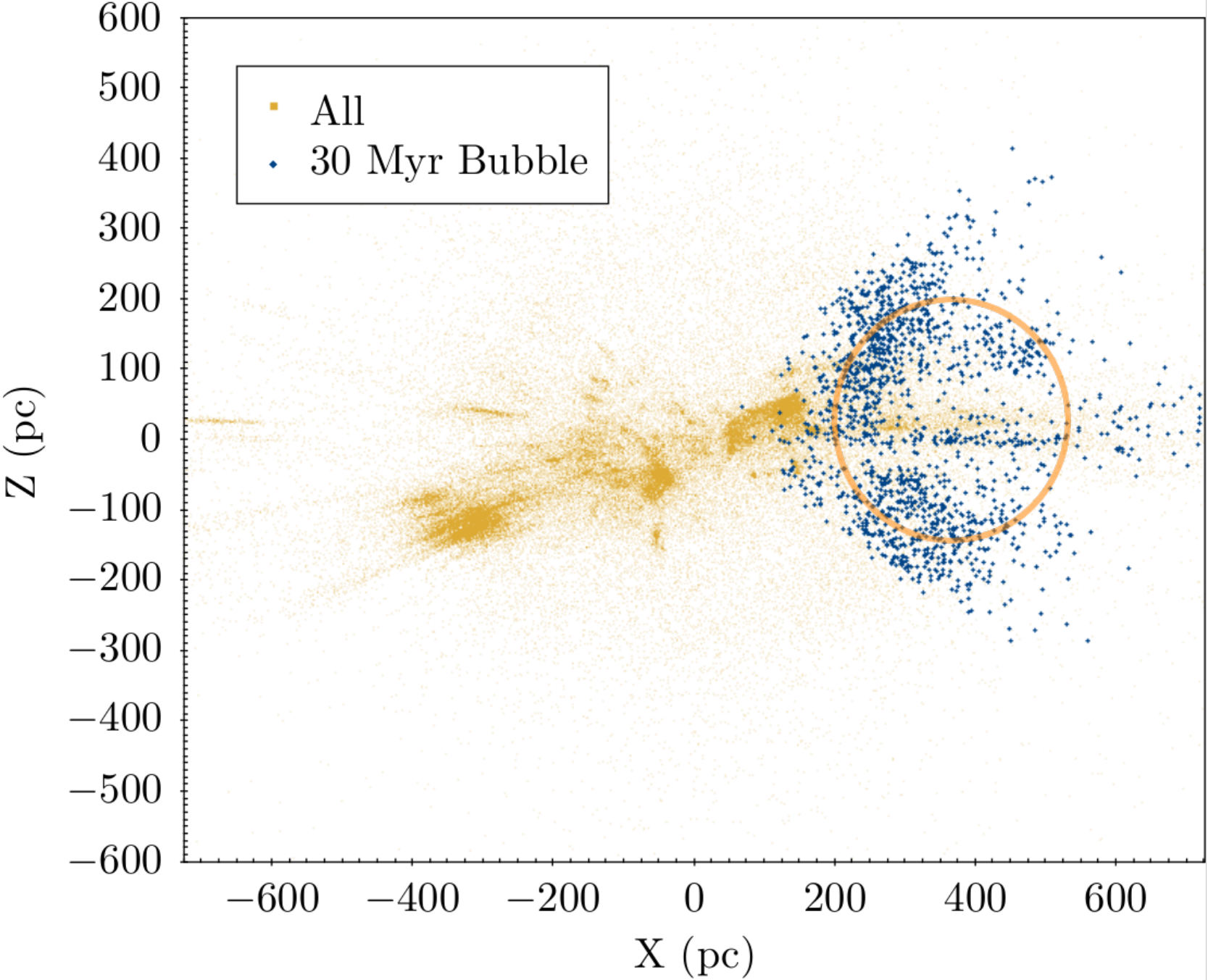}
\plottwo{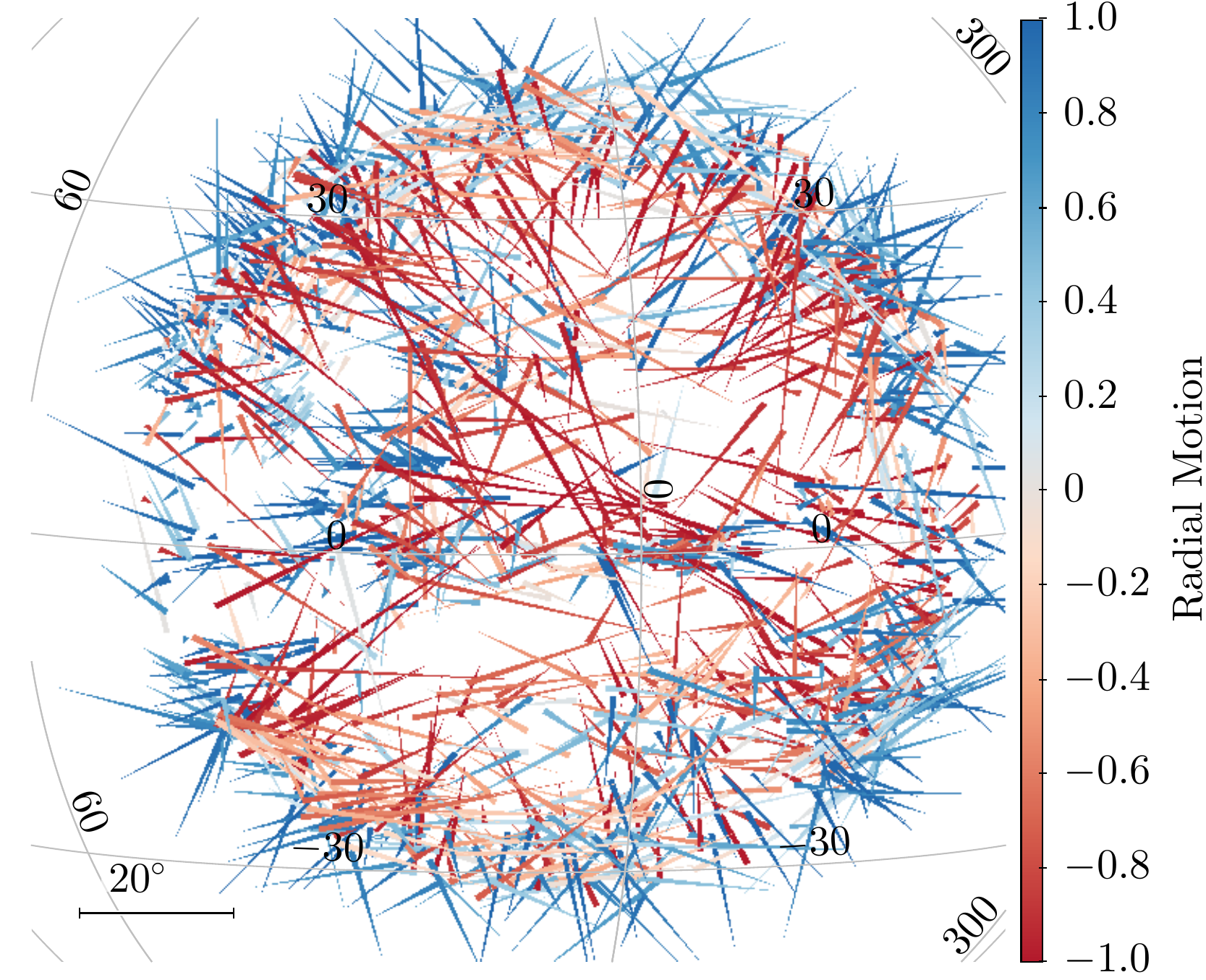}{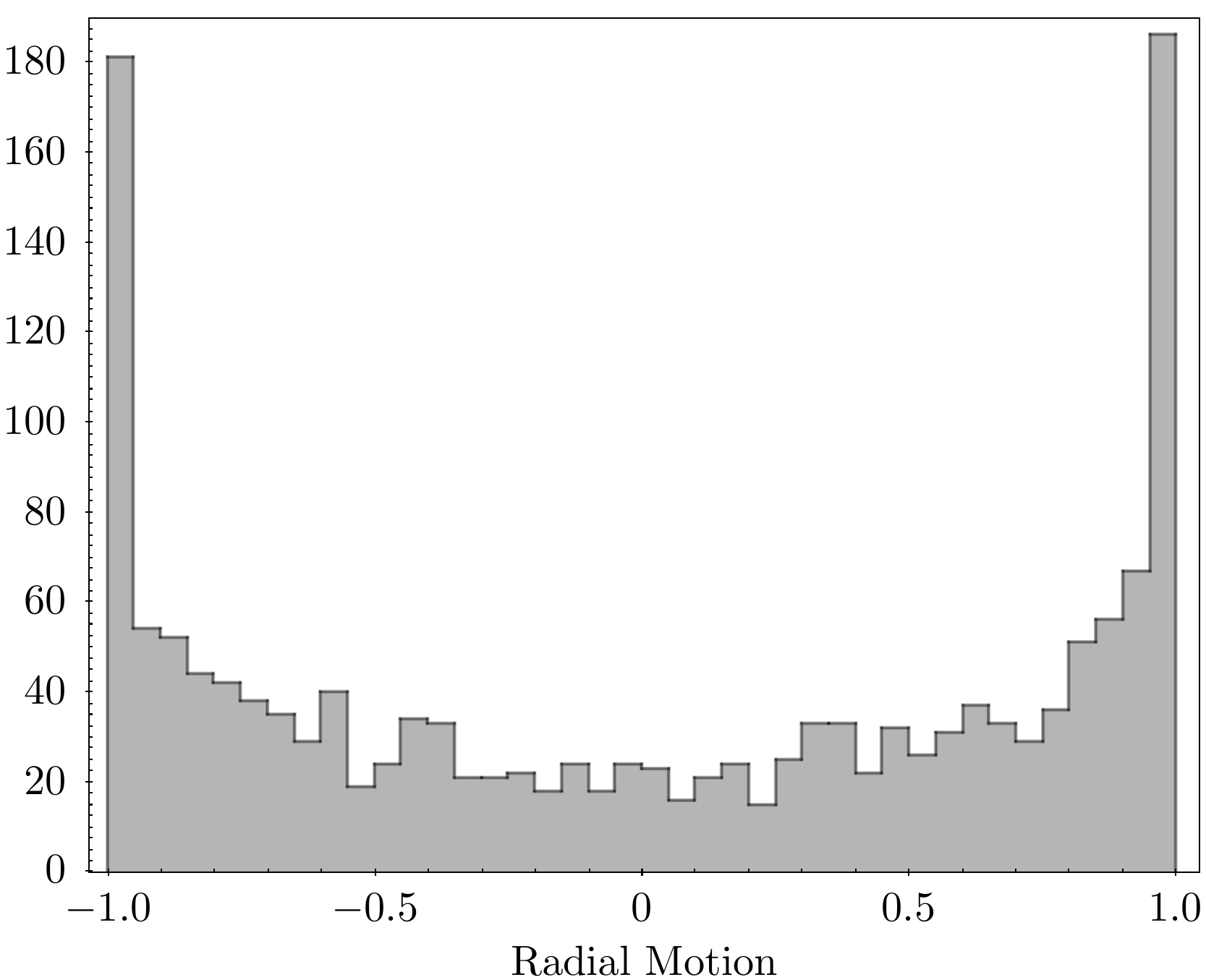}
\caption{Top left: Distribution of sources in the catalog towards the 30 Myr bubble, at the distance ranges of 200--400 pc. Top right: 3d distribution of stars of the 30 Myr bubble relative to the other stars in the catalog. Bottom left: distribution of the sources towards the bubble in the plane of the sky, with proper motions, in the local standard of rest, corrected for the average velocity of the stars, with the thick part of the arrow showing the current position of the sources. Arrows are color coded by the cosine of angle proper motions have relative to the center of the bubble, with red (-1) showing sources moving radially inward and blue (+1) showing those moving radially outward. Bottom right: a histogram showing the distribution of the aforementioned radial component of motion.} \label{fig:circ}
\end{figure*}

There is another peculiar bubble-like structure that can be identified in the data. It can be seen in the last panel of Figure \ref{fig:AgeRanges}, primarily traced by the stars older than 25 Myr, towards the direction of the Galactic center. This bubble, about $\sim40^\circ$ in radius, appears to begin at the distance of $\sim$200 pc and forming a hemisphere $\sim$400 pc in diameter (Figure \ref{fig:circ}, top). Unfortunately the distant edge of this bubble is difficult to detect as at those distances we lose sensitivity to low mass stars that can be used as tracers for this age range.

This bubble is not caused by extinction towards the Galactic center. Although there are a number of optically thick clouds in the volume of space associated with it (e.g., the Aquilla rift, for which we do recover a sizable population of stars in the 2--10 Myr range), the edge of the bubble is located far outside of those clouds with a gap between them of $>10^{\deg}$ in which only a handful of PMS candidates are present.

In analyzing the proper motions of the stars located on the other edge of the bubble, we identify a somewhat peculiar pattern corrected for the average velocity of all the stars in the sample. There is a strong preference for them to move either directly radially inwards towards the center (at $l\sim6^\circ$, $b\sim6^\circ$) or radially outward away from it, with next to no tangential component in the proper motions (Figure \ref{fig:circ}). It is unclear what could be the cause of such a signature. 

Based on the velocities of stars that are just expanding, we would estimate the expansion age of $\sim$13 Myr, or approximately half of their age.

\section{Conclusions}\label{sec:concl}

One of the outstanding questions in the star formation community is how can post T Tauri stars be identified. We present an automated method of identifying PMS stars and estimating their ages (up to $\sim$70 Myr) through \textit{Gaia} DR2 and 2MASS photometry using a neural network framework. This allows for a homogeneous analysis of large volumes of data characterizing star forming history of the solar neighborhood. Furthermore, this approach is not reliant on a kinematic selection, making it possible to search for kinematically peculiar young stars, such as runaways \citep[e.g.,][]{mcbride2019}.

Applying a classifier to a curated subset of \textit{Gaia} DR2 data with most reliable astrometry and photometry, we identify 197,315 stars as likely PMS sources with confidence of $>70$\%, and 448,824 stars in \textit{Gaia} EDR3 data. The code is made available on GitHub\footnote{\url{https://github.com/hutchresearch/Sagitta}}, to enable the usage of Sagitta outside of this curated subset.

Sagitta is robust against contamination, especially when compared to a number of previous studies that also aimed to identify young stars using optical and near infrared data. The precise confidence threshold that should be used to select candidate PMS stars in a particular region depends on the distance to and the age of the population that one seeks to characterize. 

The estimated ages that we measure are consistent with what has been previously measured in some of the better studied star forming regions. Furthermore, in many cases, they allow for a more granular look at the evolution of various populations than what was previously available. It should be noted, however, that caution should be expressed regarding the pre-main sequence binaries, as they may appear to be systematically younger than they are.

In examining the distribution of stars in the solar neighborhood, we identify various features. Most notably we identify a ring of stars at $\sim$100 pc with ages of up to $\sim$40 Myr, tracing the outer edges of the Local Bubble. It is likely that the formation of this bubble have lead to the formation of the Gould's belt. We also find a second bubble consisting of $\sim$30 Myr old stars in the direction towards the Galactic center.

In future, a follow up of the sample presented in this work by large spectroscopic surveys (such as SDSS-V APOGEE) would be of great benefit to confirming the candidates, as well as allowing for a better understanding of the dynamical and chemical evolution of PMS stars.

\software{TOPCAT \citep{topcat}, Pytorch \citep{pytorch}}

\begin{acknowledgments}

We thank Rosine Lallement for providing the latest dust map of the solar neighborhood. We thank Eleonora Zari and Keivan Stassun for giving feedback on the manuscript. A.M, M.K., and K.C. acknowledge support provided by the NSF through grant AST-1449476.
This work has made use of data from the European Space Agency (ESA)
mission {\it Gaia} (\url{https://www.cosmos.esa.int/gaia}), processed by
the {\it Gaia} Data Processing and Analysis Consortium (DPAC,
\url{https://www.cosmos.esa.int/web/gaia/dpac/consortium}). Funding
for the DPAC has been provided by national institutions, in particular
the institutions participating in the {\it Gaia} Multilateral Agreement.
Guoshoujing Telescope (the Large Sky Area Multi-Object Fiber Spectroscopic Telescope LAMOST) is a National Major Scientific Project built by the Chinese Academy of Sciences. Funding for the project has been provided by the National Development and Reform Commission. LAMOST is operated and managed by the National Astronomical Observatories, Chinese Academy of Sciences.
The authors thank the Nvidia Corporation for their donation of GPUs used in this work.

\end{acknowledgments}

\bibliographystyle{aasjournal.bst}
\bibliography{main.bbl}

\end{document}